\documentclass[letter,11pt]{article}

\usepackage{jcappub} 
\usepackage{siunitx}
\usepackage[T1]{fontenc} 
\usepackage{subcaption}
\subheader{MIT-CTP/5380}
\title{Luminosity functions consistent with a pulsar-dominated Galactic Center Excess}

\author{Jack T. Dinsmore}
\author{and Tracy R. Slatyer}
\affiliation{Center for Theoretical Physics, Massachusetts Institute of Technology, Cambridge, MA 02139, USA}

\emailAdd{jtdinsmo@mit.edu}
\emailAdd{tslatyer@mit.edu}

\newcommand{\parens}[1]{\left(#1\right)}
\newcommand{\brackets}[1]{\left[#1\right]}
\newcommand{\expp}[1]{\exp \parens{#1}}
\newcommand{\fraci}[2]{#1 / #2}

\DeclareSIUnit\erg{erg}
\DeclareSIUnit\parsec{pc}
\DeclareSIUnit\photon{photon}
\newcommand{\SIasym}[4]{#1^{+#2}_{-#3}\ \SI{}{#4}}
\newcommand{\numasym}[3]{#1^{+#2}_{-#3}}

\abstract{A new population of millisecond pulsars is a long-standing proposed explanation for the excess of GeV-scale gamma rays emanating from the region surrounding the center of the Milky Way (the ``Galactic Center excess''). We examine several simple parameterizations of possible luminosity functions for this population, as well as several benchmark luminosity functions proposed in the literature, and compare the predicted populations of resolved point sources to the {\it Fermi} 4FGL-DR2 point source catalog and a sub-population recently identified using wavelet-based methods. We provide general results that can be used to translate upper limits on the number of resolved point sources associated with the excess, and the fraction of the flux in the excess that can be attributed to resolved sources, into limits on the luminosity function parameter space. We discuss a number of important systematic uncertainties, including in the detection threshold model and the total flux attributed to the excess. We delineate regions of parameter space (containing existing benchmark models) where there is no apparent tension with current data, and the number of total pulsars needed to explain the excess is in the range of $\mathcal{O}(10^{4-5})$. Forecasting the effects of lowered point source detection thresholds, we show that novel analysis methods that probe sub-threshold point source populations can hope to resolve more than $30\%$ of the flux of the excess.}

\keywords{millisecond pulsars, gamma ray theory, dark matter experiments}

\begin{document}
\maketitle
\flushbottom

\section{Introduction}

 The Large Area Telescope (LAT) of the \textit{Fermi} Gamma Ray Space Telescope has detected an unexpected excess of gamma-rays emanating from the region surrounding the Galactic Center (GC) \cite{Goodenough:2009gk, HOOPER2011412, Gordon:2013vta}. This signal, known as the Galactic Center Excess (GCE), has a spectral energy distribution that peaks around $1-3$ GeV in $E^2 dN/dE$ 
 \cite{DiMauro:2021raz, Calore:2014xka, Zhong:2019ycb, Gordon13, Ajello:2015kwa, Ajello:2017opo}. Early studies found that the GCE spatial morphology was approximately spherically symmetric, suggesting a potential origin for the excess in dark matter annihilation, and the energy spectrum and overall rate can also be fit well by simple dark matter models \cite{Goodenough:2009gk, Gordon:2013vta, DAYLAN20161, Calore:2014nla, Abazajian:2014fta}.

However, the GCE energy spectrum and morphology also appear to be consistent with a population of largely unresolved Millisecond Pulsars (MSPs), within systematic uncertainties (e.g. \cite{Abazajian:2010zy, Gordon:2013vta, Yuan:2014rca, Calore:2014xka, Petrovic:2014xra}). A number of studies using alternate approaches to model the Galactic diffuse emission background have found that the GCE correlates better with the mass distribution of the Galactic bulge rather than the spherical distribution expected from dark matter annihilation \cite{Macias:2016nev, Bartels:2017vsx, Macias:2019omb, Abazajian:2020tww, Coleman:2019kax}, although Ref.~\cite{DiMauro:2021raz} arrives at the opposite conclusion. Studies of gamma-ray hot spots within the inner Galaxy, and/or the photon statistics of the GCE, have claimed detection of point sources (PSs) associated with the GCE \cite{Bartels:2015aea, Lee:2014mza, Lee:2015fea, Zhong:2019ycb, Buschmann:2020adf}, further supporting the MSP interpretation. However, recent work has shown that some strong earlier claims of PS detection were affected by systematic biases favoring the PS interpretation \cite{Leane:2019xiy, Leane:2020pfc, Leane:2020nmi, Collin:2021ufc}, and the interpretation of Ref.~\cite{Bartels:2015aea} as evidence for PSs that are part of the GCE has been questioned in Ref.~\cite{Zhong:2019ycb}. Machine-learning-based analyses have claimed evidence in favor of a PS component \cite{Caron:2017udl, List:2020mzd, List:2021aer, Mishra-Sharma:2021oxe}, but with modest statistical significance (and the potential for as-yet-unaccounted-for systematic errors): Ref.~\cite{Mishra-Sharma:2021oxe} attributes $38^{+9}_{-19}\%$ of the GCE to PSs (i.e. a roughly $2\sigma$ detection), and Ref.~\cite{List:2021aer} excludes a smooth (non-PS) fraction for the GCE exceeding 66\% at 95\% confidence. Alternate photon-statistics methods have found evidence for faint PSs in the inner Galaxy but cannot yet discern whether those PSs are associated with the GCE \cite{Calore:2021bty}.

Studies that aim to test the MSP explanation by searching for hot spots or examining the GCE photon statistics inherently rely on the luminosity function of the putative population of MSPs, i.e.~the number of MSPs as a function of their luminosity. There are two long-standing questions regarding the predictions made by plausible models for the MSP gamma-ray luminosity function: (1) the number of GCE PSs that should be detected individually, or in analyses that probe populations of PSs just below the sensitivity threshold of the relevant telescopes, and (2) the overall number of MSPs required to explain the excess. The first prediction depends primarily on the bright end of the luminosity function, and can be confronted with the number of observed sources or the fraction of the GCE that appears to be due to near-threshold sources. The second prediction is often controlled by the properties of low-luminosity MSPs whose emission can only be observed in aggregate (if at all), and can be compared with theoretical models or empirical inferences for the total number of MSPs ever produced in the Milky Way.

Studies using different models and parameterizations for the luminosity function have given widely varying answers to these questions, leading to differing conclusions on the viability of the MSP hypothesis. Ref.~\cite{Hooper16}, following earlier work in Refs.~\cite{Cholis:2014noa, Hooper:2015jlu}, analyzed the luminosity function for MSPs detected in globular clusters (GLCs) and found that if GCE MSPs had the same luminosity function, they could make up only a few percent of the excess. Refs.~\cite{Cholis:2014lta, Haggard:2017lyq} calibrated the anticipated number of bright MSPs to observed low-mass X-ray binaries (LMXBs), based on a scenario where MSPs form from LMXB progenitors, and found that such MSPs can contribute only $<23\%$ of the excess. More recently, however, Ref.~\cite{Ploeg:2020jeh} has argued that a physical model for the luminosity function of MSPs in the Galactic Bulge, based on observed MSPs in the Galactic disk and correlating other MSP characteristics with their luminosity \cite{Kalapotharakos:2019cio}, can fit the excess without overproducing bright sources. This work built on a previous study of MSPs in the Galactic disk \cite{Bartels:2018xom}, which constrained several simple parameterizations of the luminosity function using observational data, and found a luminosity function distinctly different from that inferred by Ref.~\cite{Hooper16}. Ref.~\cite{Gautam:2021wqn} argues that the Galactic Bulge could plausibly host $\mathcal{O}(10^5)$ MSPs formed by accretion-induced collapse (AIC), which would bypass the limit from non-observation of LMXBs, and also predicts a total flux broadly consistent with the GCE, using a luminosity function model based on the results of Ref.~\cite{Ploeg:2020jeh}.

Other studies have simply fitted a parameterized luminosity function to gamma-ray data from the inner Galaxy region. Studies using the Non-Poissonian Template Fitting (NPTF) method \cite{Lee:2015fea, Mishra-Sharma:2016gis} have generally used broken power law source count functions to describe a sub-threshold PS population associated with the GCE, and have inferred quite steeply peaked source count functions with most power in sources just below \textit{Fermi}'s PS sensitivity threshold \cite{Lee:2015fea, Buschmann:2020adf}. Ref.~\cite{Lee:2015fea} consequently requires only a small number of MSPs, $\mathcal{O}(400)$, to explain the entire GCE. Ref.~\cite{Bartels:2015aea} assumed a power-law luminosity function, $dN/dL \propto L^{-\alpha}$ with $\alpha=1.5$, with cutoffs at minimum and maximum luminosities $L_\text{min}, L_\text{max}$, and constrained $L_\text{max}$ from the data. Ref.~\cite{Zhong:2019ycb} took a similar approach, but allowed the power-law slope to vary as well as $L_\text{max}$; their preferred parameters imply $\mathcal{O}(3\times 10^6)$ MSPs are needed to explain the GCE, primarily because they consider a luminosity function with a steep power-law slope ($\alpha \gtrsim 1.8-1.9$) and extrapolate to a relatively small value of $L_\text{min}$. Non-parametric fits to the source count function have also been employed, e.g. in Ref.~\cite{List:2021aer}.

In this work, we aim to clarify the differences between these studies; study simple, commonly-used parameterizations of the MSP luminosity function to understand which regions of parameter space remain viable; and explore the resulting range of predictions for ongoing and future PS searches with increased sensitivity. A similar approach was taken by Ref.~\cite{Petrovic:2014xra}, albeit considering only power-law luminosity functions with sharp cutoffs at minimum and maximum luminosity values. We update and expand on this earlier work by considering a wider range of luminosity function parameterizations; using up-to-date PS sensitivity information given the most recent PS catalog produced by the {\it Fermi}-LAT Collaboration; and studying specific models and fits discussed in the literature in recent years.

We begin in section \ref{sec:methods} by describing our modeling of the GCE. In particular, in section \ref{sec:total-flux} we discuss the gamma-ray flux we attribute to the GCE, in order to define what we mean by ``explaining the GCE''. There are large systematic uncertainties on this total flux, at the factor-of-two level, and this may in itself be responsible for some differences in the literature; identical populations of PSs may explain $10\%$ of the GCE in one study and $20\%$ in another, due to different inferred total fluxes. We consider a range of total fluxes for the GCE drawn from Refs.~\cite{DiMauro:2021raz, Calore:2014xka, Zhong:2019ycb, Gordon13, Ajello:2015kwa, Ajello:2017opo}.

In section \ref{sec:pointsources}, we discuss the population of visible point sources we employ in this analysis and describe how we relate the properties of this population to the MSP luminosity functions (section \ref{sec:observables}). Our constraints and forecasts require understanding the sensitivity threshold for point source detection; we discuss several alternative models for this sensitivity threshold in section \ref{sec:sensitivity}. In section \ref{sec:lum-funcs} we discuss the luminosity function models we test in this work.

We present our main results in section \ref{sec:results}, and discuss some important systematic uncertainties affecting those results in section \ref{sec:further-discussion}. In section \ref{sec:future-sensitivity} we discuss the degree to which future analyses and/or observations with increased point source sensitivity would increase the fraction of resolved flux, under our various luminosity function models, and allow us to distinguish between different luminosity functions. We present our conclusions in section \ref{sec:conclusion}. Our appendices add detail on several of the intermediate steps needed for our main results, and provide supplementary results and calculations.

\section{Modeling the GCE}
\label{sec:methods}

\subsection{GCE spatial distribution}
We model the number density distribution of MSPs in the GC as the square of a generalized Navarro-Frenk-White (gNFW) profile \cite{Navarro:1995iw}, based on fits to the observed distribution of GCE flux. The gNFW profile is spherically symmetric, with radial distribution
\label{sec:spatial-distro}
\begin{equation}
    \sqrt{\rho_\text{GCE}(r)} \propto \parens{\frac{r}{r_s}}^{-\gamma}\parens{1 + \frac{r}{r_s}}^{-3+\gamma}.
    \label{eqn:nfw}
\end{equation}
Following \cite{Calore:2014xka, Gordon13, DiMauro:2021raz}, we choose $\gamma \approx 1.2$ and $r_s = \SI{20}{\kilo\parsec}$, as these parameters match the empirical data reasonably well (although $r_s$ is not strongly constrained by the GCE since the signal is only observed for $r \ll r_s$). Some references use $\gamma=1$, in which case the profile is called an ``NFW profile'' (not ``generalized''), e.g.~\cite{Zhong:2019ycb}. As discussed above, a number of studies have also found that the GCE is better described by a bulge-like density distribution than the gNFW$^2$ profile; we expect the effect of choosing a bulge-like profile instead of gNFW$^2$ to be rather small in our analysis, since the two profiles are quite similar where the GCE is bright. Furthermore, the only places we use the detailed spatial distribution of the GCE are (1) in computing the sensitivity to point sources, (2) in translating between flux and luminosity of individual sources, and (3) in computing the ratio of flux emitted between Regions of Interest (ROIs) of different size. We expect errors in the first two calculations due to an incorrect spatial distribution to be relatively small because the {\it Fermi} sensitivity map is fairly smooth, and in both models the GCE is quite concentrated in the inner Galaxy and originates from sources at a roughly constant distance from Earth. The largest effect of changing the assumed GCE spatial distribution may be via the variation in the inferred total flux, as discussed in section \ref{sec:total-flux} below; however, this quantity has other substantial systematic uncertainties related to the choice of background modeling.

We study this gNFW distribution within an ROI with $|\ell| < 20^\circ$ and $2^\circ < |b| < 20^\circ$, where we have masked the Galactic disk (consistent with e.g. Refs.~\cite{Zhong:2019ycb, Calore:2014nla}). We will generally report results over the energy range $\SI{0.1}{\giga\electronvolt} < E_\gamma < \SI{100}{\giga\electronvolt}$, for ease of comparison to point source properties reported in the 4FGL point source catalog \cite{Fermi-LAT:2019yla}.

\subsection{Total GCE flux}
\label{sec:total-flux}

In order to discuss PS populations that could potentially generate the gamma-ray flux of the GCE, we need to describe the overall brightness of the GCE. We extract the GCE flux from several previous analyses of GCE energy spectra \cite{Zhong:2019ycb, Calore:2014xka, DiMauro:2021raz, Abazajian:2014fta, Gordon13, Ajello:2015kwa, Ajello:2017opo}. Although these studies draw their data from the same source (\textit{Fermi} public data), the inferred GCE spectra differ due to choices in the fitting approach, ROI, and signal and background modeling. For example, to model the spatial distribution of the excess, Refs.~\cite{Calore:2014xka, Gordon13, Ajello:2015kwa} fix $\gamma=1.2$, while Ref.~\cite{Zhong:2019ycb} performs the analysis for both $\gamma=1.0$ and $\gamma=1.2$, and Refs.~\cite{DiMauro:2021raz, Ajello:2017opo, Abazajian:2014fta} allow $\gamma$ to float in the fit. For the cases in which $\gamma$ is fitted, typical values lie in the range 1.0--1.3.
The studies all use ROIs centered on the GC, ranging from a $40^\circ \times 40^\circ$ region without the Galactic Disk mask used in this paper, to a $7^\circ \times 7^\circ$ region. All ROIs are centered on $(\ell, b)=(0, 0)$. All studies fix $r_s=\SI{20}{\kilo\parsec}$, except for Refs.~\cite{Abazajian:2014fta, Gordon13}, which use $r_s=\SI{23.1}{\kilo\parsec}$; in any case, the GCE spectrum is rather insensitive to $r_s$ \cite{DiMauro:2021raz}. In order to compare studies with different ROIs, we re-scale the inferred flux by the method described in appendix \ref{app:roi-rescale}. The effect of varying $\gamma$ on the inferred total flux within our ROI is non-negligible --- for example, the total flux from the spectra attained by Ref.~\cite{Zhong:2019ycb} assuming $\gamma=1.0$ is $\sim40\%$ larger than the flux inferred assuming $\gamma=1.2$ --- but as we will see, there are other systematic uncertainties of comparable magnitude.

The manner in which uncertainties in the energy spectrum are reported also varies; some studies report only statistical uncertainties, and some report both statistical and systematic. Refs.~\cite{Calore:2014xka, Gordon13, DiMauro:2021raz} report both separately, and for our purposes, we add these in quadrature (this approach may lead to an overestimate of uncertainties in some cases since it neglects correlations between systematic uncertainties).

\begin{figure}
    \centering
    \includegraphics[width=0.7\textwidth]{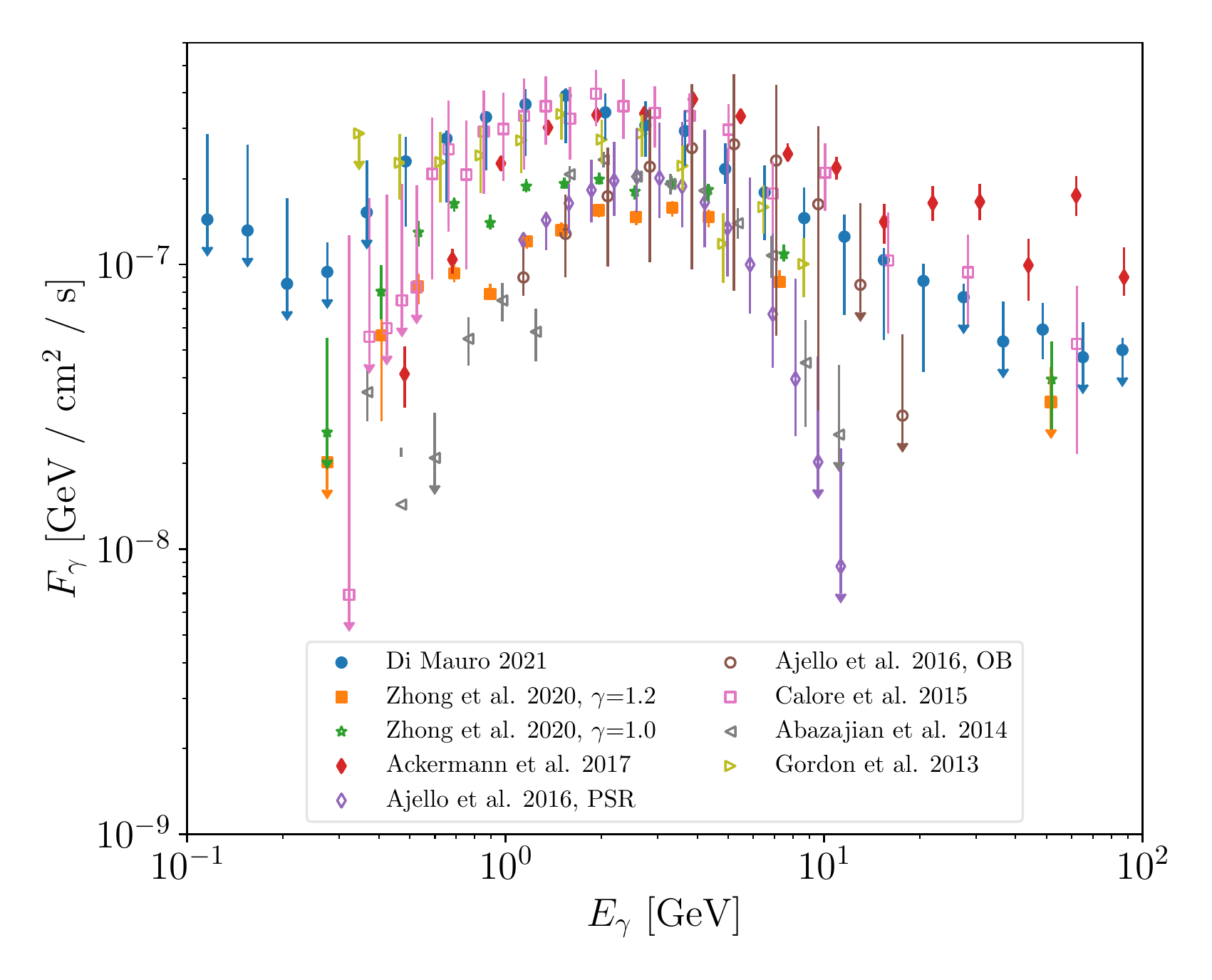}
    \caption{Previously-derived spectra of photon flux from the GCE in $F_\gamma = E^2 dN_\gamma/dE$, integrated over the ROI with $|l| < 20^\circ$ and $2^\circ < |b| < 20^\circ$, selected from nine analyses of the GCE \cite{Zhong:2019ycb, Calore:2014xka, DiMauro:2021raz, Abazajian:2014fta, Gordon13, Ajello:2015kwa, Ajello:2017opo} (note some of these references include multiple analyses). $1\sigma$ error bars are reproduced from the same references. Arrows on error bars denote upper limits (i.e. because the $1\sigma$ error bars overlap zero).}
    \label{fig:all-spectra}
\end{figure}

Figure \ref{fig:all-spectra} displays all the spectra mentioned above, with ROI rescaling included. Many studies reported flux values in units of flux per steradian; we have multiplied those fluxes by the area of their respective ROIs and then rescaled the flux as described in appendix \ref{app:roi-rescale} to attain an absolute flux from the GCE in our ROI differential in energy. We report our results in terms of $F_\gamma = E^2 dN_\gamma/dE$, where $N_\gamma$ is the number of incident photons from the ROI per unit exposure (measured in cm$^2$ s).

We compare three methods of extracting the total GCE flux, integrated over energy, from these spectrum analyses. The first method is direct numerical integration of the binned spectrum. This method is most responsive to the data measured by \textit{Fermi} and does not attempt to abstract over it with a smooth function, but it is potentially somewhat noisy and the energy range of the data varies between different studies. Therefore, we also test the effect of fitting a singly broken power law to the data, and then analytically integrating this function to infer the total integrated GCE flux in the energy range $[\SI{0.1}{\giga\electronvolt}, \SI{100}{\giga\electronvolt}]$:
\begin{equation} F_\gamma = F_0 \begin{cases}
        \parens{\fraci{E}{E_\text{b}}}^{2-n_{1}} & E < E_{b} \\
        \parens{\fraci{E}{E_\text{b}}}^{2-n_{2}} & E > E_b
    \end{cases},\end{equation}
 We perform this fit in two ways: (1) where all four parameters of the broken power law are allowed to float (the normalization constant $F_0$, the turnover energy $E_\text{b}$, and the slopes above and below the turnover energy $n_2$ and $n_1$), and (2) where all parameters are fixed except $F_0$, which is allowed to float. In the latter case we use the parameters determined by Ref.~\cite{Calore:2014xka}: $E_\text{b} = \SIasym{2.06}{0.23}{0.17}{\giga\electronvolt}$, $n_1 = \numasym{1.42}{0.22}{0.31}$, $n_2 = \numasym{2.63}{0.13}{0.095}$.

An example of the two fits applied to the spectrum of Ref.~\cite{DiMauro:2021raz} is shown in figure \ref{fig:di-mauro-example}. The best fits for all spectra are displayed in appendix \ref{app:spectra-fits}. Figure \ref{fig:total-flux-bars} displays the results for the integrated flux $F_\text{GCE}$ via the three integration methods, from each GCE spectrum studied. In general we find that there is a substantial variation in $F_\text{GCE}$ between different spectra presented in the literature; the variation associated with using different methods to model a given spectrum is comparatively small, and consistent within the nominal uncertainties. Figure \ref{fig:total-flux-bars} also shows the results of integrating the flux only up to an energy of 10 GeV, to test whether the variations between different analyses might be due to differences in the high-energy tail. We observe generally that the 10-100 GeV band provides only a subdominant component of the flux, and while this contribution is quite uncertain (being negligible in some analyses and quite substantial in others), it does not appear to be the main source of differences between analyses --- there are large variations in the flux in the 0.1-10 GeV band as well.

\begin{figure}
    \centering
    \includegraphics[width=0.7\textwidth]{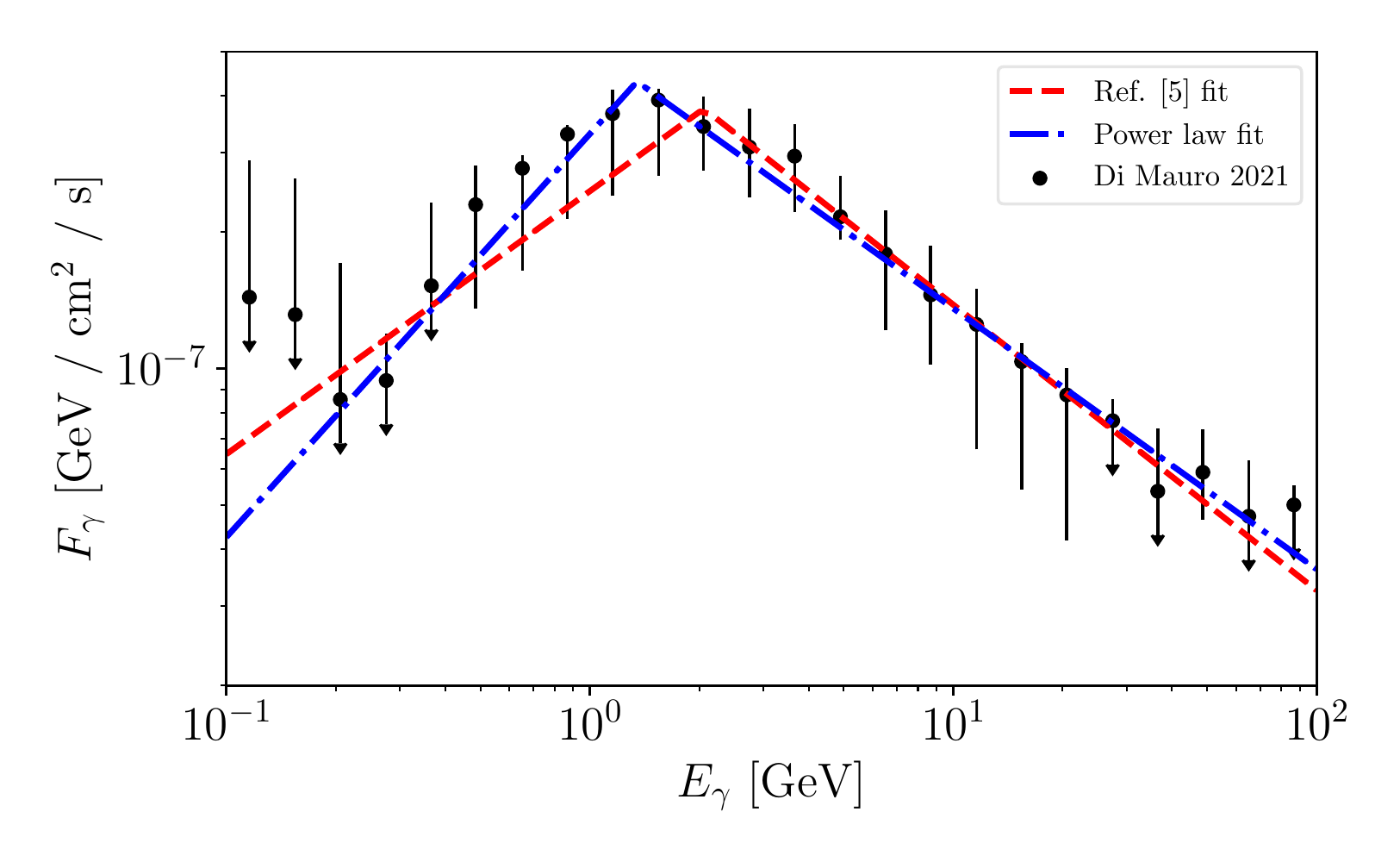}
    \caption{Spectrum produced by Ref.~\cite{DiMauro:2021raz}, with a broken power law fitted with all parameters free (green) and another broken power law using parameters given by Ref.~\cite{Calore:2014xka}, allowing only the normalization to vary. Arrows on error bars denote upper limits (i.e. because the $1\sigma$ error bars overlap zero).}
    \label{fig:di-mauro-example}
\end{figure}

The inferred flux differs by roughly a factor of two between the highest-flux and lowest-flux scenarios. It seems likely that this can be largely attributed to differences in the modeling of the background components in the fit, between the various analyses, which are known to induce substantial systematic uncertainties (e.g. \cite{Calore:2014xka}). For ease of display, we will show baseline results assuming that ``explaining the GCE'' means reproducing the full flux obtained from a broken power-law fit (with all parameters floated) to the data from Ref.~\cite{DiMauro:2021raz}, $F_\text{GCE} = \SI{1.8e-09}{\erg\per\centi\meter\squared\per\second}$. The corresponding spectral fit parameters are $E_b = \SI{1.35}{\giga\electronvolt}$, $n_1=1.11$, $n_2 = 2.58$. This is a recent analysis that agrees well with earlier results from Refs.~\cite{Calore:2014xka, Ajello:2017opo}; it corresponds to a relatively high-flux scenario. In general this means that requiring a source population to generate the GCE predicts more numerous and observable sources, compared to a lower-flux scenario, leading to stronger constraints; a luminosity function that does not overproduce known point sources using this flux choice should also be allowed in a lower-flux scenario. To account for the large uncertainty in the total flux, we will show how our results change for different assumed total fluxes in section \ref{sec:further-discussion}.

\begin{figure}
    \centering
    \includegraphics[width=0.9\textwidth]{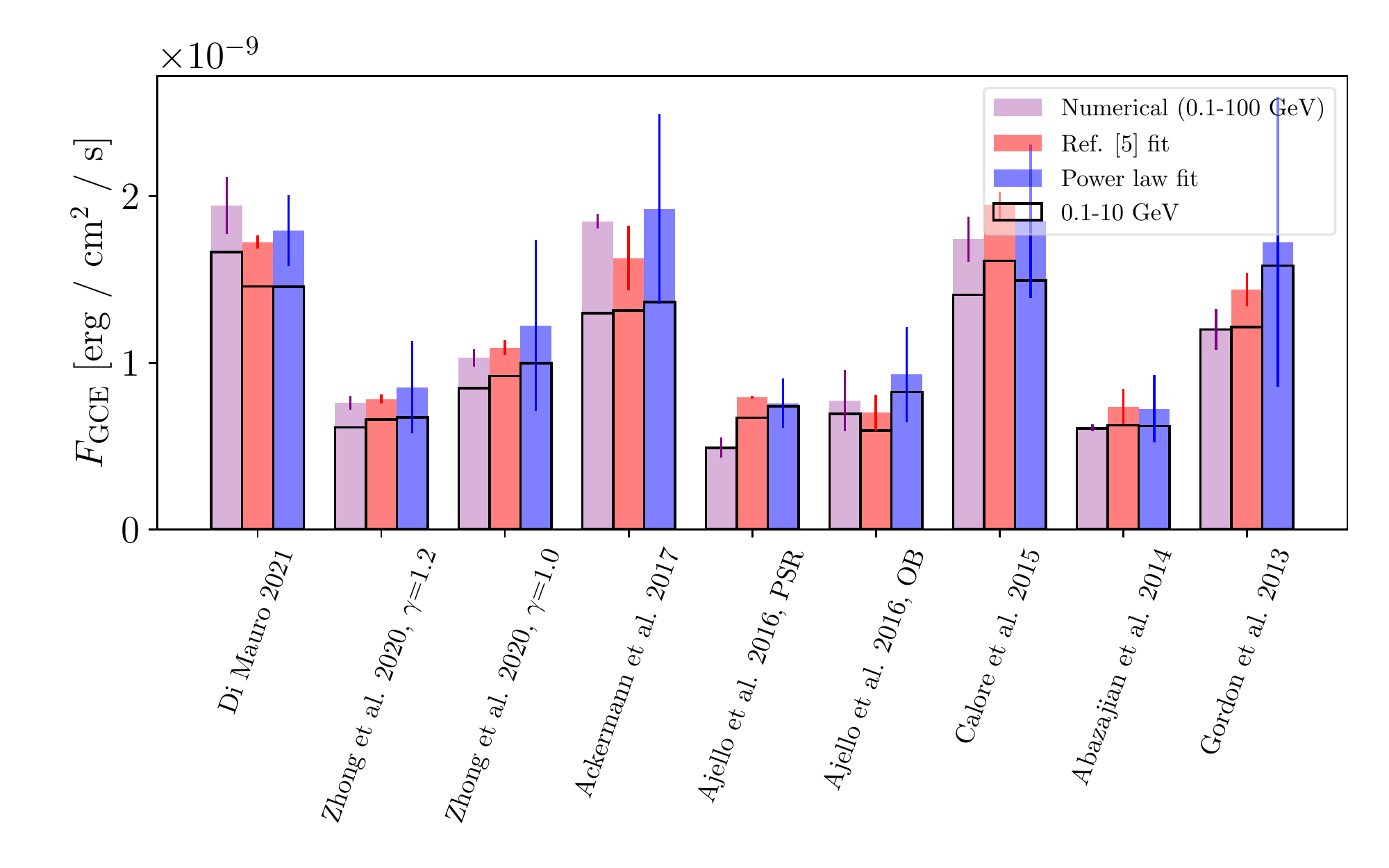}
    \caption{Total flux of the GCE in our $40^\circ \times 40^\circ$ ROI with a $|b| \leq 2^\circ$ cut around the Galactic Disk, as determined by the three integration methods discussed in the text, with spectrum range $\SI{0.1}{\giga\electronvolt}-\SI{100}{\giga\electronvolt}$. Also shown (solid black lines) are results from a fit and integral over the $\SI{0.1}{\giga\electronvolt}-\SI{10}{\giga\electronvolt}$ domain, to demonstrate that the high energy tail does not dominate the GCE's flux.}
    \label{fig:total-flux-bars}
\end{figure}

\section{Modeling detected point sources}
\label{sec:pointsources}

\subsection{Observables}
\label{sec:observables}
To constrain the luminosity function of a hypothetical MSP population responsible for the entire GCE, we track predictions of two observable properties of the population: $N_r$, the number of resolved PSs, and $R_r$, the ratio of the flux emitted by those PSs to the total flux of the GCE. Collectively, we call these two values ``the observables.'' Later, we will also discuss the distribution in flux of the resolved PSs. Predictions for the theoretical values of the observables, given a luminosity function, are discussed in section \ref{sec:sensitivity}, and in this section we discuss observational constraints.

\subsubsection{4FGL catalog}
The 4FGL Point Source Catalog tracks the positions and fluxes of resolved gamma-ray PSs in the sky, in addition to many other PS properties such as their potential origin or any associated sources in other catalogs \cite{Fermi-LAT:2019yla}. Two releases of the catalog currently exist: an 8-year release which we call Data Release 1 (4FGL-DR1)\footnote{\url{https://fermi.gsfc.nasa.gov/ssc/data/access/lat/8yr_catalog/}}, and a 10-year release called 4FGL-DR2.\footnote{\url{https://fermi.gsfc.nasa.gov/ssc/data/access/lat/10yr_catalog/}} Earlier versions of the catalog also exist --- a 4-year version labeled 3FGL and a 2-year version labeled 2FGL --- and have been employed in previous analyses of the GCE.

There are a number of approaches one might take to constraining $N_r$ from these catalogs. The most conservative (i.e.~yielding the weakest limits) would be to simply include all sources within the ROI, but it is very unlikely that all such sources are associated with the GCE. Alternatively, one could include only sources with a spectrum sufficiently consistent with the GCE (although it is possible that the spectra of individual GCE sources differ somewhat from the aggregate spectrum), only sources known to be pulsars, or some other subsample. In this work, we use for our baseline sample the full set of 4FGL-DR2 sources in the ROI, minus those which are known to be either outside the GCE region or associated with non-pulsar source classes (which are dominantly extragalactic), on the basis that this should be a true upper limit on the number of possible GCE sources detected with the sensitivity relevant to 4FGL-DR2. We then show how the constraints would change if only a fraction of these sources belong to the GCE, in the hope that this will allow for easy translation of our constraints to future analyses that eliminate a larger fraction of 4FGL-DR2 sources as possible members of a GCE source population (or identify members of that population).

Specifically, to obtain our conservative upper limit for $N_r$, we remove all PSs in the 4GFL-DR2 catalog that are outside the ROI or have a listed association with a non-pulsar source. For the PSs associated with known pulsars, we remove those known to be farther than $\SI{2}{\kilo\parsec}$ from the GC based on the ATNF Pulsar Catalog \cite{Hobbs04}. These cuts match those applied in Ref.~\cite{Zhong:2019ycb}.

There are $N_r=265$ 4FGL-DR2 PSs passing these cuts, seven of which are associated with pulsars. Together, these PSs contribute $\SI{1.6e-9}{\erg\per\second\per\centi\meter\squared}$ of flux, or $R_r = 91\%$ of the total GCE flux. This includes 4FGL-DR2 sources which have one or more analysis flags and thus should be treated with caution. Cutting the flagged PSs as well, as these are known to be affected by systematic errors and the instructions for use of the catalog indicate they should be used with great care, we have instead $N_r = 109$ and $R_r = 35\%$. In both cases, most of these sources are PSs with unknown origin and unknown distance from the GC, so only a fraction of them are likely to be GCE MSPs. Therefore, these estimates of $N_r$ and $R_r$ should be regarded as upper bounds on the true resolved source populations associated with the GCE. 

\subsubsection{Wavelet-selected subsample}

Ref.~\cite{Zhong:2019ycb} recently examined public {\it Fermi} Pass 8 data (version P8R3, recorded from 4 August 2008 to 20 February 2019), using wavelet-based methods first employed by Ref.~\cite{Bartels:2015aea} to identify a population of isolated spatial peaks. Ref.~\cite{Bartels:2015aea} had identified these peaks with a sub-detection-threshold source population that could be contributing to the GCE. Ref.~\cite{Zhong:2019ycb} found 115 significant peaks within their ROI (which matches the one used in this analysis), and then compared the locations of these peaks to the 4FGL-DR1 catalog. 107 of these peaks were within 0.3$^\circ$ of 103 PSs in the 4FGL-DR1 catalog\footnote{We further found that one of the 8 wavelet peaks not associated with a source in 4FGL-DR1 was coincident with a source in 4FGL-DR2.}, and the authors therefore concluded that the wavelet peaks could be well-approximated as a subset of the 4FGL-DR1 catalog.

\begin{figure}
    \centering
    \includegraphics[width=0.6\textwidth]{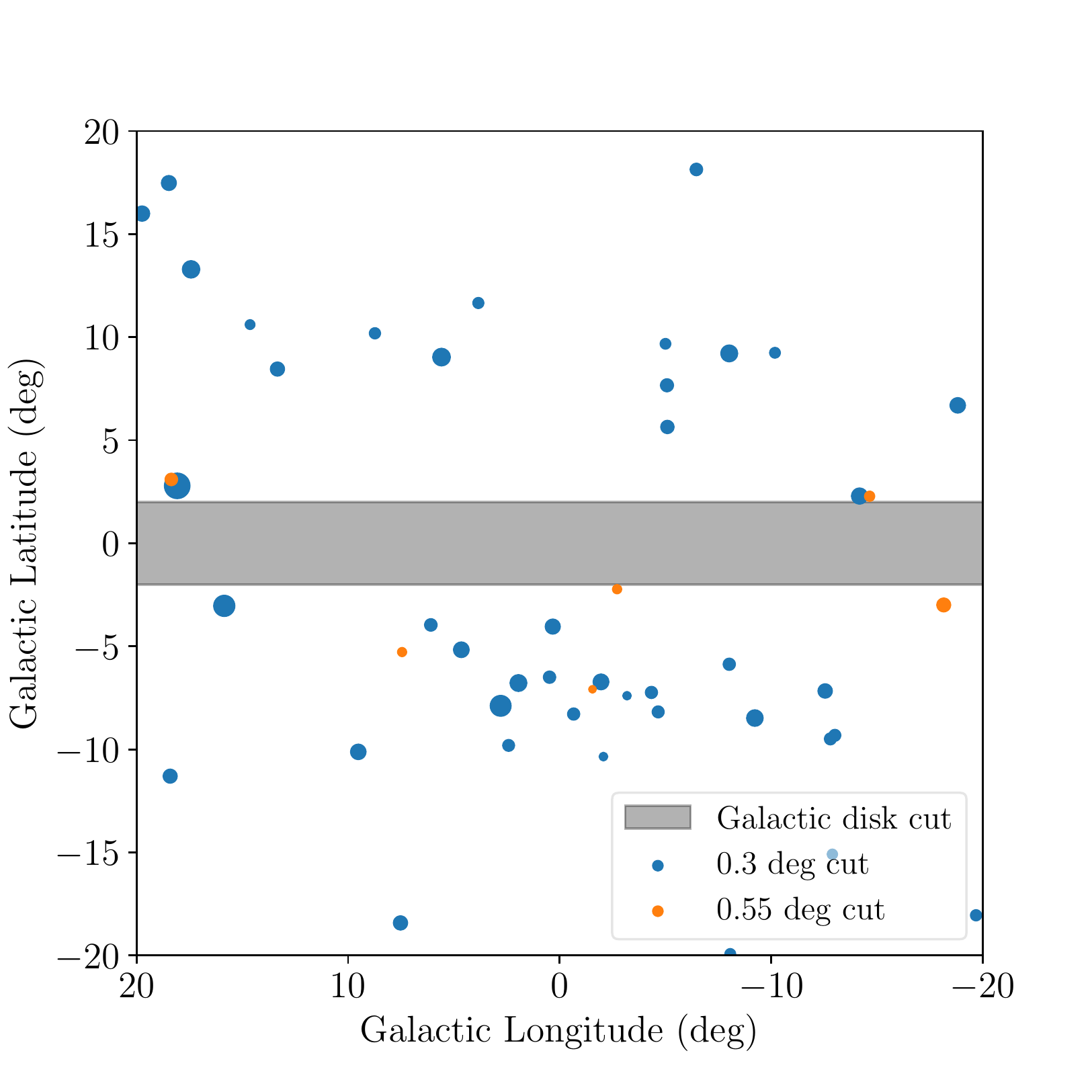}
    \caption{Positions of the 41 wavelet-selected MSP candidates within $0.3^\circ$ of a 4FGL-DR1 PS (blue circles). Also shown are the 6 PSs added when the maximum separation is increased to $0.55^\circ$  (orange circles). The PSs' radii are drawn proportional to their flux listed in the 4FGL-DR1 catalog for the 0.1-100 GeV energy bin. The gray band represents the $|b| \leq 2^\circ$ cut masking the Galactic disk.}
    \label{fig:47-sources}
\end{figure}

The authors of Ref.~\cite{Zhong:2019ycb} provided us with the locations and associations of their wavelet peaks, allowing us to reproduce their cuts. We found that of the 103 PSs associated with wavelets, 46 were excluded due to associations with non-pulsar sources (such as active galactic nuclei). A further 15 sources were associated with pulsars and excluded because of distance measures placing them outside $\SI{2}{\kilo\parsec}$ of the GC. (Pulsar J1823-3021A was the only pulsar in the list of wavelet-selected sources known to be within this radius, but because it is also a globular cluster member, it was also excluded.) This left $N_r=41$ resolved wavelet-selected Galactic MSP candidates in the 4FGL-DR1 catalog, contributing $R_r=14\%$ of our baseline GCE flux. When the proximity cut between the flux peak and the 4FGL source was extended from 0.3$^\circ$ to 0.55$^\circ$, six other sources were added, yielding $N_r=47$ PSs, contributing $R_r=17\%$ of total flux. Both $R_r$ values are slightly smaller than those quoted in Ref.~\cite{Zhong:2019ycb} because our baseline GCE flux (derived from Ref.~\cite{DiMauro:2021raz}) is larger, as discussed in section \ref{sec:total-flux}. These 41+6 PSs are shown in figure \ref{fig:47-sources}; we will use them as an example subsample of 4FGL-DR2 that could be attributed to the GCE, and compare our results in this context with those of Ref.~\cite{Zhong:2019ycb} (noting that the approach of using a wavelet-selected sample as a proxy for GCE sources was first advanced in Ref.~\cite{Bartels:2015aea}).

\subsubsection{Constraints/benchmarks for observables}

\begin{figure}
    \centering
    \includegraphics[width=0.6\textwidth]{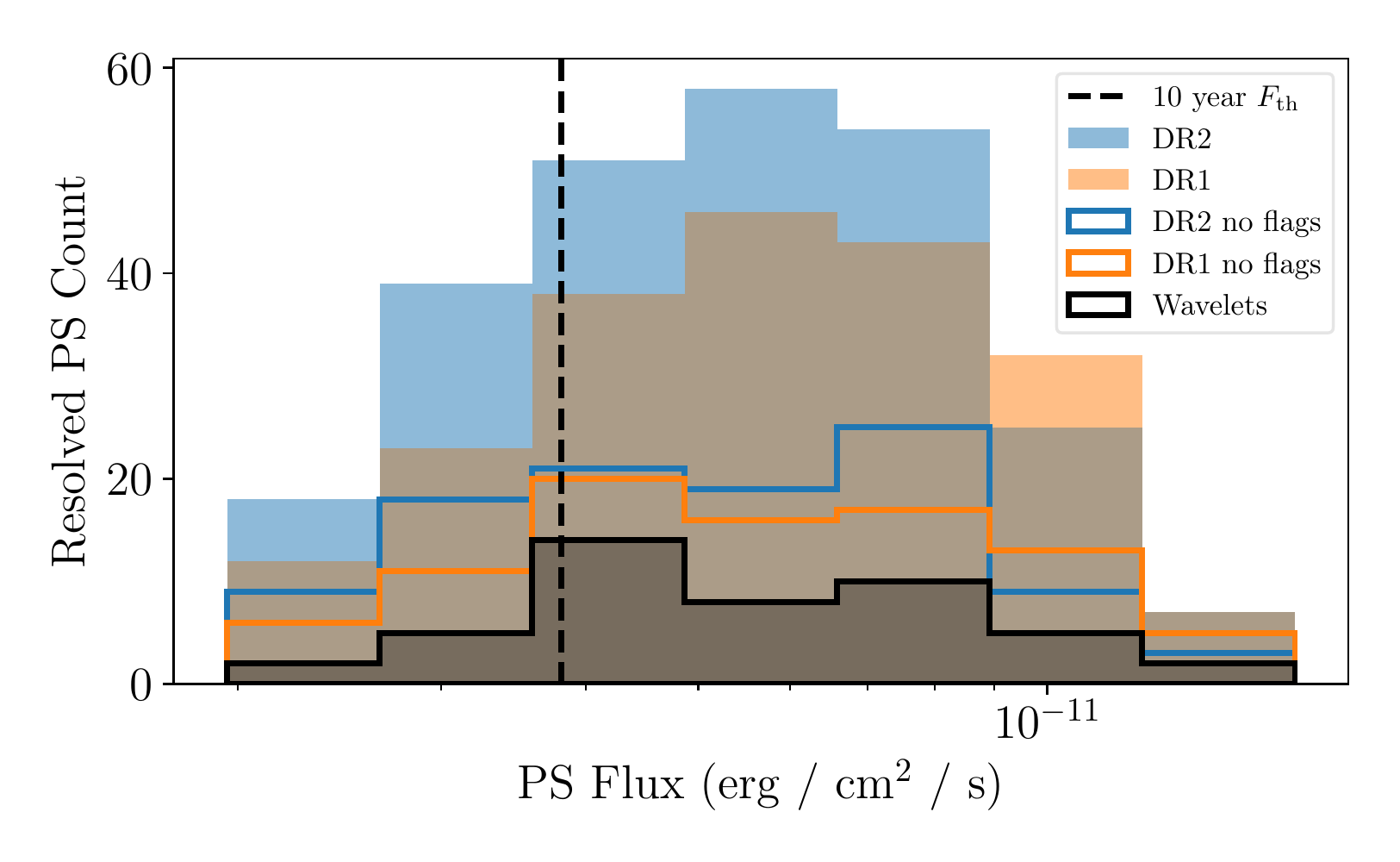}
    \caption{Comparison between the gamma-ray flux distributions of sources in the 4FGL-DR2 (blue) and 4FGL-DR1 (orange) catalogs, and the wavelet-selected 4FGL-DR1 sources of Ref.~\cite{Zhong:2019ycb}, within our ROI and after the cuts are applied (see text for details). Fluxes are obtained from the relevant catalogs and are evaluated for the 0.1-100 GeV energy band. We show results including only flagged catalog sources (solid blue/orange lines), flagged+unflagged catalog sources (shaded blue/orange regions), and wavelet-selected sources (solid black lines). The vertical line denoting $F_\text{th}$ represents the weighted average flux threshold of the 4FGL-DR2 catalog within our ROI, as computed in appendix \ref{app:step-thresh}.}
    \label{fig:fgl-to-wavelet-comparison}
\end{figure}

It is tempting to go further and use the wavelet-selected subsample of 47 PSs to define an upper limit on resolved GCE sources (or even a target number of resolved sources), as was done in Ref.~\cite{Zhong:2019ycb}. However, without reproducing the pipeline of Ref.~\cite{Zhong:2019ycb}, we do not have a good model for the sensitivity of the wavelet-based selection, which makes it difficult to predict the expected number of sources for a given luminosity function. Ref.~\cite{Zhong:2019ycb} estimates the sensitivity using a simple luminosity threshold where sources with luminosity greater than $10^{34}$ erg/s are all resolved. As a check on this approximation, we show in figure \ref{fig:fgl-to-wavelet-comparison} the flux distributions of sources in the wavelet-selected subsample and the 4FGL-DR2 and 4FGL-DR1 catalogs (in the latter case, we show separately the results where flagged sources are included or excluded).

We see that within this ROI, the wavelet-selected sources are a subsample of the 4FGL catalogs at all flux levels; i.e.~the difference in the total number of sources is not a matter of the wavelet selection having a higher sensitivity threshold relative to the 4FGL catalogs. In fact, the distribution of PSs with respect to flux is rather consistent across different cuts (note that some of the difference between e.g. 4FGL-DR1 and 4FGL-DR2 corresponds to sources moving between flux bins, not sources appearing or disappearing from the catalog), with the main difference being the total number of PSs. As a fraction $f$ of our baseline 4FGL-DR2 sample (including flagged sources), the numbers of sources detected in 4FGL-DR1, 4FGL-DR2 with unflagged sources only, 4FGL-DR1 with unflagged sources only, and the wavelet method of Ref.~\cite{Zhong:2019ycb} (with the $0.55^\circ$ radius cut), correspond respectively to $f=80\%$, 40\%, 35\%, and 18\%. As noted above, the wavelet-selected PSs contribute roughly $17\%$ of the GCE flux while the baseline 4FGL-DR2 sample (including flagged sources) contributes $91\%$ of the GCE flux. Thus the wavelet-selected PSs contribute $19\%$ of the 4FGL-DR2 sample in flux and $18\%$ in number, consistent with their flux distribution being rather similar.

Consequently, it appears that a GCE PS might well be above both the sensitivity threshold for 4FGL-DR2, and the nominal sensitivity threshold suggested in Ref.~\cite{Zhong:2019ycb}, and still fail to be detected by the wavelet method (in the sense that there are many 4FGL-DR2 and 4FGL-DR1 sources that pass all the cuts in Ref.~\cite{Zhong:2019ycb} and were not associated with a significant wavelet peak). Thus we will use the wavelet-selected subsample as an {\it example} of a resolved PS population that could be associated with the GCE, but not as a formal upper limit; to use it as a true upper limit would require an in-depth study of the sensitivity and completeness properties of the wavelet method for identifying resolved PSs.

More generally, we will consider benchmarks for $N_r = 106$, 53, 26, and 13, corresponding to populations generating 40\%, 20\%, 10\%, and 5\% of all the 4FGL-DR2 sources passing our cuts. We will separately consider benchmarks of $R_r=40\%$, 20\%, 10\%, and 5\%. In cases like the wavelet-selected population where the flux distribution of the resolved sources is similar to that of our 4FGL-DR2 sample, the $N_r$ and $R_r$ benchmarks will approximately coincide (because the 4FGL-DR2 cut PSs happen to have a similar total flux to the GCE). In particular, the $N_r=106$, $R_r=40\%$ benchmark is similar to a scenario where the resolved GCE sources coincide with all unflagged 4FGL-DR2 sources passing the cuts, and the $N_r=53$, $R_r=20\%$ benchmark is similar to the case where the resolved GCE sources coincide with the wavelet-selected sources.

There is an independent set of constraints on $R_r$ from the observation that masking all known PSs does not seem to appreciably reduce the flux of the GCE. In particular, Ref.~\cite{Zhong:2019ycb} tested the effects of masking all 4FGL-DR1 sources and found that the effect on the inferred GCE spectrum was negligible (compared to masking only 2FGL sources), reducing the GCE flux by less than $10\%$ at all energies. The recent study in Ref.~\cite{DiMauro:2021raz}, which masks all 4FGL-DR2 sources, infers a very comparable GCE flux to earlier similar studies that masked only the 3FGL \cite{Ajello:2017opo} or 2FGL \cite{Calore:2014xka}, with estimated error bars on the total flux at the 10-20\% level (although, as discussed above, the scatter between the full range of analyses is larger). The substantial systematic uncertainties in the determination of the GCE spectrum make it difficult to claim a statistically precise quantitative exclusion, but a contribution to the total GCE flux of more than $R_r\sim 20\%$ from resolved sources would appear to be in tension with the results of Ref.~\cite{Zhong:2019ycb}, and possible tension with the large GCE flux inferred by Ref.~\cite{DiMauro:2021raz} in an analysis where all 4FGL-DR2 sources were masked. Thus our $R_r=18\%$ benchmark can also be viewed as an approximate upper bound on the fraction of flux in resolved sources in order to avoid tension with observations, and the wavelet-selected source population as an example of what saturating that bound might look like.

\subsubsection{Total number of MSPs}
We will also discuss another feature of a potential MSP population in the GC: the total number of MSPs, resolved or unresolved, denoted $N_\text{GCE}$. Given a prediction for $N_\text{GCE}$, we can ask whether $N_\text{GCE}$ can reasonably be achieved in our Galaxy by existing models for the formation and evolution of MSPs. For example, it might be possible to exclude certain GCE luminosity functions on the grounds that they predict unphysically large numbers of very faint MSPs, even if those MSPs are not plausibly resolvable. As an example, the power law luminosity function proposed in Ref.~\cite{Zhong:2019ycb} produces more than three million MSPs, which is very large compared to earlier predictions from population synthesis studies. For example, Ref.~\cite{Faucher-Giguere:2005dxp} predicted a total number of radio-loud pulsars (of all periods) in the Galaxy of around $1.2$ million, with only $\mathcal{O}(10\%)$ of those beamed toward us; Ref.~\cite{Story:2007xy} predicted a birth rate of MSPs in the Galactic disk of a few $\times 10^{-4}$ per century, translating to a few $\times 10^4$ MSPs produced over the age of the Galaxy; Ref.~\cite{Gonthier:2018ymi} proposed that globular cluster disruption could source a population of 5000-16,000 MSPs in the inner Galaxy; and Ref.~\cite{Gautam:2021wqn} proposes $\mathcal{O}(10^5)$ MSPs could be produced in the Galactic bulge via accretion-induced collapse. Ref.~\cite{Zhong:2019ycb} uses this discrepancy to argue that the MSP hypothesis may be under stress, although this argument depends strongly on the assumed luminosity function. However, if non-detection of GCE PSs continues with increasing sensitivity to PSs, such that all reasonable luminosity functions would predict a number of MSPs exceeding the yield of all possible production mechanisms, then in future it might in principle be possible to exclude the MSP hypothesis.

\subsection{Sensitivity models}
\label{sec:sensitivity}
The \textit{Fermi} telescope does not detect every pulsar (or PS) in the GC; position-dependent background emission obscures dimmer PSs, and the faintest PSs may not produce a statistically significant number of photons at all. Inclusion of a source in the 4FGL catalogs generally requires a minimum value of the test statistic (TS) describing the likelihood improvement from adding the source to the model ($TS > 25$ for the likelihood test with non-curved spectra).

\begin{figure}
    \centering
    \includegraphics[width=0.49\textwidth]{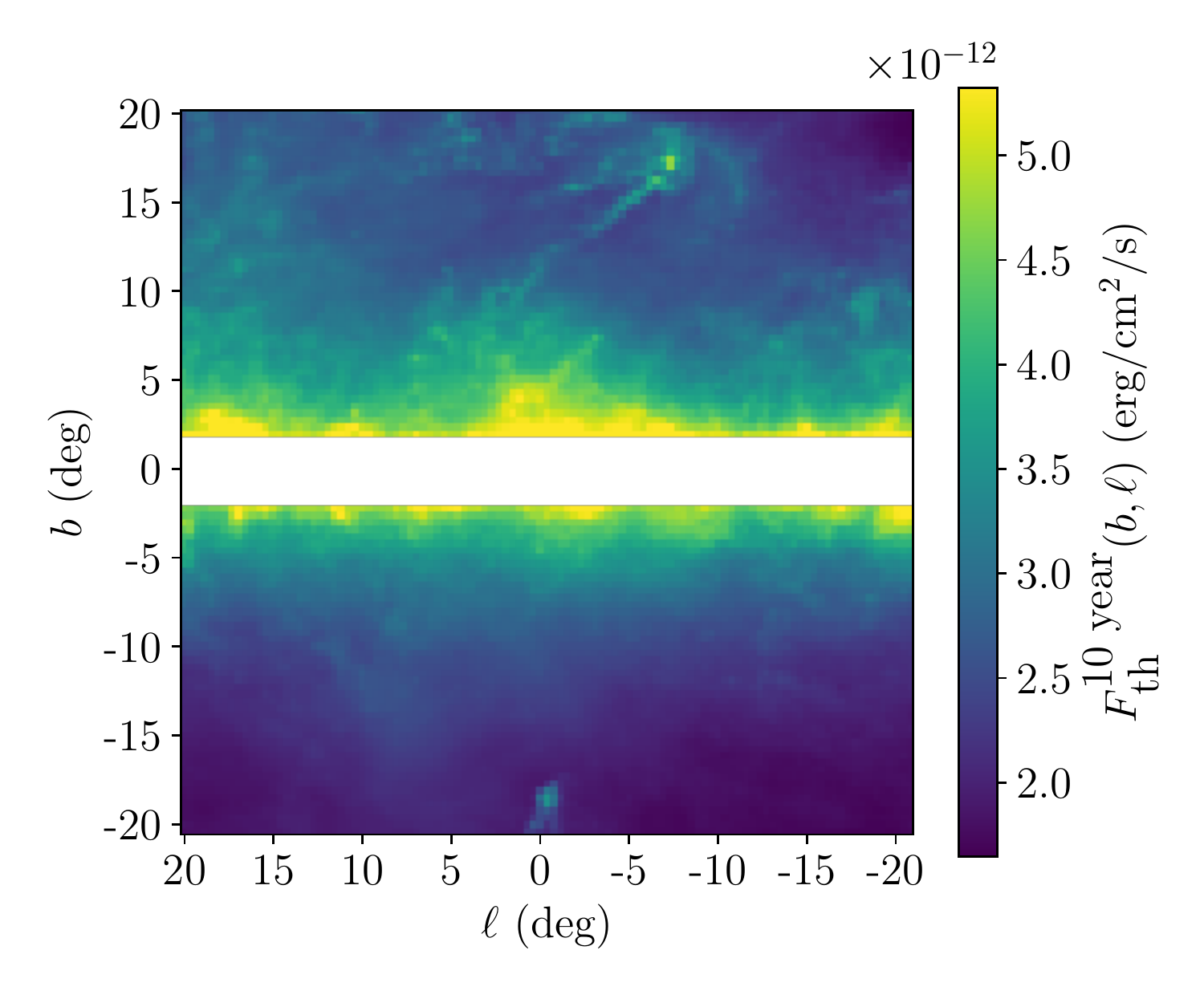}
    \hfill
    \includegraphics[width=0.49\textwidth]{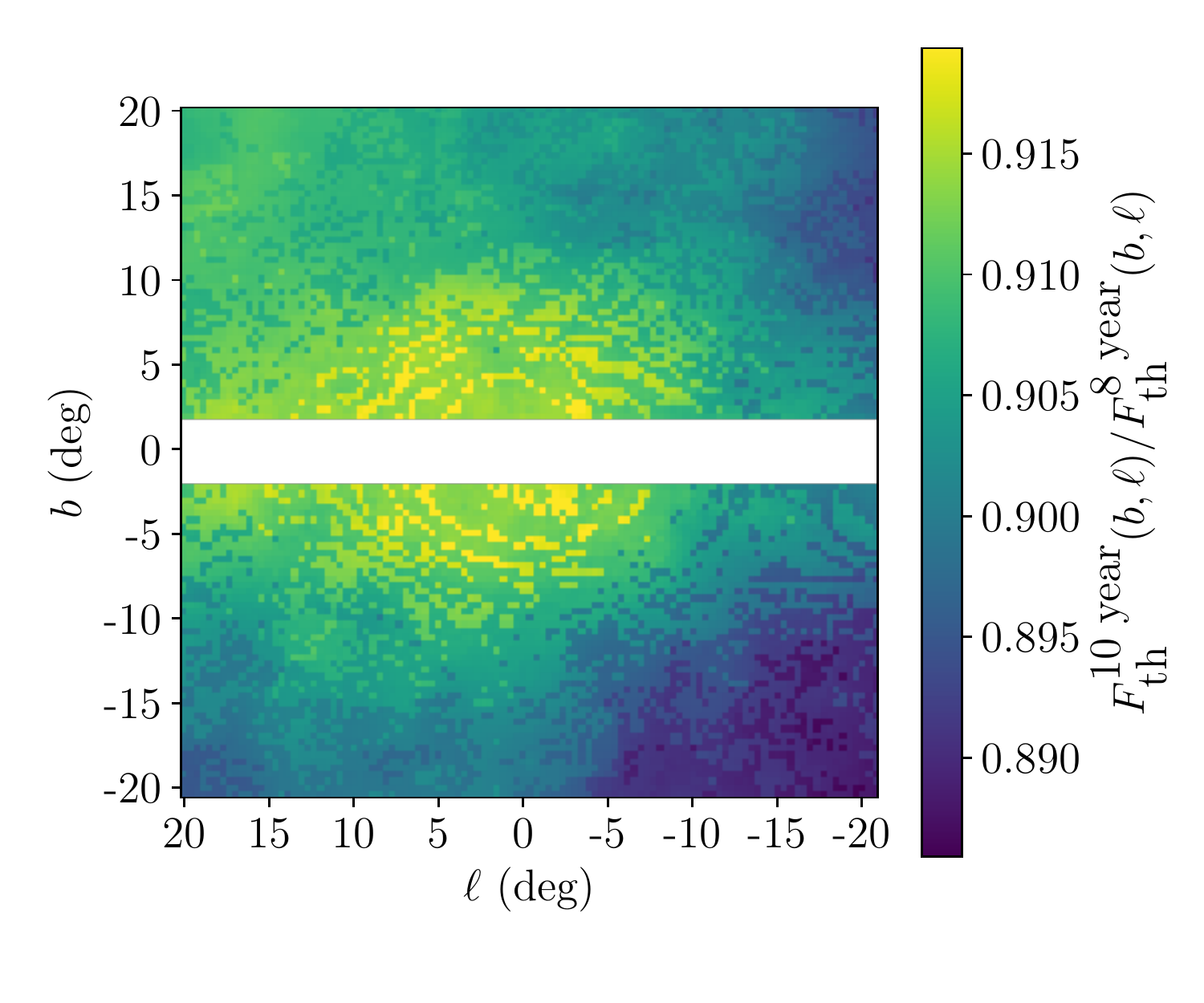}
    \caption{Flux threshold $F_\text{th}(\ell, b)$ to include a PS within our ROI in the 4FGL-DR2 catalog (left panel) \cite{Fermi-LAT:2019yla, Ballet:2020hze}, and the ratio of the 4FGL DR-2 and 4FGL DR-1 flux thresholds (right panel). Fluxes are computed in the 0.1-100 GeV energy band.}
    \label{fig:sensitivity}
\end{figure}

In our main analysis, we account for these factors by using a position-dependent flux threshold $F_\text{th}(b, l)$ published by the \textit{Fermi} team for the DR2 catalog \cite{Fermi-LAT:2019yla, Ballet:2020hze} to model the catalog's threshold sensitivity. If a PS emits flux $F > F_\text{th}(b, l)$, we model it as resolved, and if $F < F_\text{th}(b, l)$, we model it as unresolved. The position dependence of this flux threshold, restricted to our ROI and for the energy range $0.1-100$ GeV, is shown in figure \ref{fig:sensitivity}. We will refer to this sensitivity model as the standard sensitivity model.

This approach is an approximation --- in reality, Poisson fluctuations in the observed number of photons from sources or backgrounds may cause sources to move across the threshold in either direction, and the published sensitivity map also assumes a specific spectrum for the PSs --- but we expect the resulting systematic errors to be small except perhaps for PS populations with flux distributions peaked very close to the threshold. We will validate this approach by comparing the flux distribution of observed sources to the predicted distribution for plausible luminosity functions.

To calculate the required properties of GC PS populations given a luminosity function $P(L) \propto dN/dL$, we write:
\begin{equation}
    \begin{split}
        F_\text{GCE} &= \int_\Omega d\Omega \int_0^\infty s^2 ds A \rho_\text{GCE}(r)\int_{L_\text{min}}^\infty dL \frac{L}{4\pi s^2}P(L)\,, \\
        N_\text{GCE} &= \int_\Omega d\Omega \int_0^\infty s^2 ds A \rho_\text{GCE}(r)\,, \\
        F_\text{r} &= \int_\Omega d\Omega \int_0^\infty s^2 ds A \rho_\text{GCE}(r)\int_{4\pi s^2F_\text{th}(\ell, b)}^\infty dL \frac{L}{4\pi s^2}P(L)\,, \\
        N_r &= \int_\Omega d\Omega \int_0^\infty s^2 ds A \rho_\text{GCE}(r)\int_{4\pi s^2F_\text{th}(\ell, b)}^\infty dL P(L) \,. \\
        \label{eqn:observables-sens-2}
    \end{split}
\end{equation}
Here $\Omega$ denotes our $20^\circ \times 20^\circ$ ROI with $|b| < 2^\circ$ masked, and $A$ is the coefficient of the RHS of Eq.~\ref{eqn:nfw} --- i.e.~$A$ governs the number of PSs --- and is fixed by forcing $F_\text{GCE}$ to equal the observed value. In Eq.~\ref{eqn:observables-sens-2}, $r$ represents the distance to the GC from the point of integration and is determined by the law of cosines: $r^2 = s^2 + r_c^2 - 2r_c s \cos b \cos \ell$, where $r_c=\SI{8.5}{\kilo\parsec}$ is the approximate distance from the Earth to the GC (as given in e.g. \cite{Bartels:2015aea}) and $s$ is the distance between Earth and the point of integration.

We contrast the standard sensitivity model with a simplified sensitivity model used by Ref.~\cite{Zhong:2019ycb}. The simplified model assumes that all PSs with luminosity $L>L_\text{th}$ are resolved, and none with $L<L_{\text{th}}$. In Ref.~\cite{Zhong:2019ycb} and here, $L_\text{th}$ is fixed throughout the sky at $\SI{e34}{\erg\per\second}$. The four properties we intend to measure are then given by
\begin{equation}
    \begin{split}
        L_\text{GCE} &= N_\text{GCE}\int_{L_\text{min}}^\infty L P(L) dL \,, \qquad
        L_\text{r} = N_\text{GCE}\int_{L_\text{th}}^\infty L P(L) dL \,, \\
        N_r &= N_\text{GCE}\int_{L_\text{th}}^\infty P(L) dL \,,
        \label{eqn:observables-sens-1}
    \end{split}
\end{equation}
where $N_\text{GCE}$ is a normalization constant, fixed by requiring that the luminosity of the GCE $L_\text{GCE}$ reproduces the flux $F_\text{GCE}$ observed. The conversion between $L_\text{GCE}$ and $F_\text{GCE}$ required to force this equivalence is outlined in appendix \ref{app:lum-to-flux}. We call this sensitivity model the simplified model.

The threshold value $L_\text{th}$ typically has a strong impact on the values given by eq.~\ref{eqn:observables-sens-1}. A more detailed analysis of the validity and effect of the $\SI{e34}{\erg\per\second}$ estimate is carried out in appendix \ref{app:step-thresh}.

The final sensitivity model, which we label the ``smoothed model'', has been used elsewhere in the literature to model the probability for a MSP to be detected as a pulsar, not simply as a gamma-ray PS. Given a sensitivity model of this type, we could confront our predicted $N_r$ and $R_r$ values with the observed $N_r$ and $R_r$ values from {\it detected pulsars} in the ROI relevant to the GCE, potentially allowing us to discard non-pulsar background sources and leading to significantly stronger constraints. We adopt the prescription of Ref.~\cite{Ploeg:2020jeh}, 
\begin{equation}
    P_\text{r}(F) = \frac{1}{2} \parens{1 + \text{erf} \parens{\frac{\log_{10} F - (\log_{10} F_\text{th}(\ell, b) + K_\text{th})}{\sqrt{2}\sigma_\text{th}}}}.
    \label{eqn:ploeg-smoothing}
\end{equation}
Here rather than having a step-function threshold, for each source we assign a probability of detection $P_\text{r}(F)$ as a function of its flux $F$ and position. A constant parameter $10^{K_\text{th}}\ \SI{}{\erg\per\second}$ is added to the baseline 4FGL-DR2 threshold $F_\text{th}(\ell, b)$ (as used in our standard sensitivity model) to accommodate a higher threshold for pulsar identification. The width of the threshold --- accounting for uncertainties in $K_\text{th}$ or the published thresholds, statistical variations of photon counts, or temporal variation in background radiation --- is controlled by a dimensionless parameter $\sigma_\text{th}$. Ref.~\cite{Ploeg:2020jeh} extracted values for $K_\text{th}$ and $\sigma_\text{th}$ from a Markov chain Monte Carlo fit to globular cluster MSPs, finding $K_\text{th} = 0.45$ and $\sigma_\text{th} = 0.28$; while the inner Galaxy region is quite a different environment from globular clusters, we adopt these values in this work for illustration. 

In this framework, the observables $F_r$ and $N_r$ are calculated simply by multiplying the integrand of the luminosity integral in Eq.~\ref{eqn:observables-sens-2} by $P_r(L/4\pi s^2)$ for $F_\text{r}$ and $N_r$ and removing the lower limit of integration. (Note that there is no need to apply this modification to $F_\text{GCE}$ and $N_\text{GCE}$, because we do not require that the $N_\text{GCE}$ pulsars making up the total $F_\text{GCE}$ flux are resolved.)

\section{Luminosity functions}
\label{sec:lum-funcs}

\subsection{General parameterizations}

Luminosity functions for MSP populations are frequently parameterized either as a power law with a cutoff or break, or as a log-normal function with a broad peak around some characteristic luminosity. In this work we study each of these cases in general, to understand what regions of parameter space are still permitted to explain the excess and what they predict for resolved and unresolved sources, and compare with specific benchmark points given in the literature.

Our first parameterization is a power law with an exponential cutoff at high flux and a step-function cutoff at low flux (as used e.g. in Refs.~\cite{Zhong:2019ycb, Bartels:2018xom}), shown here in its normalized form:
\begin{equation}
    \frac{dN}{dL} \propto P_\text{PL}(L) = L^{-\alpha} \expp{-\frac{L}{L_\text{max}}}\brackets{\Gamma\parens{1-\alpha, \frac{L_\text{min}}{L_\text{max}}}L_\text{max}^{1-\alpha}}^{-1}.
    \label{eqn:power-law}
\end{equation}
This luminosity function restricts the range of luminosities to $[L_\text{min}, \infty)$, where $L_\text{min}$, $L_\text{max}$, and $\alpha$ are free parameters. It is in practice quite similar to a power law with a step-function cutoff or sharp break around $L_\text{max}$; the details of the cutoff at $L_\text{min}$ are not generally observable, since they describe the behavior of very faint point sources.

Our second parameterization is a log normal luminosity function (as used in e.g. Ref.~\cite{Hooper16}):
\begin{equation}
    \frac{dN}{dL} \propto P_\text{LN}(L)= \frac{\log_{10} e}{\sigma \sqrt{2\pi} L}\expp{-\frac{\parens{\log_{10} L - \log_{10} L_0}^2}{2\sigma^2}},
    \label{eqn:log-normal}
\end{equation}
where $L_0$ and $\sigma$ are free parameters. Here, all values of $L>0$ are allowed.

We also study a broken power law luminosity function:
\begin{equation}
    \frac{dN}{dL} \propto P_\text{BPL}(L) = \parens{\frac{\parens{1-n_1}\parens{1-n_2}}{L_b \parens{n_1 - n_2}}}\begin{cases}
        \parens{\fraci{L}{L_\text{b}}}^{-n_{1}} & L < L_{b} \\
        \parens{\fraci{L}{L_\text{b}}}^{-n_{2}} & L > L_b
    \end{cases},
    \label{eqn:nptf}
\end{equation}
with parameters $L_\text{b}$, $n_1$, and $n_2$. We use this luminosity function to benchmark our results, but do not scan the parameter space as we do for the power law and log normal functions.

\subsection{Benchmark configurations}

 To better interpret these functional forms, we further study seven benchmark configurations which have been proposed in the literature.

We use two benchmarks for the power law luminosity function from Refs. \cite{Zhong:2019ycb} and \cite{Bartels:2015aea}. Both sources perform a wavelet search on GCE data, isolate point sources in the GCE, and test what properties are required for a power-law luminosity function to match these observations.

Ref.~\cite{Zhong:2019ycb} found that $L_\text{min}=\SI{e29}{\erg\per\second}$, $L_\text{max}=\SI{e35}{\erg\per\second}$, and $\alpha =1.94$ exactly reproduces the observed number of resolved wavelet-selected PSs $N_r=47$ and the corresponding flux fraction $R_r$, while producing the full GCE flux. (As noted above, this prediction employs the simplified sensitivity model and a lower GCE flux than our baseline analysis.) They find that this model predicts approximately three million MSPs (almost entirely unresolved) in the GCE. Although this luminosity function matches observations, it is unusual in that it has a very steep slope / large spectral index $\alpha$ with a low $L_\text{min}$, corresponding to an average luminosity $\SI{1.9e30}{\erg\per\second}$ that is much lower than the other six benchmarks discussed here. We label this benchmark as ``Wavelet 1''.

Ref.~\cite{Bartels:2015aea} included in their analysis 13 resolved PSs in the 3FGL catalog, together with a population of wavelet peaks not associated with 3FGL sources. They assumed a power-law slope of 1.5 and a step-function cutoff at $L=L_\text{max}$. Fitting to this population (assuming all wavelet peaks were associated with the GCE) yielded a best-fit value for the cutoff of $L_\text{max}\approx\SI{7e34}{\erg\per\second}$. We adopt these parameters, but with an exponential cutoff at the same $L_\text{max}$. This configuration has average luminosity $\SI{7.4e31}{\erg\per\second}$, and we label this benchmark ``Wavelet 2''.

We benchmark the log normal luminosity function with three configurations found in the literature. Ref.~\cite{Hooper16} fits a log normal model to data from globular cluster MSP populations, yielding values $L_0 = 8.8^{+7.9}_{-4.1}\times \SI{e33}{\erg\per\second}$ and $\sigma=0.62^{+0.15}_{-0.16}$. We label this luminosity function the ``GLC'' benchmark, since it is based on observations of globular clusters.

Ref.~\cite{Ploeg:2020jeh} proposes a more intricate luminosity function, derived from a model of the relationship between MSP spectrum and physical properties. They split the MSP population into three components: a Disk component containing most of the Galactic resolved MSPs, and a Boxy Bulge and Nuclear Bulge component which both contained unresolved MSPs and make up the GCE. A luminosity function is generated for each. We use the Boxy Bulge luminosity function, since the Boxy Bulge extends out to $\SI{}{\kilo\parsec}$ scales whereas the Nuclear Bulge is smaller in radial extent and lies mostly within our mask; that said, the luminosity functions are very similar.  While this luminosity function is generated numerically, it very closely resembles a log normal curve with $L_0 = \SI{1.3e32}{\erg\per\second}$ and $\sigma=0.70$ (these values were obtained from a simple least-squares fit, detailed in appendix \ref{app:lum-func-fit}). We label this luminosity function the ``GCE'' benchmark, as it is fitted in part based on observations of the Bulge and is intended to fit the GCE.

Another recent study analyzes the possibility that the hypothetical MSP population in the GCE is generated not via the spin-up of old pulsars in low-mass X-ray binary progenitor systems, but directly via accretion-induced collapse (AIC) \cite{Gautam:2021wqn}. The population synthesis model developed in this work yields a numerical luminosity function for MSPs in the GC, reported as a function of flux. It predicts hundreds of thousands of low-flux MSPs in the GC with very few being bright enough for detection. For ease of comparison to the other luminosity functions discussed here, we convert the reported flux distribution to a luminosity distribution as described in appendix \ref{app:lum-to-flux}, and show in appendix \ref{app:lum-func-fit} that this distribution closely resembles a log normal function with $L_0 = \SI{4.3 \pm 0.2e+30}{\erg\per\second}$ and $\sigma = 0.94$. These parameters were obtained via a least squares fit in appendix \ref{app:lum-func-fit}. We label this luminosity function the ``AIC'' benchmark, as it employs an AIC-based population synthesis model.

Our first broken power law benchmark is derived from Ref.~\cite{Bartels:2018xom}, which uses a self-consistent Bayesian analysis to fit a luminosity function, spatial distribution, and other properties to observed MSPs in the Galactic Disk. That work derives parameters of $n_1=0.97$, $n_2=2.60$, and $L_\text{b}=\SI{1.7e33}{\erg\per\second}$, additionally impose upper and lower limits on the luminosity of MSPs of $\SI{e37}{\erg\per\second}$ and $\SI{e30}{\erg\per\second}$ respectively. We make use of these limits as well by modifying the normalization of Eq.~\ref{eqn:nptf}. We call this benchmark the Disk benchmark.

Our second broken power law benchmark is derived from Ref.~\cite{Lee:2015fea}'s search for point sources in the GCE via Non-Poissonian Template Fitting (NPTF). In this paper the flux distribution of sources was modeled as a broken power law, with parameters fitted to the data assuming a gNFW-squared-distributed population of MSPs (in addition to separate extragalactic and disk PS populations with different luminosity functions), and data restricted to a $\sim 2-12$ GeV energy range. We discuss the conversion to our energy range and to luminosity rather than flux in appendices \ref{app:lum-to-flux} and \ref{app:photon-energy}. The resulting best-fit parameters are $n_1=-0.66$, $n_2=18.2$, and $L_b = \SI{2.5e+34}{\erg\per\second}$; due to the size of $n_2$, this luminosity function behaves similarly to a power law with a step-function cutoff at $L_b$, and predicts that most of the power in the GCE arises from MSPs with flux comparable to \textit{Fermi}'s sensitivity. We call this luminosity function the ``NPTF'' benchmark.

\begin{figure}
    \centering
    \includegraphics[width=0.49\textwidth]{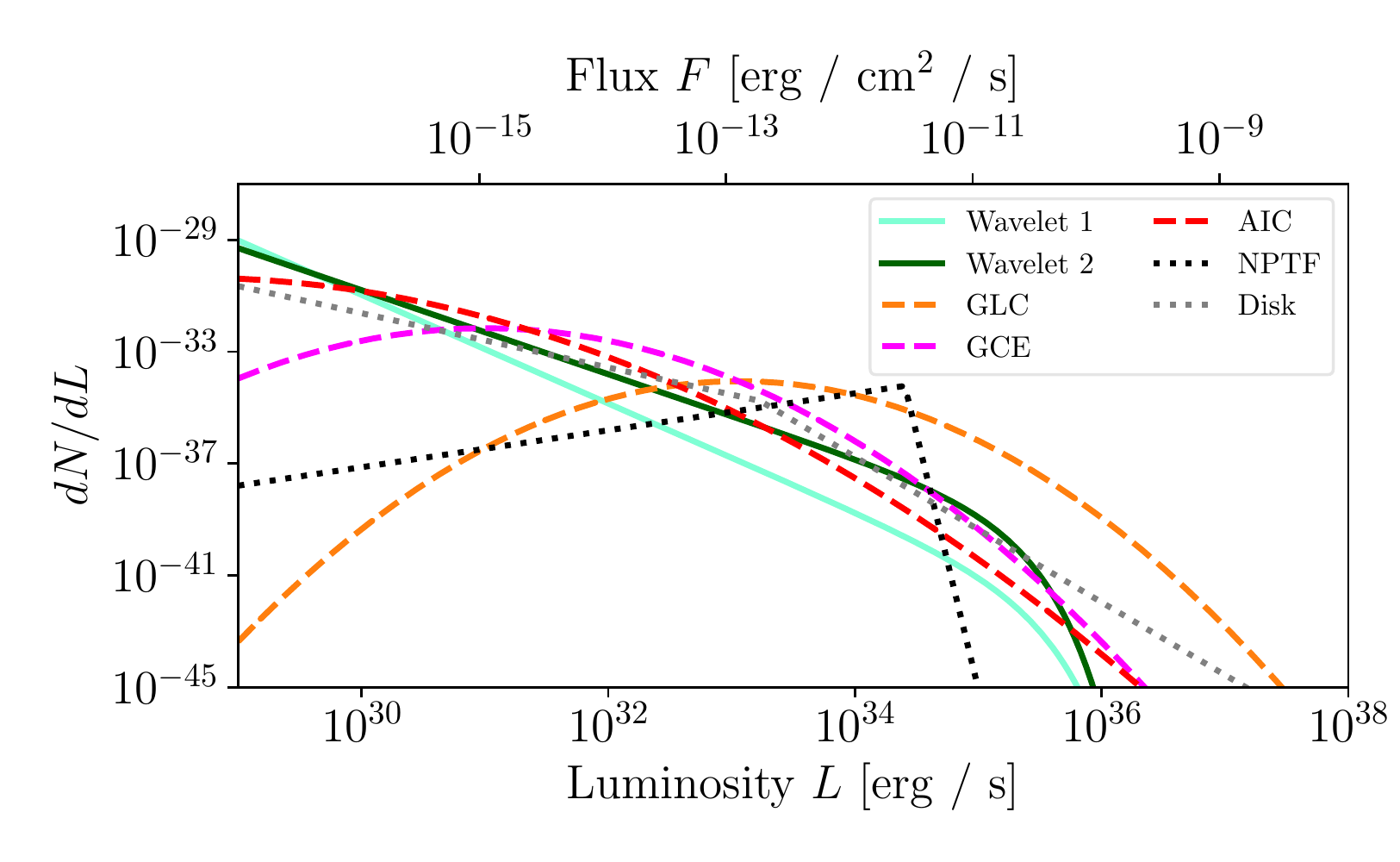}
    \includegraphics[width=0.49\textwidth]{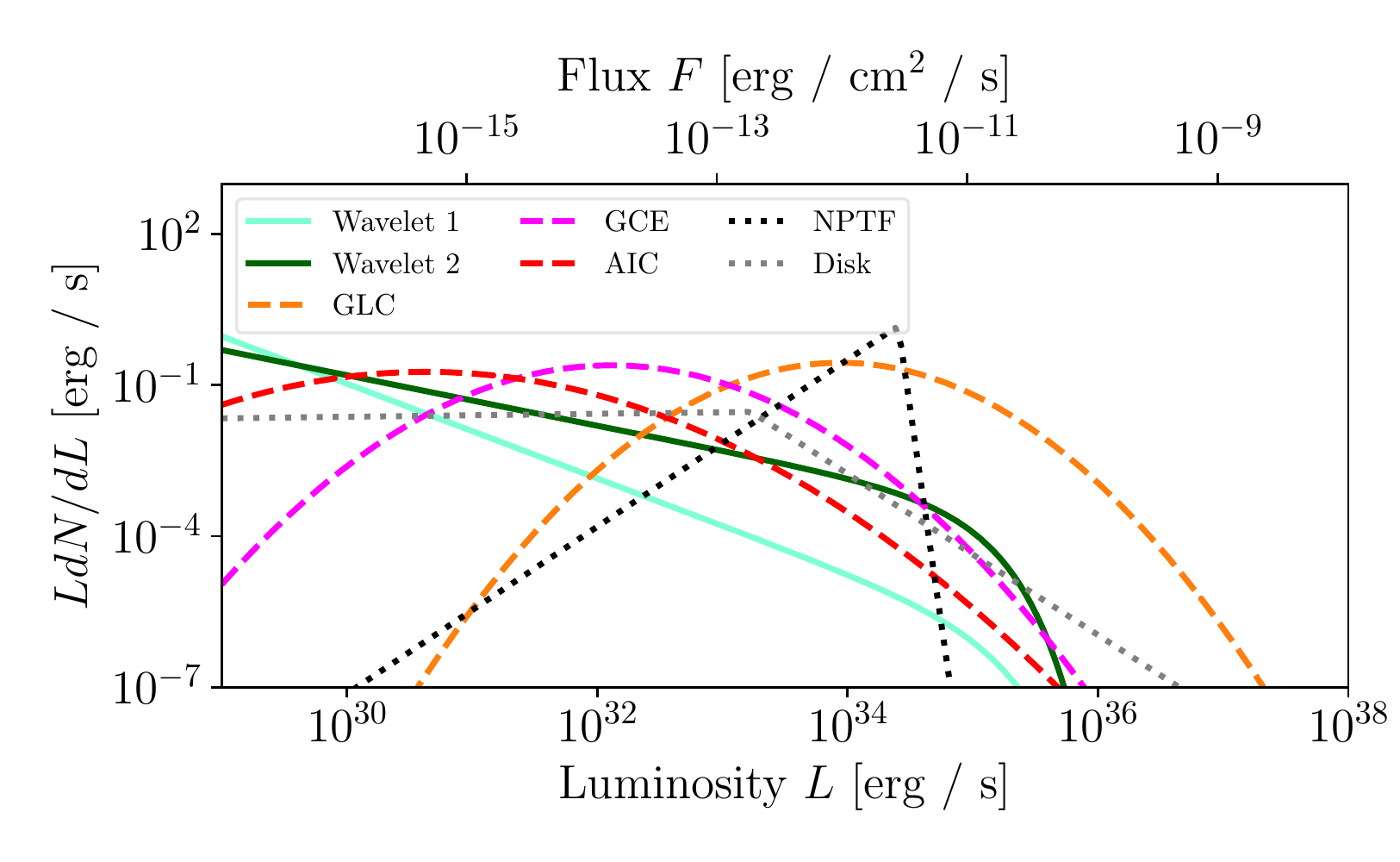}
    \includegraphics[width=0.49\textwidth]{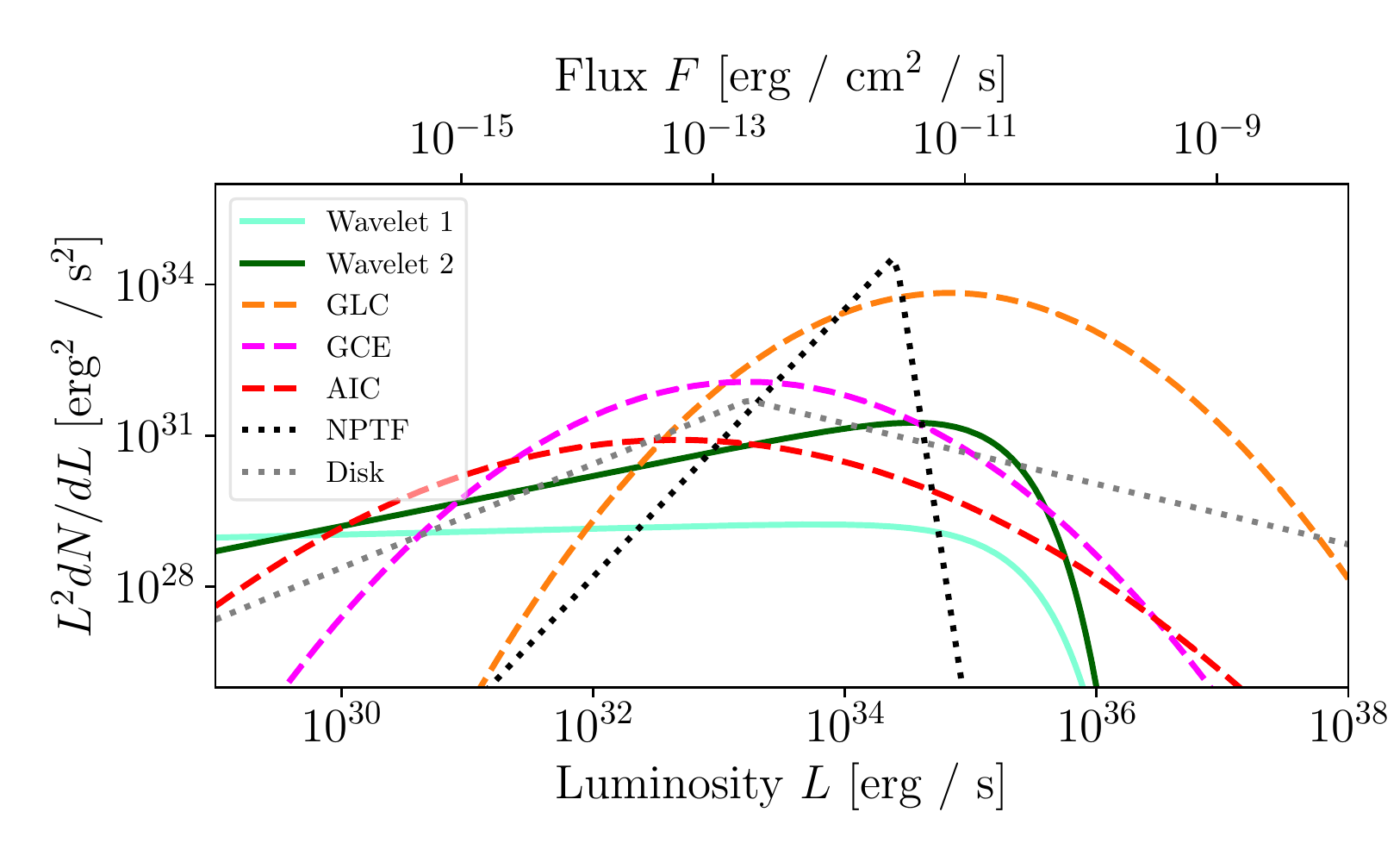}
    \caption{\textit{Left:} Seven benchmark luminosity functions for GCE MSP luminosity functions, normalized to give the luminosity probability density as in equations \ref{eqn:power-law} to \ref{eqn:nptf}, for (left panel) $dN/dL$, (right panel) $L dN/dL$, and (bottom panel) $L^2 dN/dL$. The references for each benchmark are: Wavelet 1 \cite{Zhong:2019ycb}, Wavelet 2 \cite{Bartels:2015aea}, GLC \cite{Hooper16}, GCE \cite{Ploeg:2020jeh}, AIC \cite{Gautam:2021wqn}, NPTF \cite{Lee:2015fea}, and Disk \cite{Bartels:2018xom}.}
    \label{fig:lum-funcs}
\end{figure}

All the above-mentioned luminosity functions are shown in figure \ref{fig:lum-funcs}; we plot $dN/dL$, $dN/d\ln(L) = L dN /dL$ (which shows how the total number of pulsars is distributed across e-folds in $L$), and $L dN /d\ln(L)=L^2 dN/dL$ (which shows how the total luminosity is distributed across e-folds in $L$) for clarity.
We note that despite the different parameterizations, several of these benchmarks are roughly similar in shape; the main exceptions are the GLC and NPTF luminosity functions, which have much more power near the high-luminosity cutoff than others, and the ``Wavelet 1'' benchmark, which is flatter than the others.

It is important to note that all these luminosity functions are peaked at flux values below the \textit{Fermi} sensitivity threshold around a few times $\SI{e34}{\erg\per\second}$ --- well below, in most cases. The high luminosity tail therefore dominates the resolvable MSPs and hence our observables. Our analysis is thus primarily sensitive to the luminosity function parameters insofar as they affect this tail, and to the sensitivity model, which dictates how much of the tail is exposed to \textit{Fermi}.

\section{Results}
\label{sec:results}
\subsection{Allowed luminosity function configurations}
\label{sec:allowed-configurations}
\begin{figure}
    \centering
    \begin{subfigure}[b]{0.49\textwidth}
     \includegraphics[width=\textwidth]{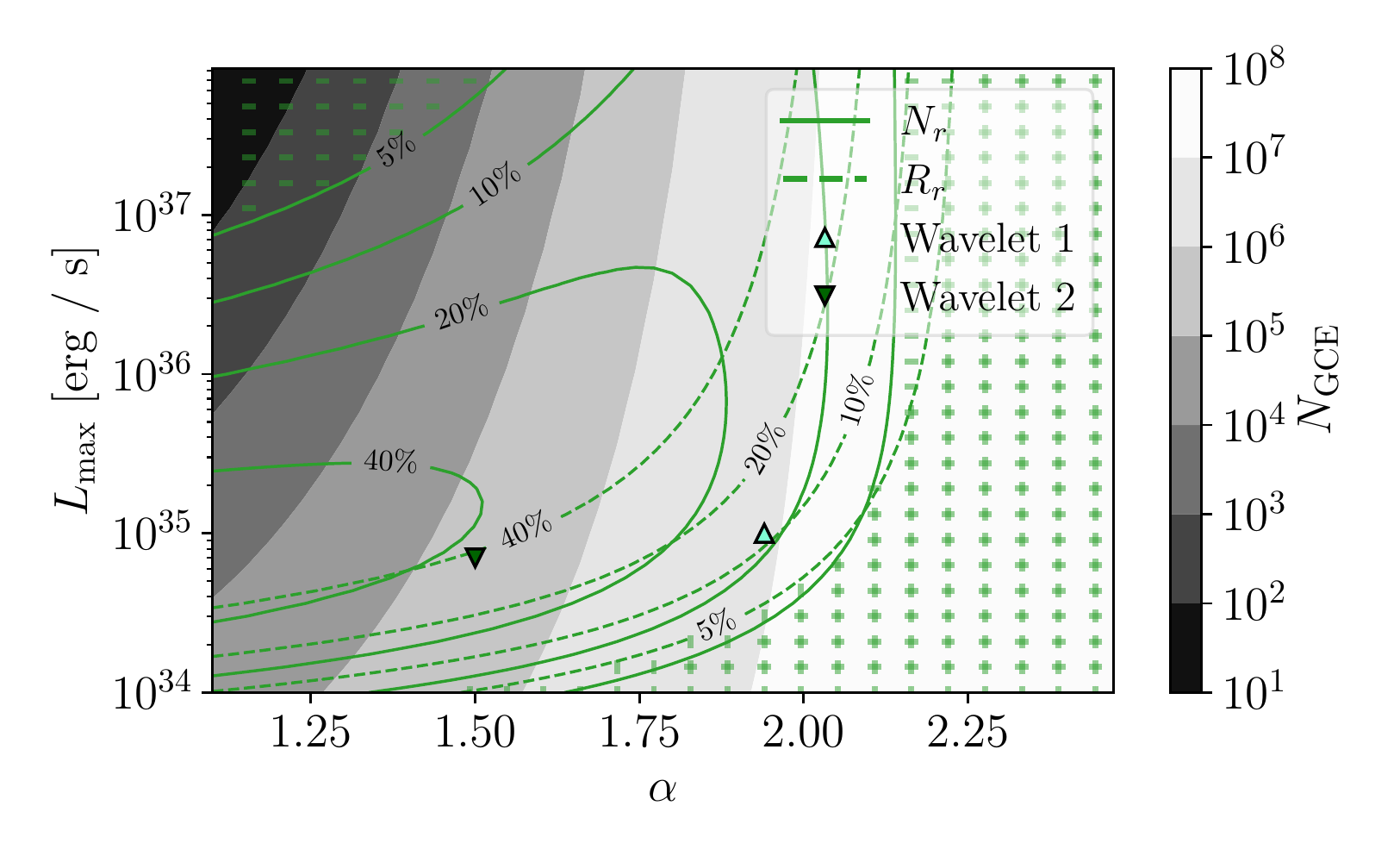}
        \includegraphics[width=\textwidth]{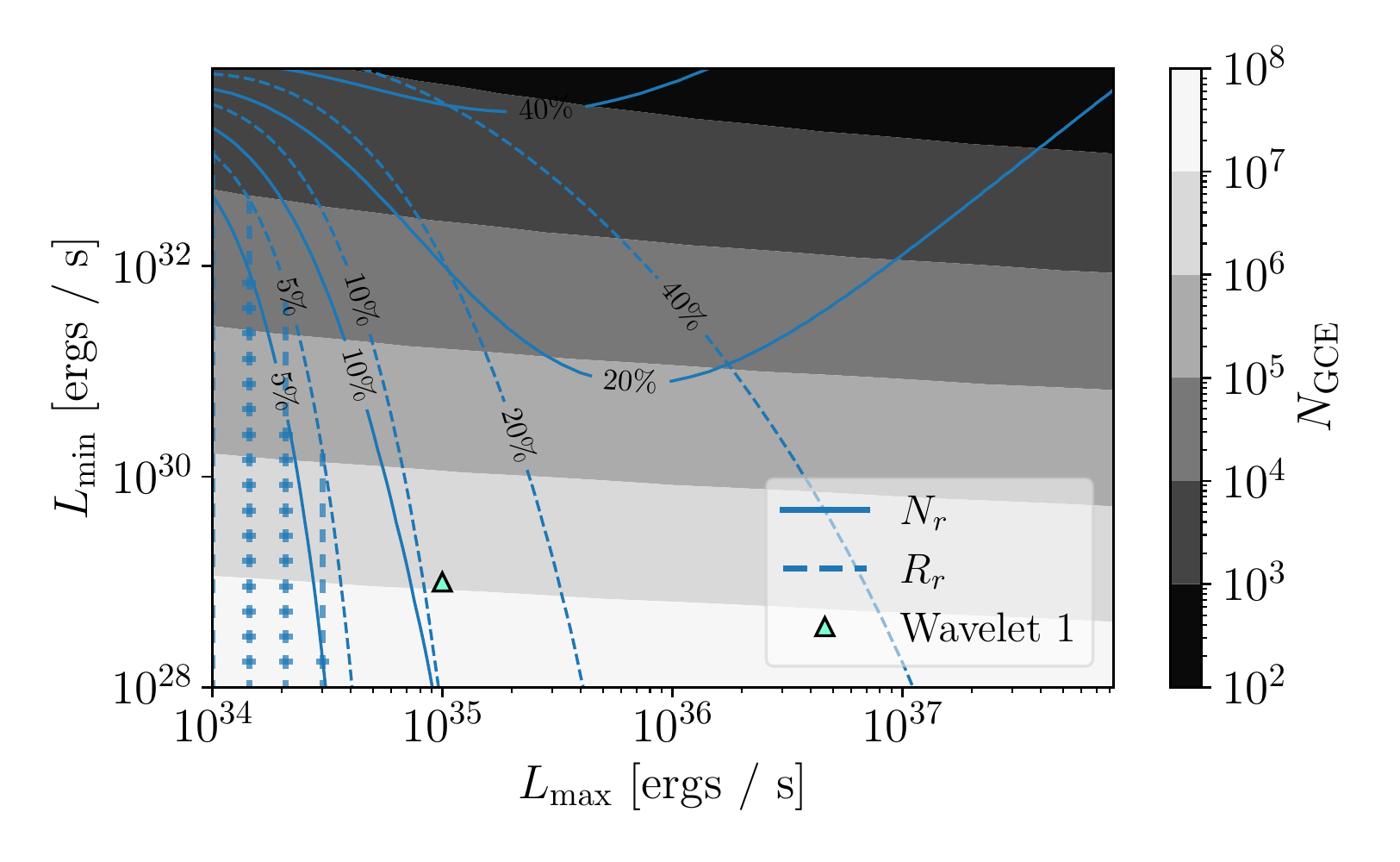}
        \includegraphics[width=\textwidth]{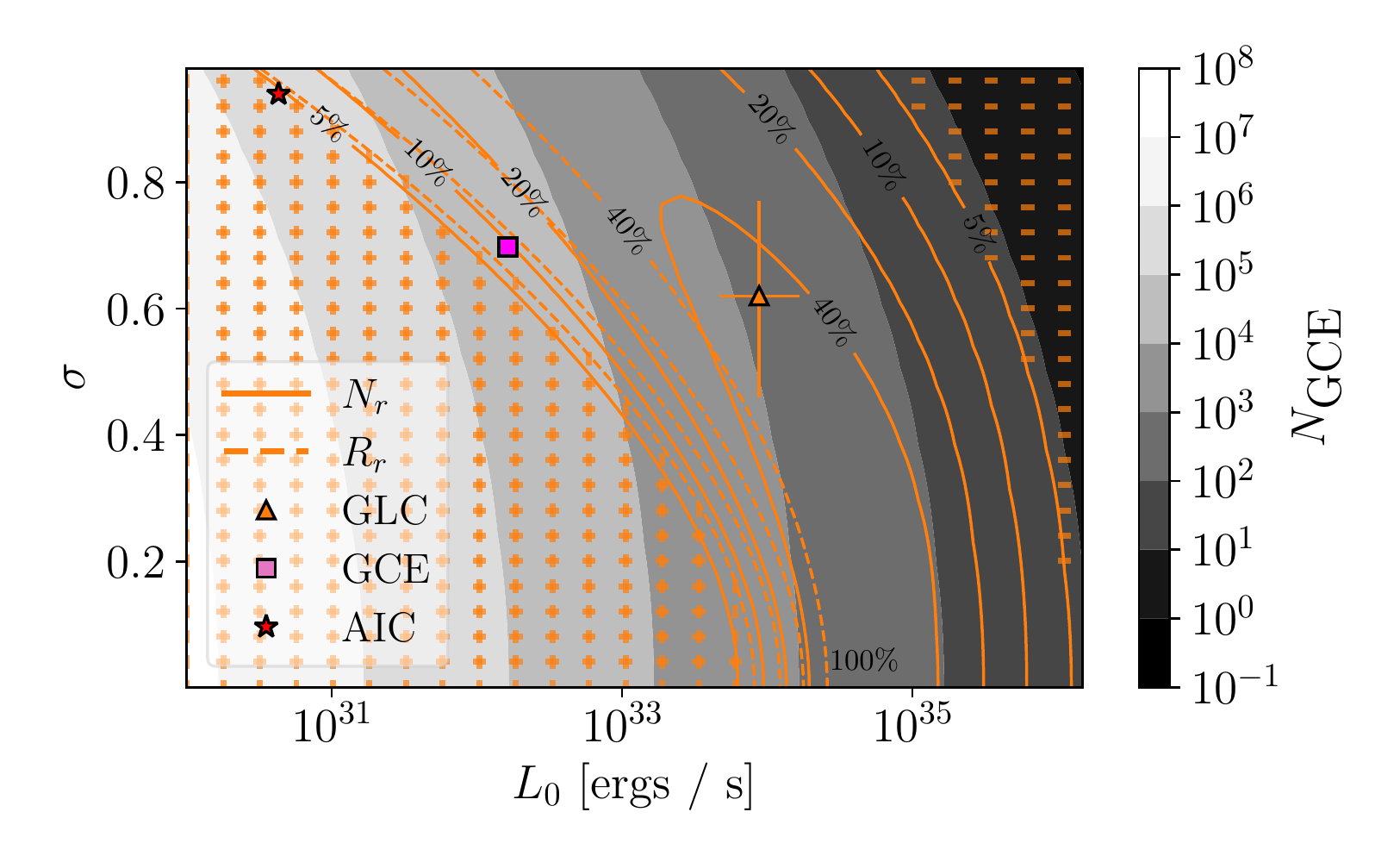}
        \caption{Standard sensitivity model}
        \label{fig:position-dependent}
    \end{subfigure}
    \hfill
    \begin{subfigure}[b]{0.49\textwidth}
            \includegraphics[width=\textwidth]{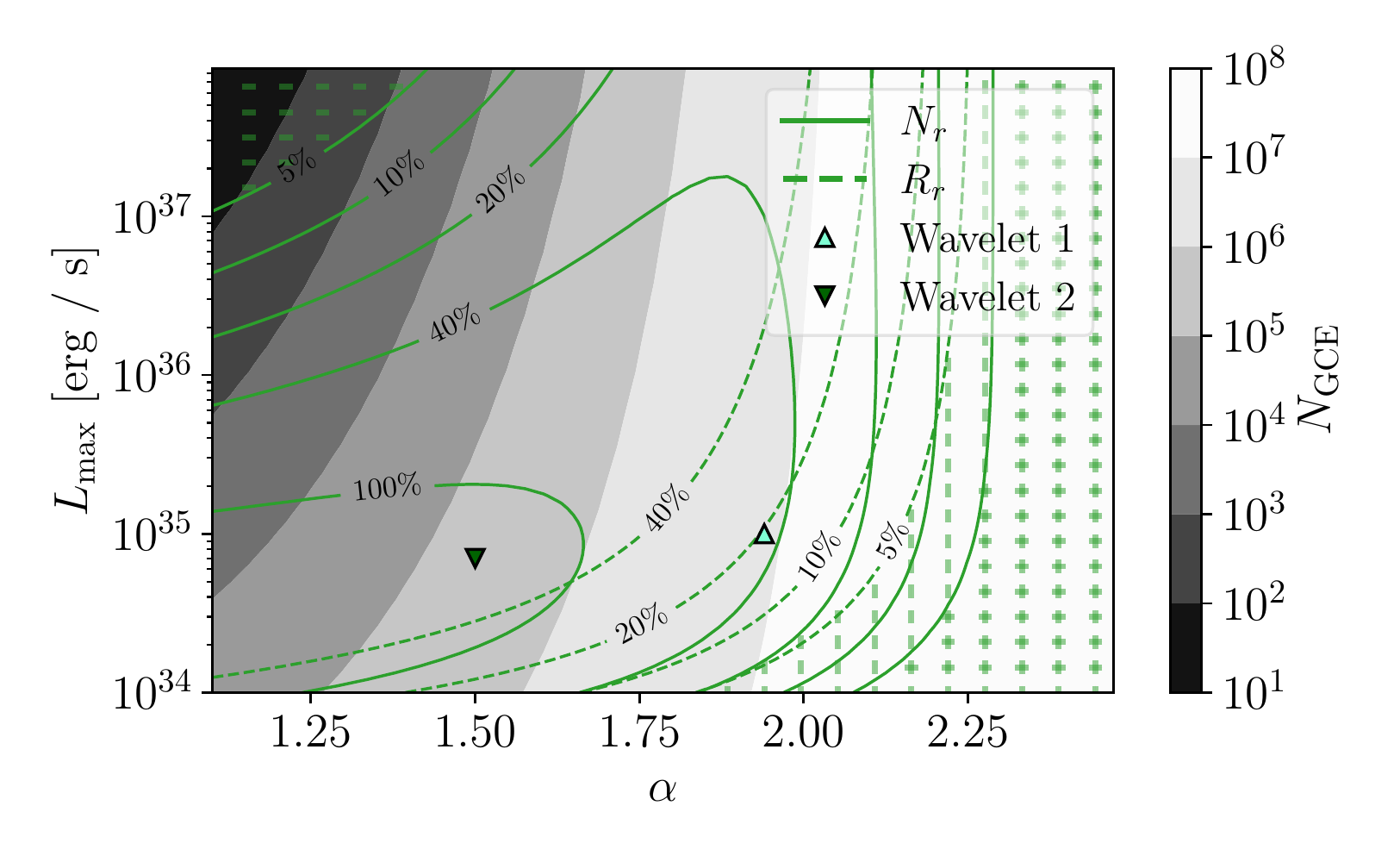}
        \includegraphics[width=\textwidth]{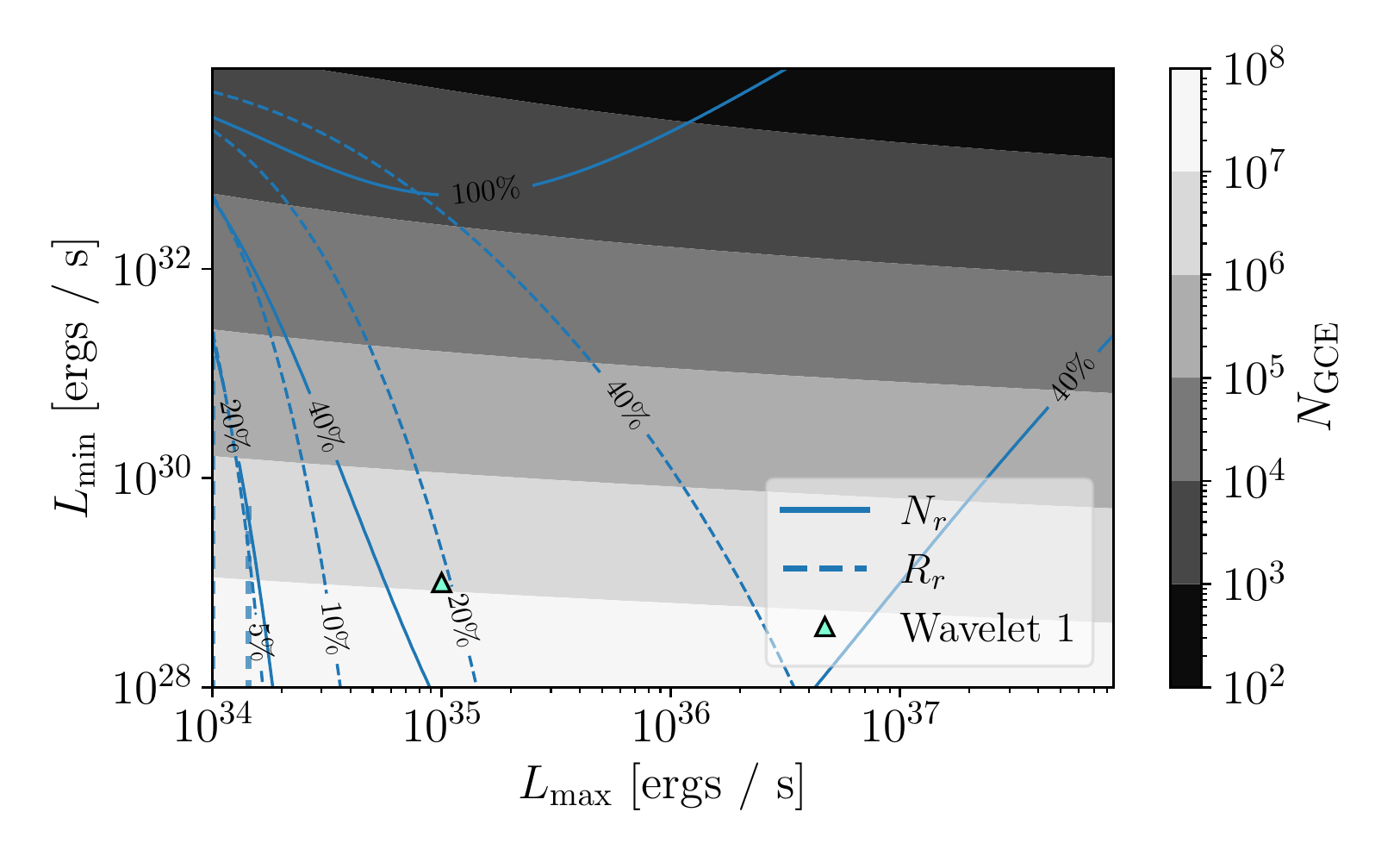}
        \includegraphics[width=\textwidth]{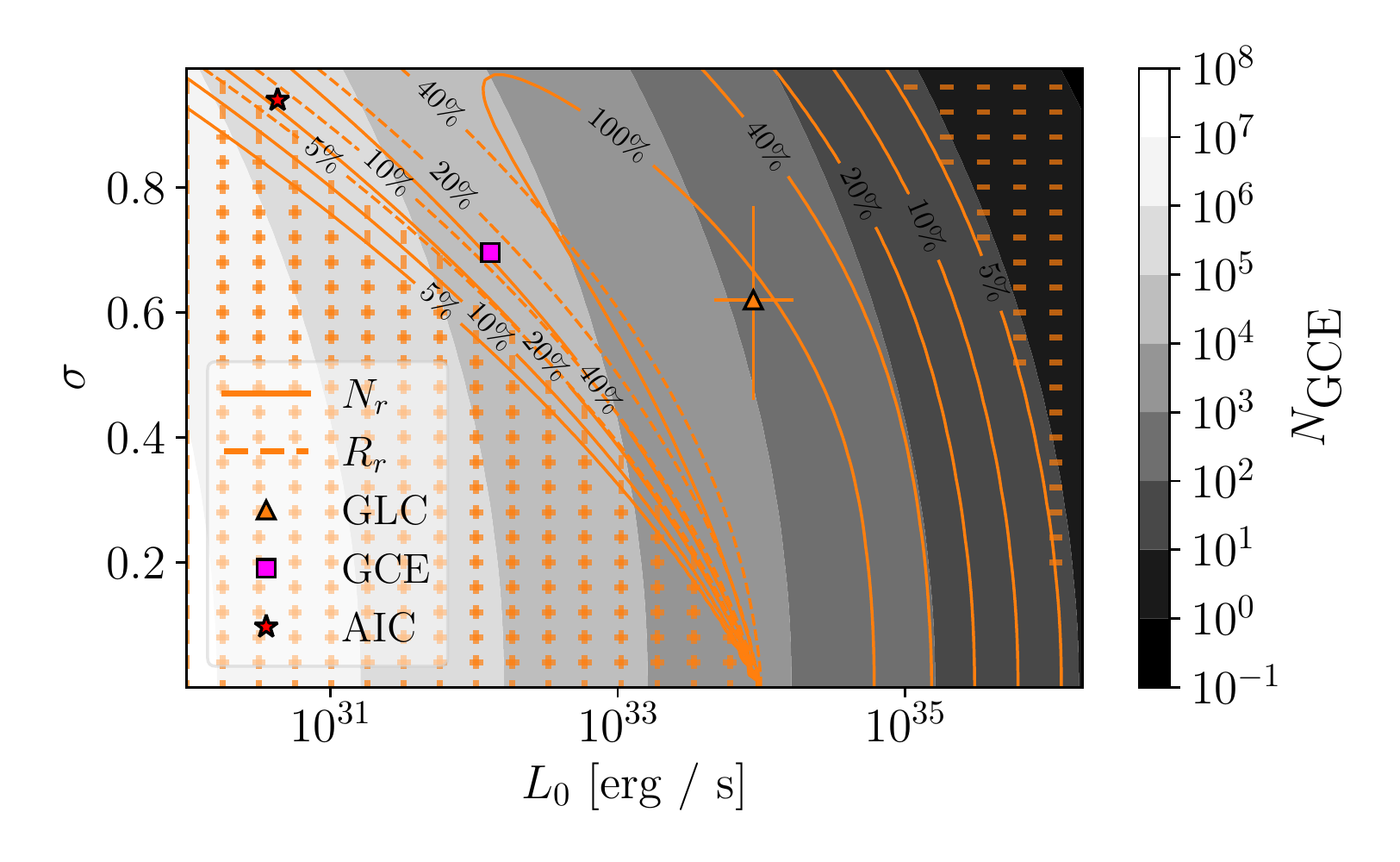}
        \caption{Simplified sensitivity model}
        \label{fig:step-function}
    \end{subfigure}
    \caption{$N_r$, $R_r$, and $N_\text{GCE}$ values for power-law (top two rows) or log normal (bottom row) luminosity functions, normalized to explain the full GCE flux as determined by Ref.~\cite{DiMauro:2021raz}. The benchmark configurations are marked (see text for details). The $N_r$ values (solid) are marked as fractions of the 4FGL-DR2 PSs passing our cuts (see text); the $R_r$ values (dashed) are marked as fractions of the total GCE flux. These fractions are labeled on the contours. The regions where $N_r$ ($R_r$) is less than $5\%$ of the number of observed resolved sources (GCE flux) are marked with \texttt{|} (\texttt{-}), while regions where both numbers are below this threshold are marked with \texttt{+}. The gray-scale contours indicate the total number of sources required to explain the GCE. The top row (green) fixes $L_\text{min}=\SI{e29}{erg\per\second}$ while the second row (blue) fixes $\alpha=1.94$.}
    \label{fig:step-and-pos}
\end{figure}

Using each of the luminosity function parameterizations and benchmarks outlined in section \ref{sec:lum-funcs}, we extract the total expected number of resolved MSPs $N_r$, the ratio of the flux received from those resolved MSPs to the total flux of the GCE $R_r$, and the total number of MSPs $N_\text{GCE}$ by forcing each function to reproduce the observed flux of the GCE as discussed in section \ref{sec:total-flux}. This is done for the parameter space of the power law and log normal luminosity functions (Eqs.~\ref{eqn:power-law} and \ref{eqn:log-normal}), with $N_\text{GCE}$ displayed as a contour map in figures \ref{fig:position-dependent} and \ref{fig:step-function}, using the standard and the simplified sensitivity models respectively.

The first two rows of these figures explore the parameter space for the power law luminosity functions with a cutoff at $L_\text{max}$. In the first row we hold $L_\text{min}$ fixed at $L_\text{min}=\SI{e29}{\erg\per\second}$, and vary $L_\text{max}$ and $\alpha$. In the second row we instead hold $\alpha$ fixed at the value corresponding to the Wavelet 1 benchmark ($\alpha=1.94$), to explore the effects of varying $L_\text{min}$. The third row explores the parameter space for log normal luminosity functions, in terms of $L_0$ and $\sigma$.

The observational constraints of $N_r$ constant and $R_r$ constant each trace out a one-parameter family of luminosity functions that are also displayed for various reference values. For $N_r$, we display $f=100\%$, 40\%, 20\%, 10\%, and 5\%, where $f$ is the fraction of the DR2 catalog modeled as part of the GCE. For $R_r$, we display the fraction of GCE modeled as produced by resolved sources, with reference values $R_r=40\%$, 20\%, 10\%, and 5\%. The regions with even less $N_r$ and $R_r$ are marked with $+$ symbols to denote that they are still consistent with observations. Luminosity functions slightly above the observational constraints may still be marginally allowed, given both the Poisson error bar in $N_r$ and systematic uncertainties arising e.g. from the choice of GCE flux. The observational curves are intended to indicate the regions of parameter space where tension with observational constraints starts to become a concern.

Relevant luminosity function benchmarks are also marked in figure \ref{fig:step-and-pos}; only the Wavelet 1 benchmark is shown in the second row because the choice of $\alpha$ in the second row is inconsistent with other benchmarks. The NPTF benchmark is not shown in the first row of power-law plots because its best-fit value for $\alpha$ is well off the left-hand side of the plot, near the bottom of the figure. The expected values of $N_r$, $R_r$, and $N_\text{GCE}$ at the benchmark configurations are given in table \ref{tab:specific-results} for the standard sensitivity model and the alternatives.

Focusing on the standard sensitivity model, we observe that in the power-law case, requiring that resolved sources make up less than $\sim 10-20\%$ of the GCE would drive us either to large values of $\alpha \gtrsim 2$ or values of $L_\text{max} \lesssim 10^{35}$ erg/s. In other words, if $L_\text{max}$ is substantially larger than the typical sensitivity threshold, a steep power-law slope is required to avoid overproduction of bright sources, and furthermore this implies a very large number of point sources ($\gtrsim 10^7$). This case corresponds to the solution identified in Ref.~\cite{Zhong:2019ycb}. If we only restricted the {\it number} of bright sources and not their flux, there would be another possible solution where a small number of very bright sources explain the whole GCE, but this is observationally excluded (as masking the bright sources does not remove the GCE). Increasing the sensitivity (as shown in the right-hand panels) pushes the threshold down and hence reduces the value of $L_\text{max}$ compatible with a wide range of choices for the power-law index, where the number of required MSPs is modest. In the second row of figures, we see that when $\alpha$ is held fixed at 1.94, demanding that the resolved sources make up less than $\sim 10-20\%$ of the GCE would primarily constrain $L_\text{max}$ (to a value between $\sim 10^{34} - 2\times 10^{35}$ erg/s), whereas the total number of sources is primarily controlled by $L_\text{min}$; if a larger flux fraction in resolved sources is permitted, the limit on $L_\text{max}$ will similarly rise, but without meaningfully constraining $L_\text{min}$ and the total number of required sources.

Figure \ref{fig:step-function} shows that the Wavelet 1 power law benchmark used by Ref.~\cite{Zhong:2019ycb}, which matches the properties of the wavelet-selected subsample exactly when that reference's GCE luminosity and the simplified sensitivity model are used ($N_r=47$ and $R_r=0.2$), fails to match these properties in our analysis even when the simplified sensitivity model is used. The reason is the larger GCE flux value we have taken from Ref.~\cite{DiMauro:2021raz}. This flux is somewhat more than twice that extracted from Ref.~\cite{Zhong:2019ycb}, and therefore more than twice as many MSPs are required to reproduce it; with the simplified sensitivity model, this scenario would thus predict more than double the number of resolved sources that occur in the wavelet-selected subsample (while closely matching the {\it fraction} of the GCE flux attributed to that subsample), since it was designed to match that subsample with a smaller GCE flux. However, using the standard sensitivity model, we see that this benchmark point is actually quite consistent with the wavelet-associated subsample when our standard sensitivity model is applied (see table \ref{tab:specific-results}); the simplified sensitivity model used in Ref.~\cite{Zhong:2019ycb} overestimates the number of detected sources by a larger factor than the effect of varying the flux.

We can ask where the point is that would correspond to matching both $N_r$ and $R_r$ for the wavelet-selected subsample, as an example of a possible resolved-source population yielding roughly 20\% of the total GCE flux (and roughly 20\% of the 4FGL-DR2 sources), if we use the standard sensitivity model and the flux from Ref.~\cite{DiMauro:2021raz} (while keeping in mind that there are substantial systematic uncertainties that can modify this value). Figure \ref{fig:position-dependent} indicates that in this case the configuration that matches both $N_r$ and $R_r$ values moves to $\alpha \approx 1.8$ and $L_\text{max} \approx  \SI{9e34}{\erg\per\second}$ if $L_\text{min}$ is fixed, or to $L_\text{max} \approx \SI{1e35}{\erg\per\second}$ and $L_\text{min} \approx \SI{8e31}{\erg\per\second}$ if $\alpha$ is fixed instead. 

These configurations possess average luminosities of $\sim\SI{5e30}{\erg\per\second}$ and $\SI{3e33}{\erg\per\second}$ respectively, compared with $\SI{1.9e30}{\erg\per\second}$ of the original benchmark, and millions and hundreds of thousands of MSPs respectively. The change in average luminosity and total number of MSPs demonstrates the strong dependence of the power law configuration satisfying observational constraints, and the associated total number of sources, on the sensitivity model used (as well as the assumed GCE flux) --- the large number of MSPs inferred by Ref.~\cite{Zhong:2019ycb} is not a requirement, even when a power-law luminosity function model is imposed.

As shown in table \ref{tab:specific-results}, Wavelet 2 and NPTF benchmarks both require high $N_r$ values --- over 100 resolved sources --- to reproduce the excess when using our standard sensitivity model; in other words, if these scenarios were realized, more than 40\% of all 4FGL-DR2 sources (including flagged sources) should be real and associated with the GCE source population. These scenarios also predict that a relatively large fraction of the flux of the excess should originate from resolved sources (38\% for Wavelet 2 and 26\% for NPTF), which already appears to be in some tension with observations, and this particular tension is not dependent on the chosen GCE flux value. The smoothed sensitivity model (used to estimate the detection threshold for MSPs elsewhere in the sky) predicts tens of detected inner-Galaxy MSPs for both of these scenarios.

A dedicated analysis of the effects of masking 4FGL-DR2 sources, the sensitivity to pulsars in this population, and/or the spatial distribution of these sources, could potentially sharpen these statements. Note that while both of these benchmarks were derived from examination of older inner Galaxy gamma-ray data, it is still quite reasonable for them to be in tension with the data now that the sensitivity of the catalogs have improved; the NPTF analysis may also have overestimated near-threshold PSs due to systematics from background mismodeling \cite{Leane:2019xiy, Leane:2020pfc, Leane:2020nmi}.

\begin{table}
    \centering
    \begin{subtable}[h]{\textwidth}
        \centering
        \begin{tabular}{|p{4cm} | >{\centering\arraybackslash}p{2cm} >{\centering\arraybackslash}p{2cm} >{\centering\arraybackslash}p{2cm}|}\hline
            Luminosity function & $N_r$ & $R_r$ & $N_\text{GCE}$\\
            \hline \hline
            Wavelet 1 & 31 & 0.11 & $\num{8.5e6}$ \\
            Wavelet 2 & 98 & 0.38 & $\num{2.2e5}$ \\
            GLC & 124 & 0.72 & 670 \\
            GCE & 20 & 0.059 & $\num{3.5e4}$ \\
            AIC & 12 & 0.039 & $\num{3.6e5}$ \\
            NPTF & 111 & 0.26 & 970 \\
            Disk & 30 & 0.13 & $\num{2.6e4}$ \\ \hline
        \end{tabular}
        \caption{Standard sensitivity model}
        \label{tab:position-dependent-results}
    \end{subtable}

    \vspace{2em}

    \begin{subtable}[h]{\textwidth}
        \centering
        \begin{tabular}{|p{4cm} | >{\centering\arraybackslash}p{2cm} >{\centering\arraybackslash}p{2cm} >{\centering\arraybackslash}p{2cm}|}
            \hline
            Wavelet 1 & 120 & 0.19 & $\num{8.5e6}$ \\
            Wavelet 2 & 320 & 0.59 & $\num{2.2e5}$ \\
            GLC & 310 & 0.91 & 660 \\
            GCE & 120 & 0.13 & $\num{3.4e4}$ \\
            AIC & 61 & 0.078 & $\num{3.6e5}$ \\
            NPTF & 770 & 0.93 & 960 \\
            Disk & 140 & 0.22 & $\num{2.5e4}$\\ \hline
        \end{tabular}
        \caption{Simplified sensitivity model}
        \label{tab:step-function-results}
    \end{subtable}

    \vspace{2em}

    \begin{subtable}[h]{\textwidth}
        \centering
        \begin{tabular}{|p{4cm} | >{\centering\arraybackslash}p{2cm} >{\centering\arraybackslash}p{2cm} >{\centering\arraybackslash}p{2cm}|}\hline
            Wavelet 1 & 9.6 & 0.053 & $\num{8.5e6}$ \\
            Wavelet 2 & 32 & 0.18 & $\num{2.2e5}$ \\
            GLC & 50 & 0.45 & 670 \\
            GCE & 5.5 & 0.024 & $\num{3.5e4}$ \\
            AIC & 3.5 & 0.018 & $\num{3.6e5}$ \\
            NPTF & 25 & 0.077 & 970 \\
            Disk & 10 & 0.078 & $\num{2.6e4}$\\ \hline
        \end{tabular}
        \caption{Smoothed sensitivity model}
        \label{tab:smoothed-results}
    \end{subtable}
    \caption{Number of resolved PSs, ratio of resolved flux to total flux, and total number of PSs predicted to make up the GCE based on seven luminosity function benchmarks (see section \ref{sec:lum-funcs}) and the requirement that the PSs reproduce the entire flux of the GCE, for three different sensitivity models (see section \ref{sec:sensitivity}).}
    \label{tab:specific-results}
\end{table}

For the log normal luminosity functions, we see that there are two broad regions of parameter space consistent with a small value of $N_r$, but one of these regions --- where $L_0$ is very large and so the number of sources required to explain the GCE is small --- is excluded by any plausible limit on $R_r$ (this again corresponds to the scenario where a handful of bright resolved sources explain the GCE; the fact that masking 4FGL sources does not eliminate the GCE excludes this region). In the remaining region of parameter space, the $N_r$ and $R_r$ constraint lines are quite similar, suggesting that log normal luminosity functions generating a fraction close to $R_r$ of the GCE will also quite generically predict a number of observed sources comparable to $N_r$ (and vice versa). However, the constraint lines spans several orders of magnitude in the total number of allowed sources, from around $10^3$ up to $10^5$ or more.

When we consider specific benchmarks studied in the literature, figure \ref{fig:step-and-pos} demonstrates that the GLC log normal benchmark predicts a GCE dominated by resolved sources ($R_r = 0.72$), which is quite inconsistent with observations; this is consistent with Ref.~\cite{Hooper16}, which proposed this luminosity function and argued it could not explain the GCE. On the other hand, the GCE and AIC benchmarks show no tension with observations, predicting a modest but not dominant contribution to the resolved sources that pass all our cuts (10-20 sources, yielding 3-5\% of the GCE flux). Note that the predictions from these benchmarks for resolved sources are rather similar (differing by a factor of about 1.5); the main difference in these luminosity functions occurs at the low-luminosity end and is responsible for a roughly one-order-of-magnitude difference in the required number of sources ($3.5\times 10^4$ for the GCE benchmark and $3.6\times 10^5$ for the AIC benchmark).

Because the smoothed sensitivity model has an overall offset in the detection threshold (controlled by the parameter $K_\text{th}$), we expect it to yield lower $N_r$ and $R_r$ values, and identical $N_\text{GCE}$, since the sensitivity model does not affect the total number of pulsars. From table \ref{tab:smoothed-results} we see that values of $N_r$ fall by a factor of $2-3$ in most cases compared to the standard sensitivity model, and $R_r$ falls by a factor of around two, demonstrating that all seven configurations studied are roughly equally responsive to the difference between the smoothed and standard sensitivity models.

\subsection{Number of MSPs in the GCE}
\label{sec:num-msps}
For the power law luminosity function, Ref.~\cite{Zhong:2019ycb} determined that about three million MSPs are required to reproduce the GCE using a simplified sensitivity model. Using the total GCE flux inferred from Ref.~\cite{DiMauro:2021raz}, which is 2--3 times higher, but the same parameters, that number inflates to 8.5 million. As discussed above, power-law configurations that match the wavelet-selected subsample of Ref.~\cite{Zhong:2019ycb} can be consistent with only $\mathcal{O}(10^4)$ MSPs if we use the flux value of Ref.~\cite{DiMauro:2021raz} and the standard sensitivity model (figure \ref{fig:position-dependent}), largely because $L_\text{min}$ is not observationally well-constrained but has a large impact on the total required number of sources.

For log normal luminosity functions, figure \ref{fig:step-and-pos} demonstrates that $N_\text{GCE}$ is much more sensitive to $L_0$ than to $\sigma$ for $\sigma < 1$. This is intuitively sensible, as to the degree that $L_0$ determines the average luminosity and hence the average flux of sources, it should simply be inversely related to the required number of sources. We can also read off from the figure that $N_\text{GCE} < 1000$ is generally inconsistent with requiring $R_r \lesssim 20\%$, and $N_\text{GCE} \ll 10,000$ is only achievable with low $\sigma$.

\subsection{Flux distribution}
\label{sec:flux-distribution}
Until this point, we have compared predictions for the population of MSPs in the inner Galaxy to observations only via the number and flux of resolved PSs. We may expand this analysis by comparing the predicted flux distribution of resolved MSPs to the observed distribution for resolved PSs in the inner Galaxy and the 0.1-100 GeV energy band (as described in figure \ref{fig:fgl-to-wavelet-comparison}). This will also serve as a cross-check on our sensitivity threshold estimate, as we should see both the observed and predicted sources fall off at low fluxes.

\begin{figure}
    \centering
    \begin{subfigure}[b]{0.47\textwidth}
        \includegraphics[width=\textwidth]{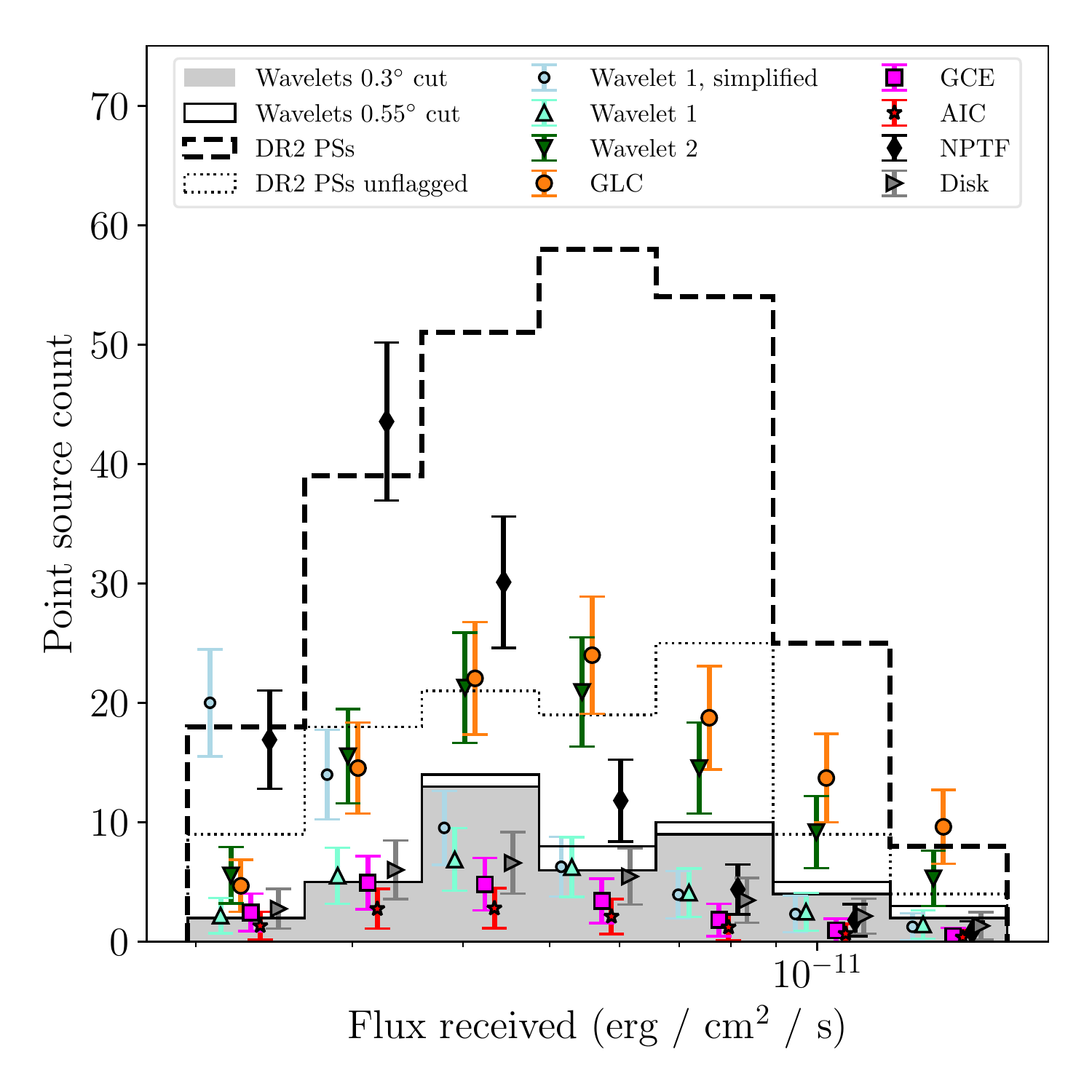}
        \caption{Each luminosity function is forced to reproduce the flux of the GCE.}
        \label{fig:flux-distro-fix-flux}
    \end{subfigure}
    \hfill
    \begin{subfigure}[b]{0.47\textwidth}
        \includegraphics[width=\textwidth]{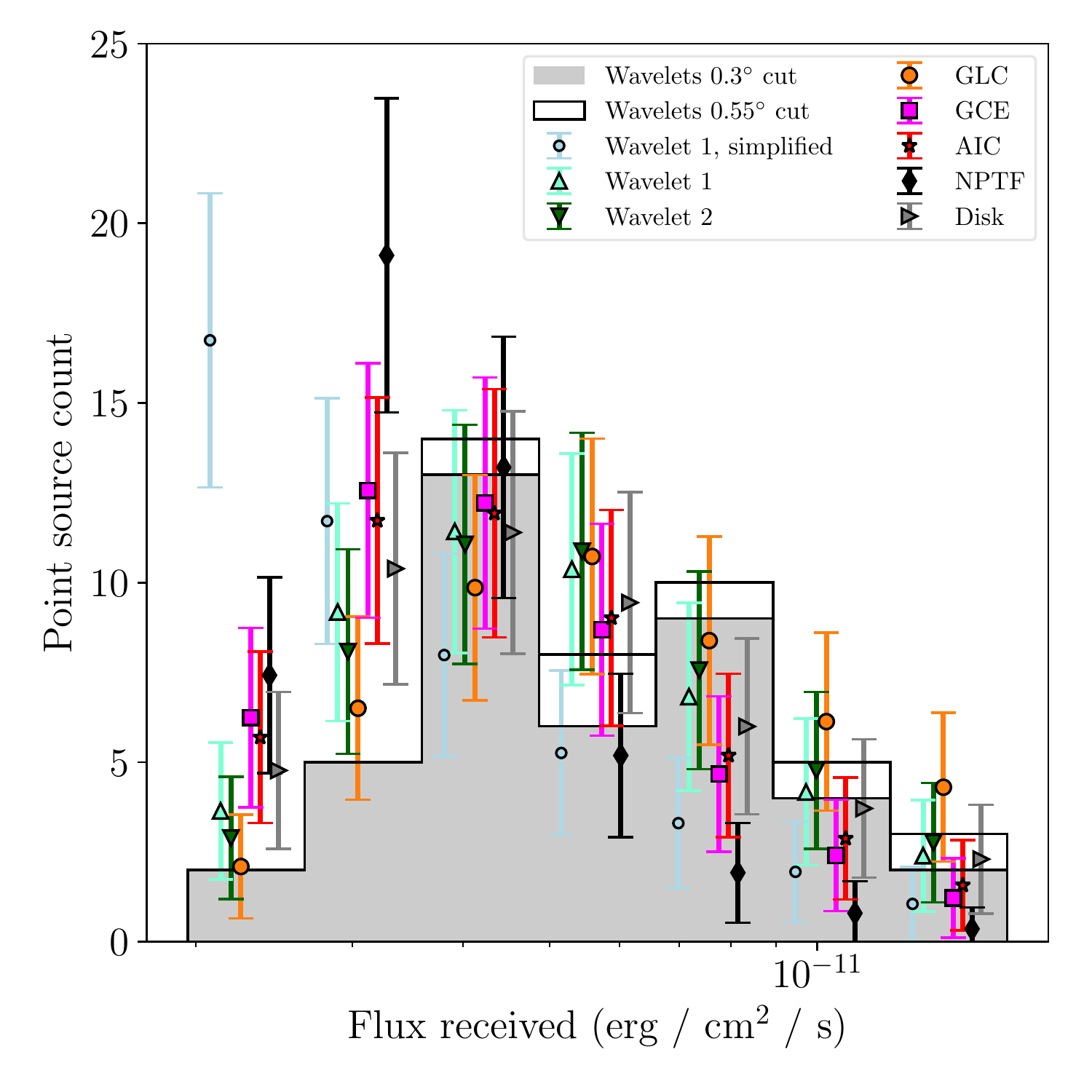}
        \caption{Each luminosity function is forced to reproduce 47 PSs, which is the observed number of sources in the wavelet-selected subsample.}
        \label{fig:flux-distro-fix-count}
    \end{subfigure}
    \caption{Distribution of resolved PSs predicted by each luminosity function with the standard sensitivity model, compared to the observed distribution of sources passing our cuts in the 4FGL-DR2 catalog (left panel, black dashed line), in the 4FGL-DR2 catalog with flagged sources removed (left panel, black dotted line), and in the wavelet-selected subsample (both panels) with a proximity cut of 0.3$^\circ$ (gray histogram) and 0.55$^\circ$ (hollow histogram). The predicted resolved PS distribution is also shown for the Wavelet 1 benchmark using the simplified sensitivity model (light blue points).}
    \label{fig:flux-distro}
\end{figure}

The left panel of figure \ref{fig:flux-distro} shows the observed flux distributions for 4FGL-DR2 sources (with flagged sources included/excluded) and the wavelet-selected subsample of 47 sources (or 41 with a more stringent proximity requirement), together with the predicted distributions for the seven benchmark luminosity functions, normalized to explain the total GCE flux. The right panel shows the wavelet-selected subsample of observed sources, and the predicted source distributions based on the benchmark luminosity functions, normalized to match the total number of sources (47) in this population. In some cases this latter normalization would require overproducing the full GCE; the purpose is to test the similarity in the distributions of resolved sources between models, when they are forced to have a common normalization. Both panels are calculated with the standard sensitivity model; for comparison, we also display the results of using the Wavelet 1 luminosity function with the simplified sensitivity model, as in Ref.~\cite{Zhong:2019ycb}.

From the left panel, we observe that binning in flux clarifies the degree of consistency between various luminosity function models and the data. Consistency of the NPTF model with the data would require essentially all 4FGL-DR2 sources (including flagged sources) with a flux below $\SI{4e-12}{\erg\per\centi\meter\squared\per\second}$ to be attributed to the GCE population; the tension arises from the lowest-flux bins. In contrast, the GLC model consistently attributes an appreciable fraction of the 4FGL-DR2 sources at all flux levels to the GCE, including slightly overpredicting the total sources in the highest-flux bin; the Wavelet 2 model has very similar behavior to the GLC model, but with a slightly lower normalization at high fluxes.  As expected from the total flux and source counts calculated earlier, the Wavelet 1, GCE, AIC and Disk models all predict quite low numbers of counts and will be difficult to exclude observationally with current data. Finally, the simplified sensitivity model clearly does not capture the observed drop-off in the number of sources at low fluxes due to decreased sensitivity, signaling that the threshold is underestimated.

This last point is even more apparent in the left panel, where the observed wavelet-selected population has a decline at fluxes below $\sim 3-4\times 10^{-12}$ erg/cm$^2$/s that is also observed in all the predictions using the standard sensitivity model; the example with the simplified sensitivity model matches the overall number of sources but overpredicts low-flux sources at the expense of high-flux ones. By eye, the distributions of resolved sources for all the luminosity function benchmarks with the standard sensitivity model are rather similar (once their normalization is fixed): we can quantify their differences by looking at the $p$-values associated with the fit of each benchmark resolved-PS flux distribution to the observed distribution of sources in the wavelet-selected subsample. (Note that a poor $p$-value should not be taken to exclude a specific model, since there is no guarantee that the wavelet-selected subsample is a good match to the true GCE population.)

In table \ref{tab:histogram-fits}, we show the $p$-values associated with the fit of each benchmark resolved flux distribution to the observed distribution shown in figure \ref{fig:flux-distro-fix-count}. Specifically, we let  $\chi^2 = -2\ln(\mathcal{L} / \mathcal{L}_0)$ where $\mathcal{L}$ gives the probability of achieving the binned data given the predicted resolved flux distribution, defined via a multinomial distribution (since the total sources are fixed). $\mathcal{L}_0$ is the probability of achieving exactly the model expectation values in each bin based on the analytically extended multinomial-distributed probability distribution function. We then extract a $p$-value using a $\chi^2$ distribution with degrees of freedom equal to the number of bins minus one.

\begin{table}
    \begin{tabular} {|l | c |}
    \hline
    Luminosity function & Number fixed $p$-value\\ \hline \hline
    Wavelet 1 & 0.64 \\
    Wavelet 2 & 0.86 \\
    GLC & 0.89 \\
    GCE & 0.0080 \\
    AIC & 0.046 \\
    NPTF & $\num{2.2e-10}$ \\
    Disk & 0.30 \\
    Wavelet 1, simplified model & $\num{7.3e-8}$ \\ \hline
    \end{tabular}
    \centering
    \caption{$p$-values comparing the predicted resolved flux distributions for each benchmark luminosity function to the observed distribution, where the overall normalization of the predicted distributions is chosen to be equal to the number of observed resolved PSs in the wavelet-selected subsample of 4FGL-DR2 sources.}
    \label{tab:histogram-fits}
\end{table}

We find that all of our re-scaled benchmark flux distributions, using the standard sensitivity model, are quite consistent in shape with the observed flux distribution of wavelet-selected PSs, except for the NPTF benchmark and (at much lower significance) the GCE benchmark. Even in the scenario where all these PSs, and only these PSs, were associated with the GCE (where we would have maximal signal to noise), it would therefore be challenging to differentiate luminosity functions similar to these benchmarks based on the observed flux distribution of the resolved sources. The NPTF is the one major exception: such a strongly peaked luminosity function could potentially be clearly distinguished from other benchmarks.


\section{Systematic uncertainties}
\label{sec:further-discussion}

\subsection{Dependence on assumed GCE total flux}

\begin{figure}
    \centering
    \includegraphics[width=0.49\textwidth]{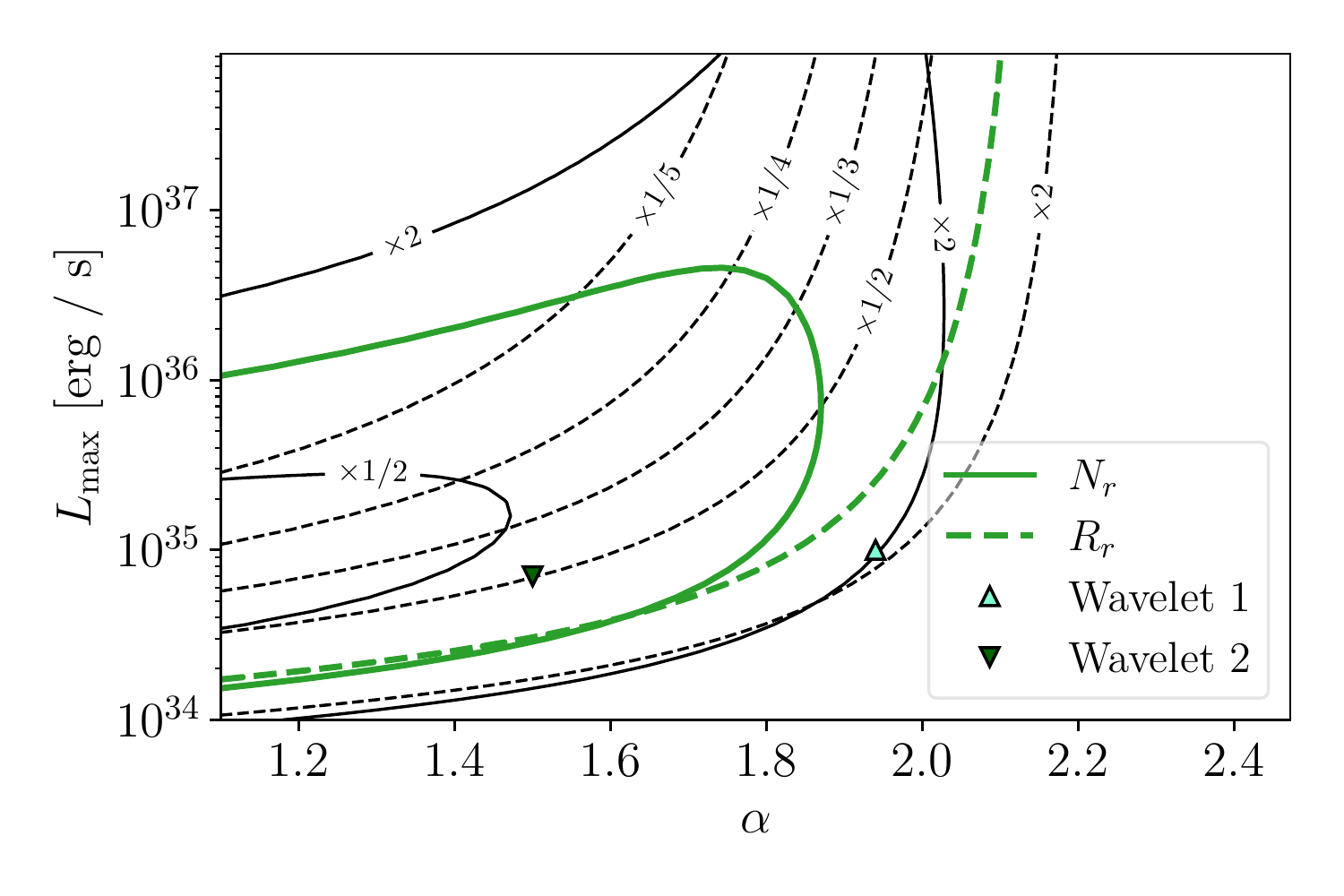}
    \includegraphics[width=0.49\textwidth]{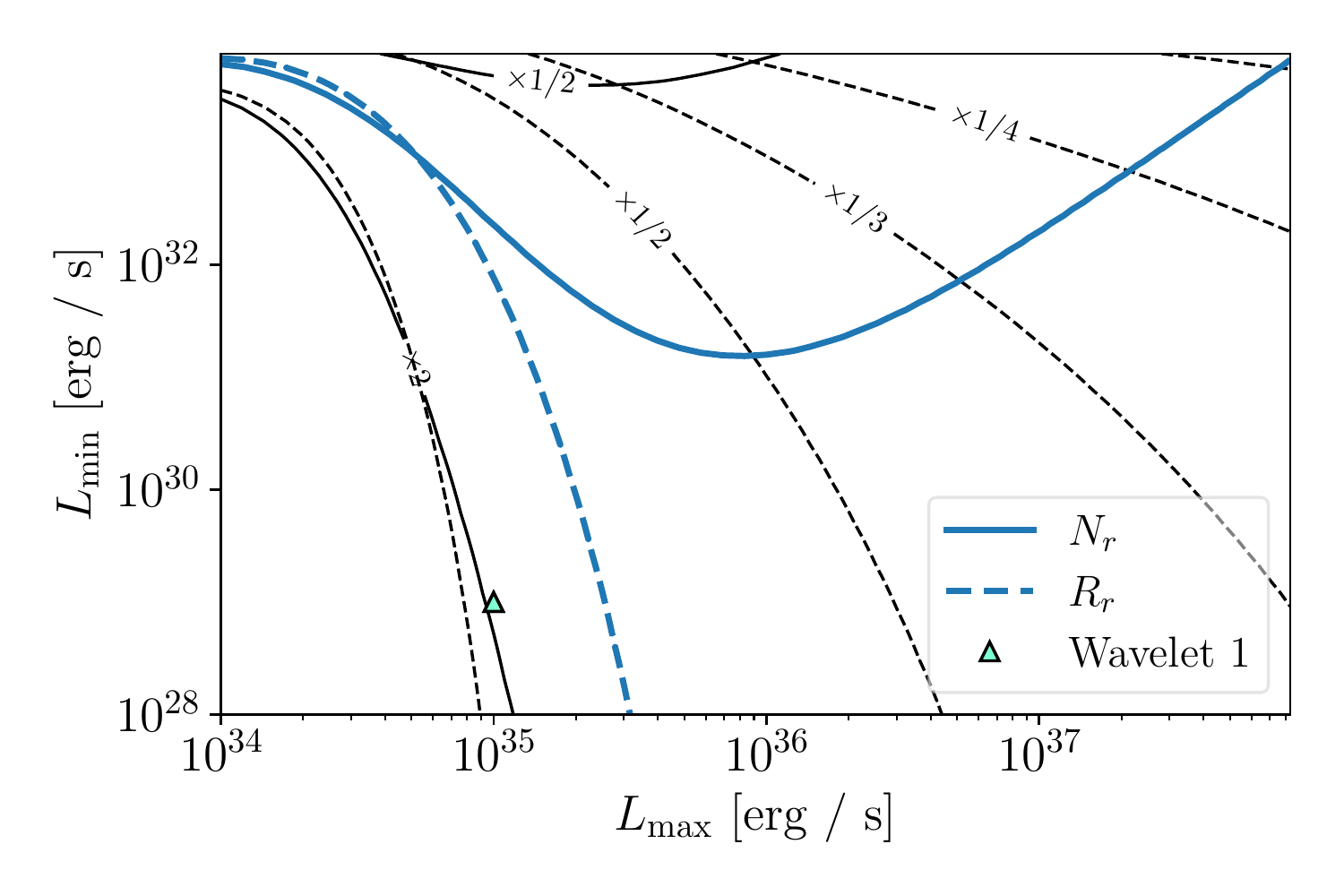}
    \includegraphics[width=0.49\textwidth]{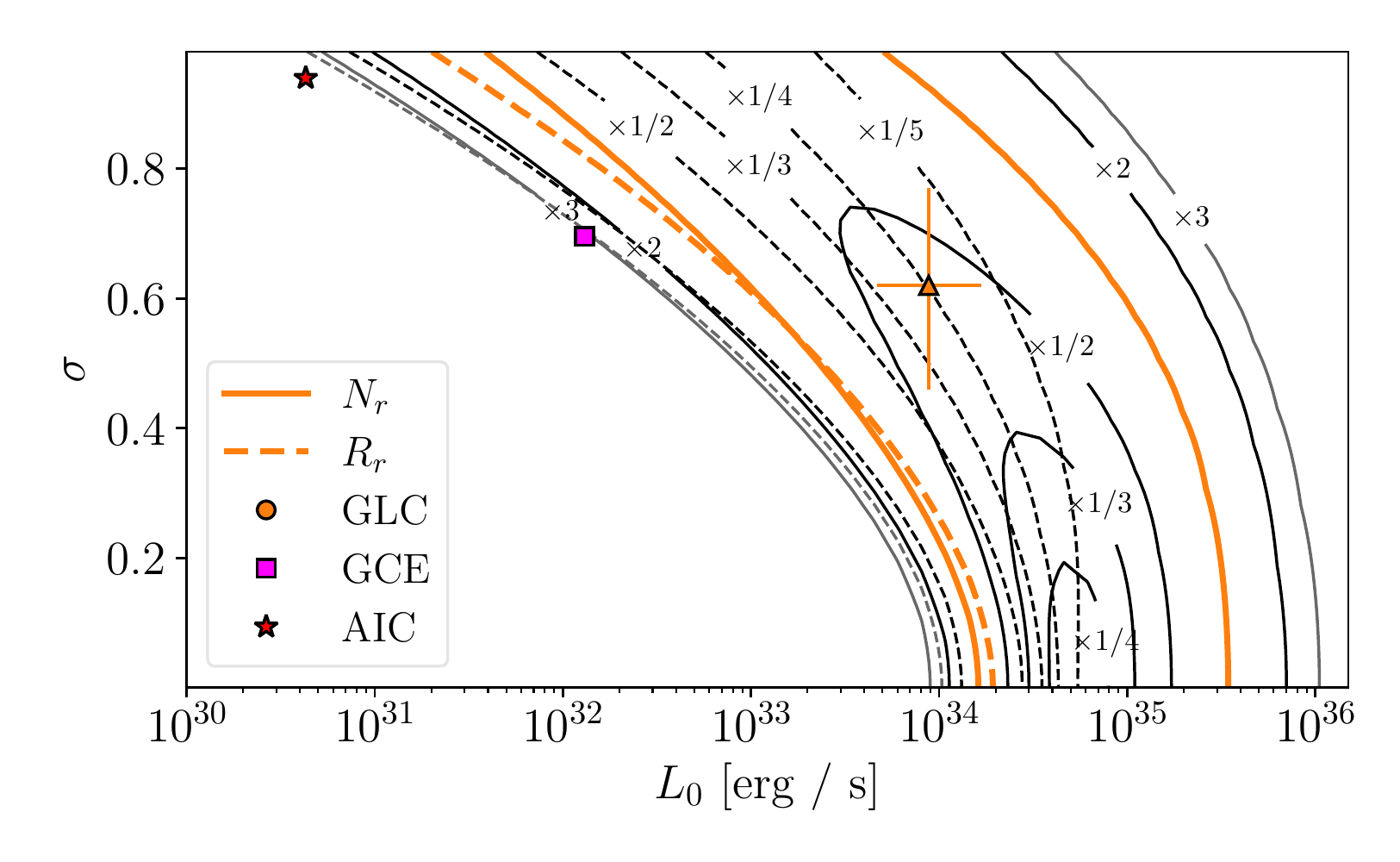}
    \caption{The parameter space that does not overproduce the wavelet-selected subsample of 4FGL-DR2 PSs ($N_r \leq 47$ and $R_r \leq 0.17$) for the power law luminosity function (top two plots) and log normal luminosity function (bottom plot). The total GCE flux has been increased by factors of 2--5 and decreased by factors of 2 and 3; contours of fixed $N_r$ (solid) and fixed flux in resolved PSs (dashed) are labeled with the factor by which $F_\text{GCE}$ has been multiplied. The colored lines exactly match the wavelet-selected subsample for the nominal GCE flux value (extracted from Ref.~\cite{DiMauro:2021raz}). The left power law plot fixes $\alpha=1.94$ while the right fixes $L_\text{min}=\SI{e29}{\erg\per\second}$.}
    \label{fig:multiple-fluxes}
\end{figure}

Figure \ref{fig:multiple-fluxes} shows luminosity function configurations that follow the observational constraints at different GCE fluxes, demonstrating how sensitive the regions corresponding to specific $N_r$ and $R_r$ values are to the total flux. To make the plots easier to follow, we have fixed the baseline $N_r$ and $R_r$ contours at the values corresponding to the wavelet-selected subsample of PSs, $N_r=47\approx 18\%$ of the 4FGL-DR2 PSs, and $R_r \approx 0.17$. We then ask the question: suppose we wish to maintain the predicted number of sources and their fluxes, then how does the region of parameter space that does not overproduce those sources/fluxes evolve as $F_\text{GCE}$ changes? In particular, if $F_\text{GCE}$ doubles, $R_r$ must halve, to maintain the total flux in resolved sources. The standard sensitivity model is used in all cases.

In all three parameter spaces investigated, the allowed region expands (contracts) when $F_\text{GCE}$ is decreased (increased). The change is especially rapid in the power law case. An increase by a factor of two in $F_\text{GCE}$ will cause the Wavelet 1 benchmark to overpredict the wavelet-selected subsample. A decrease by a factor of two implies that $L_\text{min} \lesssim \SI{e33}{\erg\per\second}$ satisfies $N_r \leq 47$ for a wide range of $L_\text{max}$ values, so that nearly all luminosity functions produce fewer PSs than observed in the wavelet-selected subsample. The region allowed by the $R_r$ constraint also changes rapidly; if $L_\text{min}$ and $\alpha$ are fixed at the Wavelet 1 benchmark, then a reduction in $F_\text{GCE}$ by a factor of two allows $L_\text{max}$ to increase by a factor of ten and remain in the allowed region. A decrease in $F_\text{GCE}$ by a factor of 2--3 would also bring the Wavelet 2 benchmark into consistency with the observables associated with the wavelet-selected subsample.

For the log normal luminosity function, neither constraint is as sensitive to increases in $F_\text{GCE}$. The $R_r$ constraint is not very sensitive to decreases either, but a decrease in $F_\text{GCE}$ by a factor of 2--3 will cause the GLC benchmark configuration to satisfy the $N_r\leq 47$ condition to not overproduce the wavelet-selected subsample. Using another of the GCE spectra analyzed in section \ref{sec:total-flux} could achieve this amount of decrease. However, the GLC configuration would still be disallowed by the $R_r$ constraint unless $F_\text{GCE}$ were decreased by a factor of $\gtrsim 5$.

\subsection{Dependence on sensitivity threshold}
Figure \ref{fig:step-and-pos} and table \ref{tab:specific-results} both demonstrate that the simplified and standard sensitivity models yield starkly different predictions for the observed PSs. The allowed parameter regions (for given choices of $N_r$ and $R_r$) differ markedly, especially in the power law case, between the two models, and predictions for $N_r$ can differ by as much as a factor of $\sim 7$ for benchmarks. Predictions of $R_r$ differ by as much as a factor of three. This is mostly because the simplified flux threshold $L_\text{th} = \SI{e34}{\erg\per\second}$ used in Ref.~\cite{Zhong:2019ycb} is low, as demonstrated by the fact that the lowest value in \textit{Fermi}'s sensitivity map in the ROI corresponds to a luminosity of $\SI{1.7e34}{\erg\per\second}$ according to the luminosity-to-flux conversion outlined in appendix \ref{app:lum-to-flux}. Ref.~\cite{Zhong:2019ycb} also states results with a threshold of $L_\text{th}=\SI{3e34}{\erg\per\second}$, which appears more accurate.

We recompute results for the simplified sensitivity model with a larger threshold in appendix \ref{app:step-thresh} and show that they mirror results from the standard model much better. But differences remain between the standard and simplified model, caused by the fact that bright pulsars distant from Earth have lower flux and therefore can appear unresolved in the standard sensitivity model, while the simplified sensitivity model always marks these pulsars as resolved. Also, the threshold is correlated with MSP population density due to the larger backgrounds close to the Galactic plane, which is not accounted for in the simplified model.

\subsection{$N_r$, $R_r$, and $N_\text{GCE}$ with cut around Galactic Center}
In obtaining our 4FGL-DR2 baseline sample, we cut all MSPs known to be $>\SI{2}{\kilo\parsec}$ from the GC as likely unrelated to the GCE, following Ref.~\cite{Zhong:2019ycb}. This leaves many PSs of unknown distance from the GC, which might pass or fail the $\SI{2}{\kilo\parsec}$ cut if their positions were known. When predicting $N_r$, we do not cut MSPs more than $\SI{2}{\kilo\parsec}$ away from the GC in our model. This failure to cut potentially gives rise to a discrepancy between our signal prediction and the data we use for comparison.

Specifically, if the list of observed PSs contains all pulsars in the ROI, regardless of their distance from the GC, then our predicted $N_r$ values can be compared to the observed $N_r$ values. However, if the observed sources are all within $\SI{2}{\kilo\parsec}$ of the GC, then we should cut all predicted MSPs more distant than $\SI{2}{\kilo\parsec}$ from the GC. In practice, we know that the observed $N_r$ values almost certainly lie somewhere between these two cases --- we have cut pulsars with known galactocentric distances greater than 2 kpc, but there remain many sources whose distances are unknown and which may lie more than 2 kpc from the GC. In this subsection, we therefore compute how much predicted $N_r$ and $R_r$ change if distant MSPs are cut, to bracket the associated systematic uncertainties (an alternative approach would be to remove the cut and see how much the number of sources changes).

Our default treatment assumes this cut is quite inefficient at removing resolved sources more than 2 kpc from the GC, and thus does not modify the predicted number of sources. In the opposite limit where we treat this cut as perfectly efficient at removing resolved sources more than 2 kpc from the GC, $R_r$ and $N_r$ fall by 55-75\% for most of the luminosity function models (excluding the NPTF model), and for the NPTF model, both $N_r$ and $R_r$ fall by 80\%. The strength of this effect for the NPTF model is due to its rather narrow luminosity function and the proximity of the luminosity function peak to the threshold, meaning that cutting sources that lie closer to Earth may make the difference between the population being detectable and undetectable.

This raises the question of the degree to which our predictions rely on extrapolating the GCE into regions of the sky where it may not have been significantly detected. In the main analysis, these regions are modeled with a gNFW$^2$ distribution for the source density, but this distribution may not be accurate at high galactocentric distance. To test the importance of the contributions from these regions, we can look at the effect of truncating the source density distribution at $\SI{2}{\kilo\parsec}$ from the GC, excluding both resolved sources {\it and} GCE flux from more distant points. For a fixed point in parameter space, the result is a decrease in $F_\text{GCE}$ and $N_\text{GCE}$ of about 30\%; consequently, to match the GCE flux requires a higher normalization for the source population within $\SI{2}{\kilo\parsec}$ of the GC. In this case the predicted values for $N_r$ and $R_r$ decrease by 40-65\% and 5-45\% respectively for non-NPTF benchmarks, and by 75\% for the NPTF benchmark.

The effect of either truncating the source density distribution or assuming a perfect efficiency for the 2 kpc cut is thus to reduce the number and flux of predicted sources and move all benchmarks away from the regions where there may be tension with the data, in figure \ref{fig:step-and-pos}, with a particularly strong effect for the NPTF benchmark.


\section{Future sensitivity}
\label{sec:future-sensitivity}
In this section, we determine the capability of an improved $\gamma$-ray telescope to constrain the luminosity function parameters of an MSP population in the GC. We simulate an increase in sensitivity of GCE measurements by reusing the same $F_\text{th}(\ell, b)$ sensitivity map (figure \ref{fig:sensitivity}) used in our standard sensitivity model, but with an overall multiplicative decrease. In particular, we study cases where the sensitivity threshold is decreased by a factor of two, five, ten, and twenty and reproduce some of the analyses described earlier in this paper.

\subsection{Resolved PS flux distributions at higher sensitivity}

\begin{table}[t]
    \centering
    \small
    \begin{tabular}{|l | c c | c c | c c | c c | c c |}
        \hline
        Luminosity function & $N_r^{\times 1}$ & $R_r^{\times 1}$ & $N_r^{\times 2}$ & $R_r^{\times 2}$ & $N_r^{\times 5}$ & $R_r^{\times 5}$ & $N_r^{\times 10}$ & $R_r^{\times 10}$ & $N_r^{\times 20}$ & $R_r^{\times 20}$\\ \hline \hline
        Wavelet 1 & 31 & 0.11 & 77 & 0.17 & 230 & 0.25 & 470 & 0.31 & 955 & 0.37 \\
        Wavelet 2 & 98 & 0.38 & 210 & 0.53 & 490 & 0.69 & 830 & 0.78 & 1300 & 0.84 \\
        Log normal, GLC & 124 & 0.72 & 220 & 0.85 & 380 & 0.95 & 490 & 0.98 & 570 & 0.99 \\
        Log normal, GCE & 20 & 0.059 & 73 & 0.12 & 340 & 0.25 & 930 & 0.39 & 2200 & 0.55\\
        Log normal, AIC & 12 & 0.039 & 41 & 0.071 &180 & 0.14 & 520 & 0.22 & 1400 & 0.32\\
        NPTF & 111 & 0.26 & 460 & 0.70 & 810 & 0.96 & 907 & 0.99 & 940 & 0.999 \\
        Disk & 30 & 0.13 & 89 & 0.20 & 370 & 0.34 & 1032 & 0.50 & 2500 & 0.69 \\
        \hline
    \end{tabular}
    \caption{Number of resolved MSPs $N_r$ and ratio of resolved flux to total flux $R_r$ for seven luminosity function benchmarks. The flux threshold of the telescope, $F_\text{th}(\ell, b)$, has been decreased by a factor of one (i.e., sensitivity is at its current value), two, five, ten, and twenty, with the factor of increase given as superscripts in the header. Apart from the change in threshold, the standard sensitivity model is used.}
    \label{tab:sensitivity-values}
\end{table}

Table \ref{tab:sensitivity-values} indicates that even a doubling in sensitivity of \textit{Fermi} is expected to greatly increase the number of resolvable MSPs in the GC; $N_r$ increases by a factor of 2-4 across our benchmark luminosity functions, and the smallest change is for the GLC model which is already in severe tension with the data due to its large $R_r$ value. $R_r$ increases by a similar factor for models where most of the GCE flux is not already resolved. Greater sensitivity increases, at the $10-20\times$ level, would be expected to resolve at least $20-30\%$ of the GCE even in the most pessimistic benchmark cases.

\begin{figure}
    \centering
    \begin{subfigure}[b]{0.49\textwidth}
        \includegraphics[width=\textwidth]{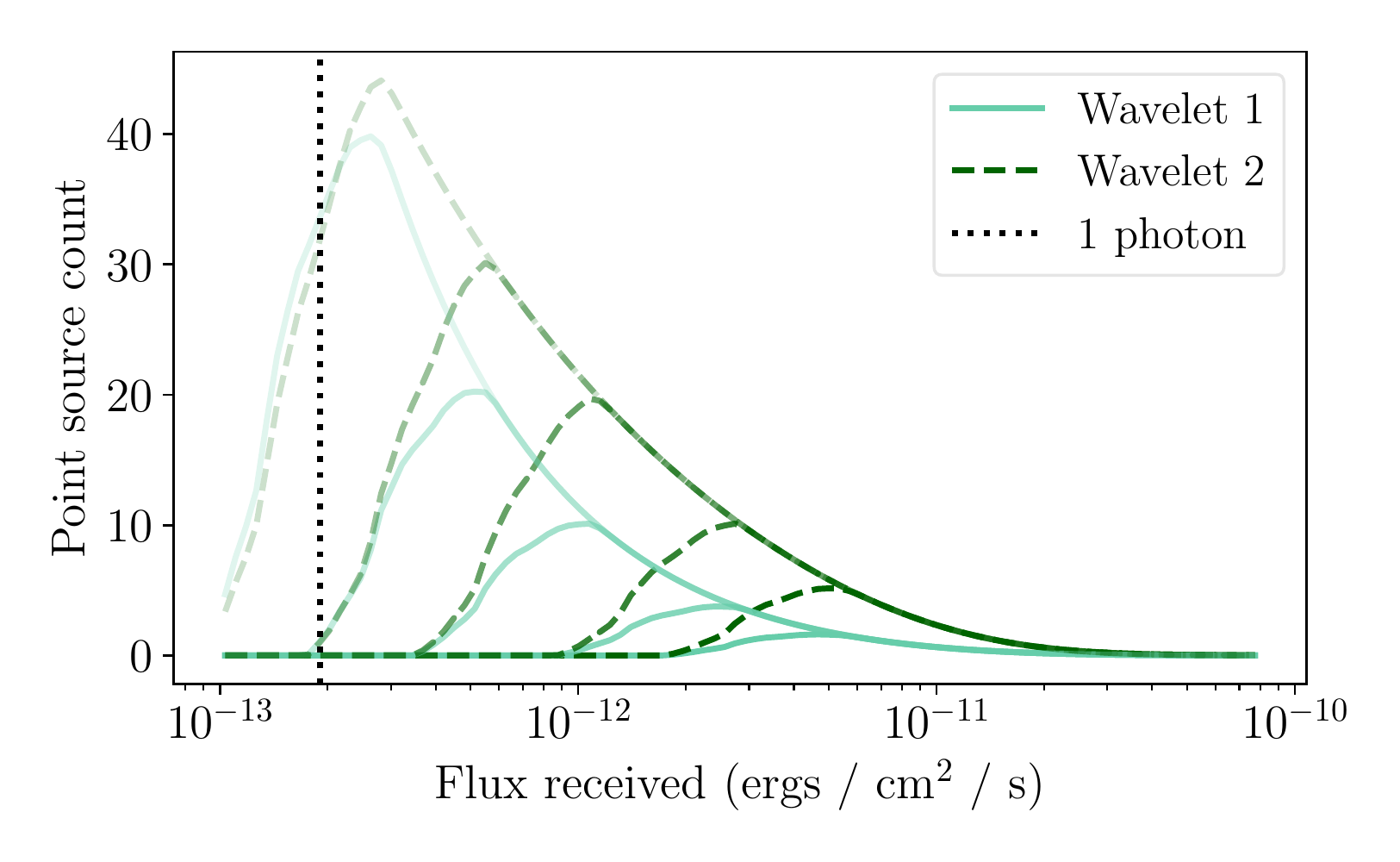}
        \caption{Power law luminosity function}
    \end{subfigure}
    \hfill
    \begin{subfigure}[b]{0.49\textwidth}
        \includegraphics[width=\textwidth]{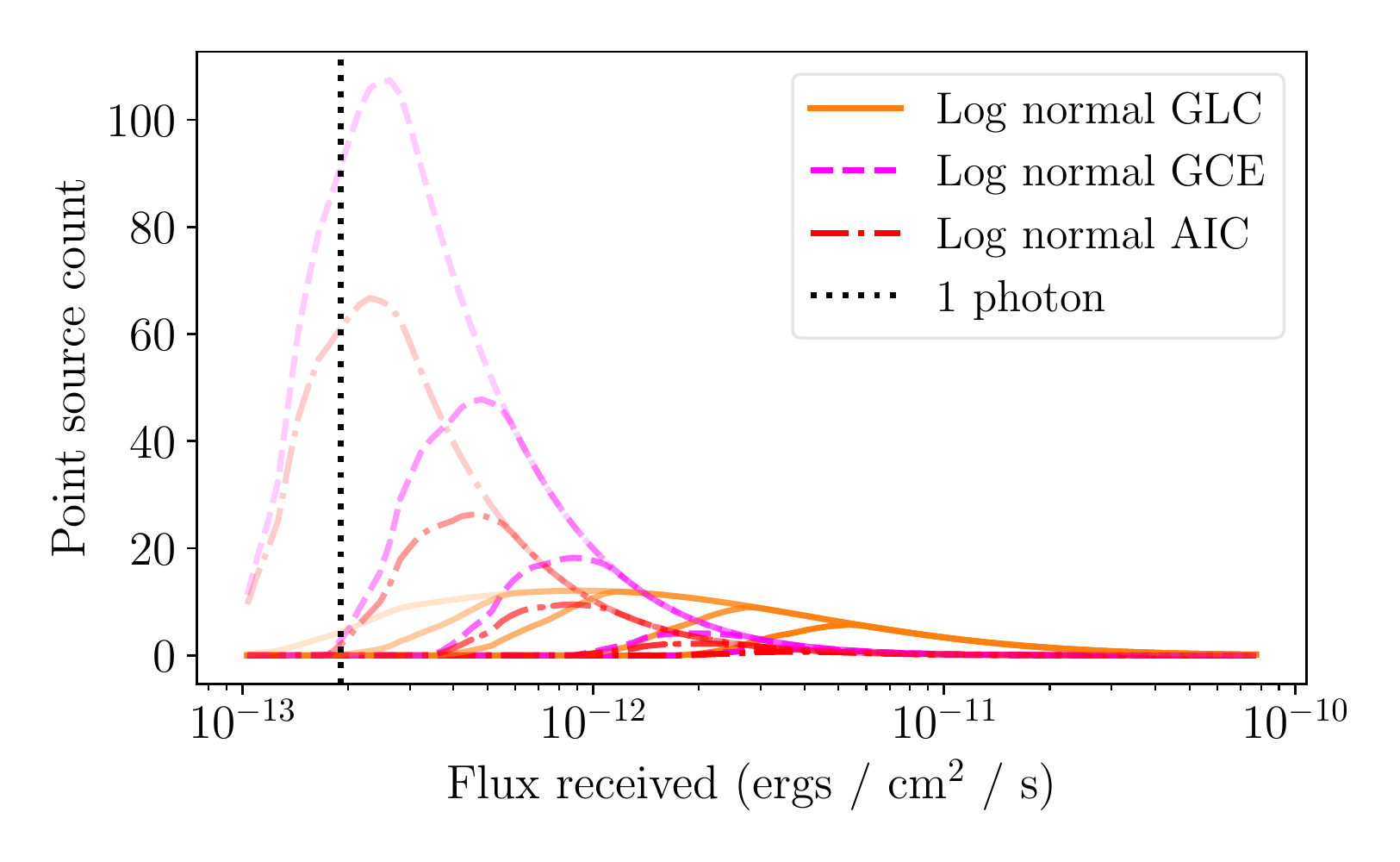}
        \caption{Log normal luminosity functions }
    \end{subfigure}
    \hfill
    \begin{subfigure}[b]{0.49\textwidth}
        \includegraphics[width=\textwidth]{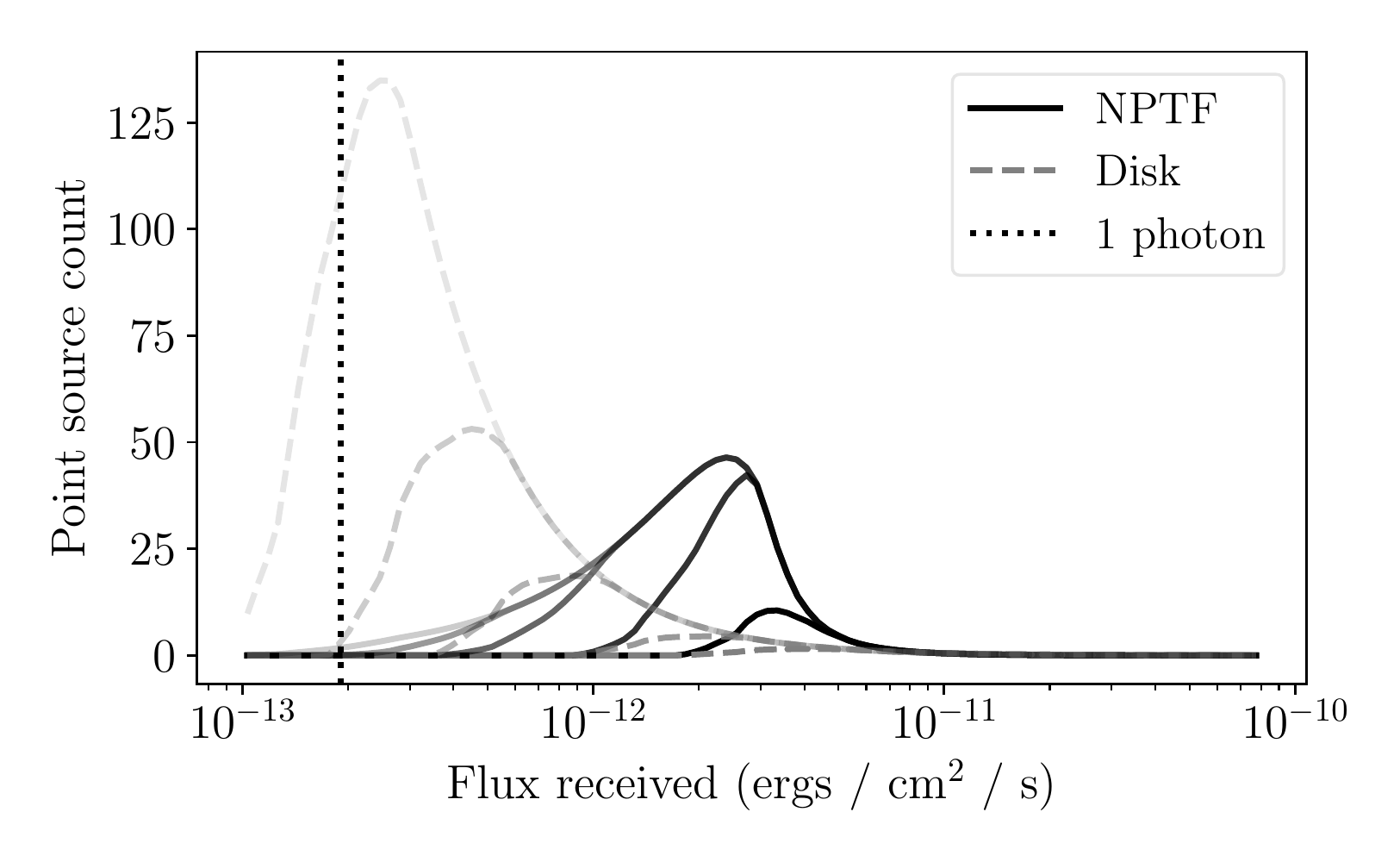}
        \caption{NPTF luminosity function}
    \end{subfigure}
    \caption{Histograms of the predicted number of resolvable pulsars required by each luminosity function benchmark studied here to reproduce the GCE, for different models of \textit{Fermi} sensitivity. Darker lines indicate the current sensitivity; lighter lines represent increases in sensitivity by factors of two, five, ten, and twenty. The vertical dotted line represents the energy flux corresponding to a source with an expected photon count of one in the part of the \textit{Fermi} data set used by Ref.~\cite{List:2021aer}. The bins are spaced evenly in log flux space, with width $\Delta F / F = 0.069$.}
    \label{fig:sensitivity-results}
\end{figure}

In figure \ref{fig:sensitivity-results}, we show the forecast histograms for the expected number of resolved sources as the PS sensitivity increases. MSPs with flux $F \gtrsim \SI{5e-12}{\erg\per\second\per\centi\meter\squared}$ are predicted to already be visible in the \textit{Fermi} data set, so that the MSPs resolvable only by a sensitivity increase are low luminosity. With a five-, ten-, or twenty-fold sensitivity increase, we also see that a large fraction of MSPs (approaching 1) are resolved for the GLC and NPTF benchmarks. This fact is quantitatively visible by comparing table \ref{tab:sensitivity-values} to the $N_\text{GCE}$ entries of table \ref{tab:position-dependent-results}, and qualitatively visible in figure \ref{fig:sensitivity-results} because the distributions begin to take on the shape of the true luminosity function, with the peak of the GLC log normal curve and the low-luminosity branch of the NPTF broken power law visible for high sensitivities.

One might ask whether the highest sensitivities studied here are ever plausibly achievable. While achieving these sensitivities purely through increased integration time with \textit{Fermi} does not seem feasible, novel analyses of the existing \textit{Fermi} data set can shed light on the properties of source populations below the current detection threshold. Ref.~\cite{List:2021aer} claims sensitivity to the source count function (closely related to the luminosity function) down to fluxes corresponding to a single photon per source (or even lower in principle), using a neural-network-based analysis, within the subset of \textit{Fermi} data used in that work (which is restricted to a narrower range of energies, and includes other cuts, relative to the data used to generate the 4FGL-DR2 catalog). We convert the one-photon threshold flux in the analysis of Ref.~\cite{List:2021aer}  to our 0.1-100 GeV energy band and find it corresponds to a PS flux of \SI{1.91e-13} {\erg\per\centi\meter\squared\per\second}, which is roughly twenty times lower than the approximate current \textit{Fermi} threshold we calculate in appendix \ref{app:step-thresh}. This ``one photon'' line is marked on figure \ref{fig:sensitivity-results}.

Consequently, if neural-network-based analyses can achieve sensitivity to PS populations down to their one-photon threshold, it is intriguing that recent studies of this type appear to detect a PS fraction around the 30-40\% level; Ref.~\cite{List:2021aer} claims to exclude a non-PS contribution greater than 66\% at 95\% confidence, while Ref.~\cite{Mishra-Sharma:2021oxe} finds a best-fit PS contribution of 38\% (which differs from zero by $2\sigma$).

Figure \ref{fig:sensitivity-results} also demonstrates that flux histograms predicted by different luminosity function configurations begin to visibly diverge at higher sensitivity. The log normal populations in particular are visibly distinct; the GLC luminosity function predicts a flatter distribution of MSP flux with peak near $\SI{1e-12}{\erg\per\centi\meter\squared\per\second}$ while the GCE and GLC benchmarks predict larger and thinner peaks near $F=\SI{3e-13}{\erg\per\second\per\square\centi\meter}$ at twenty-times-greater sensitivities. We perform an initial quantitative analysis of this point below.

\subsection{Shape differences between resolved PS flux distributions at higher sensitivity}

The flux distribution of a population of resolved sources could potentially be a more powerful consistency test of luminosity function models, compared to simply checking the total flux and number of sources, as discussed in section \ref{sec:flux-distribution}. However, as demonstrated in that section, the similarity of the high-flux tails of different benchmark luminosity functions, and their tendency to underpredict currently observed sources, makes this test rather uninformative at the present level of sensitivity. In this section we discuss the degree to which improved sensitivity could allow us to distinguish benchmark luminosity function models, in the idealized case where we can accurately identify a subpopulation of inner Galaxy pulsars.

To determine our ability to differentiate between the flux distribution of resolved PSs predicted by different luminosity functions, for every pair of luminosity functions, we take one to be the hypothesis while the other is the true luminosity function. We then draw a mock data set $D$ of observed source fluxes $F_i$ from the true (predicted) flux distribution, and determine the unbinned likelihood for the mock data set given the hypothesis luminosity function $P_\text{hyp}$. We draw a number of data points equal to the expected number of resolved MSPs $N_r$ predicted by the true luminosity function, under the assumption that the source population explains 100\% of the GCE (these values are given in table \ref{tab:sensitivity-values}). Because we are interested in discriminating between the high-flux distributions of different source populations based on their shape, not just their normalization, we normalize the hypothesis luminosity function so that it produces the same expected number of resolved PSs as the assumed-true luminosity function. (Note that this does mean that some hypotheses could be independently excluded by the fact that they overproduce the total GCE flux; in this section we are just using the benchmark luminosity functions as examples of scenarios with somewhat different forms for the high-flux tail.)

We define the likelihood of the mock data given the hypothesis to be:
\begin{equation}
\mathcal{L}_\text{hyp}=\prod_{F_i\in D}P(F_i), \qquad P(F_i) \propto P_\text{r}(F_i)\int_\Omega d\Omega \int s^4 ds \rho_\text{GCE}(r) P_\text{hyp}(4\pi s^2 F_i),
\label{eqn:histogram-likelihood}
\end{equation}
where $P(F_i)$ is the probability density function describing the probability that a PS is detected with flux $F_i$. Here, as in Eq.~\ref{eqn:observables-sens-2}, $\Omega$ is the region of interest, $s$ is the distance from the point of integration to Earth, and $r$ is the distance from the point of integration to the GC, which is given by $r^2=s^2+r_c^2-2sr_c\cos b \cos \ell$, where $r_c=\SI{8.5}{\kilo\parsec}$ is the distance between Earth and the GC. The probability of a certain PS with flux $F$ being resolved by \textit{Fermi} given that it exists is denoted by $P_\text{r}(F)$, which is the sensitivity model (e.g. eq.~\ref{eqn:ploeg-smoothing}).

To test the similarity of the two flux distributions, we calculate the test statistic,
\begin{equation}
\lambda=-2\ln(\mathcal{L}_\text{hyp}/\mathcal{L_\text{true}}),
\label{eqn:lambda}
\end{equation}
where $\mathcal{L_\text{true}}$ is obtained by substituting $P_\text{hyp}$ for $P_\text{true}$ in eq.~\ref{eqn:histogram-likelihood}. We expect this number to be low for similar luminosity functions (zero for identical luminosity functions) and higher for different ones.

\begin{figure}
    \centering
    \includegraphics[width=\textwidth]{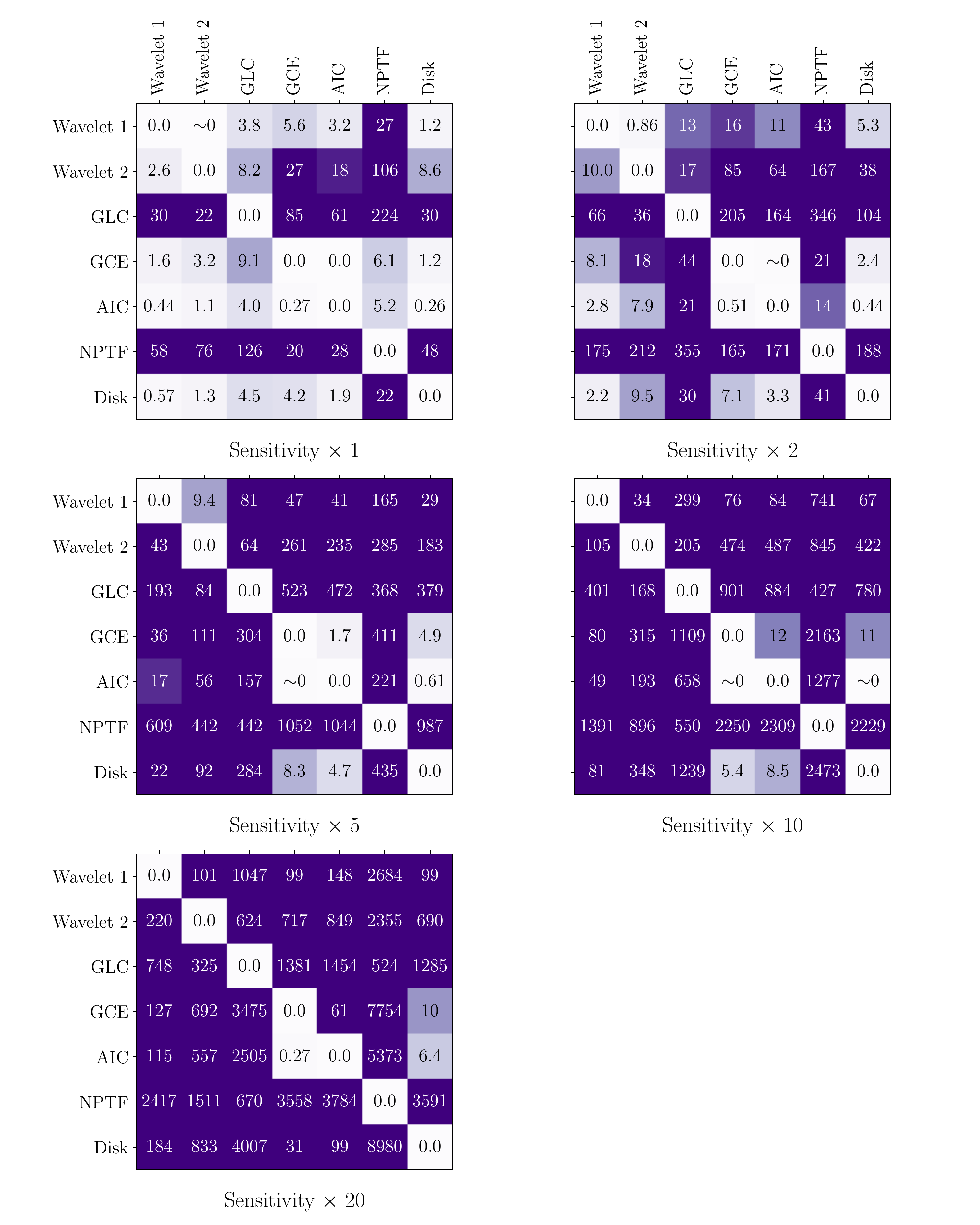}
    \caption{Map of the expected value of $\lambda$ resulting from assuming one luminosity function (horizontal axis) as the hypothesis while the other (vertical axis) is the true luminosity function. Each luminosity function is normalized to produce the same number of resolved PSs as the assumed-true luminosity function. Each plot represents a different multiplicative increase in the \textit{Fermi} sensitivity level. Diagonals indicate the fit of a luminosity function to itself. The standard sensitivity model is used.}
    \label{fig:hist-fitting-sensitivity}
\end{figure}

Figure \ref{fig:hist-fitting-sensitivity} displays the $\lambda$ values for each pair of luminosity functions, averaged over 10,000 sample data sets drawn from each luminosity function, for each sensitivity level studied. As previously, a rescaled version of the standard sensitivity model is used.

As expected, the flux distributions become more distinct as sensitivity increases, as demonstrated by increasing values of $\lambda$. This is largely due to the fact that $N_r$ increases when sensitivity increases, thereby amplifying the likelihood ratio. Note that our procedure has the effect that scenarios where the true model has many bright resolved sources allow models to be more easily distinguished. 
For example, the GCE and AIC luminosity functions are hard to distinguish from the Disk luminosity function even at high sensitivity, when the GCE or AIC functions (with relatively few bright sources) are assumed true. However, when the Disk luminosity function is assumed true, the difference with GCE/AIC increases due in part to the large number of resolved sources produced by the Disk luminosity function. Other similar pairs of luminosity functions highlighted in figure \ref{fig:hist-fitting-sensitivity} are the GCE and AIC functions which share the same log normal functional form, and the power law functions Wavelet 1 and 2 to a lesser extent. The NPTF luminosity function benchmark is distinctly different from the others, even at low sensitivity, as discussed above.

Until now, we have avoided showing the resolved flux distribution predicted by the smoothed sensitivity model because that model is designed to predict the probability that a PS is resolved as a MSP specifically. It therefore predicts many fewer resolved sources than the standard model and the threshold is shifted to higher flux. But the shape and slope of the smoothed cutoff is largely independent of this shifting parameter $K_\text{th}$, so the above analysis, including figures \ref{fig:sensitivity-results} and \ref{fig:hist-fitting-sensitivity}, may be informative when recalculated for the smoothed model; it may also be easier in practice to avoid contamination from non-GCE sources when focusing on resolved MSPs rather than just resolved PSs. This exercise is done in appendix \ref{app:smooth-hists}.


\section{Conclusions}
\label{sec:conclusion}
We have explored the total number of MSPs, number of resolved MSPs, and the fraction of flux emitted by resolved MSPs required to produce the observed GCE flux. We extract these properties, for a detection threshold corresponding to the 4FGL-DR2 \textit{Fermi}-LAT source catalog, in a broad scan over the parameter space of the commonly-used power-law and log-normal luminosity function parameterizations. We also benchmark these functions and a broken power law luminosity function with seven configurations found in the literature. We compared the predicted resolved source populations with sources in the 4FGL-DR2 catalog that pass cuts to reject extragalactic sources, as well as with a subpopulation of such sources identified by the wavelet-based analysis of Ref.~\cite{Zhong:2019ycb} (following up on the earlier study of Ref.~\cite{Bartels:2015aea}).

We found that observationally viable luminosity functions can produce between hundreds and millions of MSPs in the GCE without being excluded by overproduction of resolved sources. The high end of this range corresponds to luminosity functions with very low average gamma-ray luminosities ($\lesssim \SI{e31}{\erg\per\second}$ in the 0.1-100 GeV energy band) compared to physically or observationally motivated benchmarks. The low end of this range, with less than $\mathcal{O}(10^4)$ MSPs, generally requires a strongly peaked luminosity function with relatively high average luminosity. Luminosity functions derived from physical models of MSPs and fitted to GCE data can conform to observations and generally predict tens to hundreds of thousands of MSPs, without any tension with studies finding that masking resolved PSs does not significantly reduce the GCE \cite{Zhong:2019ycb}.

Of the benchmarks we tested, the luminosity functions inferred from early non-Poissonian template fitting analyses \cite{Lee:2015fea}, from observations of globular clusters \cite{Hooper16}, and from a search for wavelet peaks in older {\it Fermi} data \cite{Bartels:2015aea} appear to produce more resolved sources/flux than are contained in the subpopulation identified via wavelet methods in Ref.~\cite{Zhong:2019ycb}, while the other benchmarks are consistent with this subpopulation (underproducing it). However, it is not clear whether this subpopulation can truly be used as an upper bound on resolved GCE sources, as there are a large number of sources in 4FGL-DR2 which satisfy the same cuts as the wavelet-selected population (imposed to reject extragalactic sources) except that they were not significantly detected by the wavelet analysis. This is not an issue of the wavelet analysis simply having a higher sensitivity threshold than the 4FGL-DR2 analysis; many of the 4FGL-DR2 sources not detected by the wavelet method are quite bright. To strengthen the constraints on the GCE PS population, it would be helpful to better understand the completeness properties of the two source lists above the 4FGL-DR2 sensitivity threshold, and to further study the properties of 4FGL-DR2 sources to separate possible GCE candidate sources from others. We have provided estimates of how the constraints on PS populations would behave as a function of the fraction of resolved PSs attributed to the GCE, or the fraction of the total GCE flux attributed to resolved sources, both with current data and with an improved PS sensitivity threshold.

We also tested the ability of current and future analyses to distinguish between the benchmark luminosity functions, finding that the high-flux tails of these luminosity functions are rather similar and would currently be difficult to distinguish even if we could correctly identify all the GCE PSs above threshold. However, improvements in sensitivity --- either from future data or from improved analyses --- could change this conclusion. We explored the fraction of the GCE that would be resolved by analyses with point-source sensitivity down to the 1-photon threshold of Ref.~\cite{List:2021aer}, and found that the benchmark scenarios generally predicted that fractions exceeding 30\% of the GCE flux would be resolved in this case. This estimate is intriguingly similar to the claimed fraction of the GCE tentatively detected as PSs in Ref.~\cite{Mishra-Sharma:2021oxe}, and consistent with the limit in Ref.~\cite{List:2021aer}.

We found there are a number of significant systematic uncertainties which must be taken into account when claiming to exclude or match specific luminosity function models. For example, the total flux attributed to the GCE varies by up to a factor of two between different studies; we choose a benchmark value based on a recent analysis \cite{DiMauro:2021raz} and then show the effects of modifying this choice. Varying the total GCE flux modifies the required properties of a PS population explaining some fraction of the GCE; for a fixed luminosity function, a higher total flux implies a higher number of resolved sources (although they will yield the same total fraction of the GCE flux). Furthermore, especially for steeply peaked luminosity functions where much of the power is in sources that are barely resolved or barely unresolved, the distribution of the GCE at large distances from the GC is potentially very important --- while most of the total flux of the GCE may originate from the region around the GC, the resolved sources can dominantly originate from regions closer to Earth and away from the line of sight to the GC, where the sensitivity is improved. Finally, the sensitivity modeling is important; we have demonstrated that a simple sensitivity estimate employed in Ref.~\cite{Zhong:2019ycb} predicts quite a different distribution for the fluxes of resolved sources compared to observations, and suggest an improved prescription.

\acknowledgments
The authors thank Ilias Cholis, Yi-Ming Zhong, and Sam McDermott for providing us with the results of their wavelet search to the GCE and the names of the 47 resolved PSs in the 4FGL catalog as detailed in Ref.~\cite{Zhong:2019ycb}, for very helpfully answering follow-up questions about their work, and for their valuable feedback. We also thank Roland Crocker for elucidating a figure in Ref.~\cite{Gautam:2021wqn}, Ballet Jean and Seth Digel for consultation about the 4FGL-DR2 sensitivity map, and Nicholas Rodd for helpful comments. This work made heavy use of the {\it Fermi}-LAT 8-year and 10-year 4FGL catalogs and the ATNF Pulsar catalog. This material is based upon work supported by the U.S. Department of Energy, Office of Science, Office of High Energy Physics of U.S. Department of Energy under grant Contract Number  DE-SC0012567. The work of JD was funded by the Massachusetts Institute of Technology Undergraduate Research Opportunities Program (MIT UROP) office.

\bibliographystyle{unsrt}
\bibliography{gce.bib}


\appendix
\section{Scaling of ROIs and spectral ranges}
\label{app:roi-rescale}
To establish a value for the total flux of the GCE, we draw on several analyses of the GCE spectrum in section \ref{sec:total-flux}. However, not all the analyses we study use the same ROI as ours. To convert between our ROI and others', we assume a gNFW squared spatial distribution of MSPs in the GCE as discussed in section \ref{sec:spatial-distro}. Then we calculate the ratio of flux in our region of interest $F_\Omega$ to flux in another analysis's region of interest $F_{\Omega'}$ via
\begin{equation}
    \frac{F_{\Omega'}}{F_\Omega} = \brackets{\int_{\Omega'}d\Omega\int_0^\infty ds \rho_\text{GCE} (r)}\brackets{\int_{\Omega}d\Omega\int_0^\infty ds \rho_\text{GCE} (r)}^{-1}.
    \label{eqn:roi-conversion}
\end{equation}
Here, $s$ represents the distance between Earth and the point of integration, and $r$ represents the galactocentric distance. They are related by $r^2 = s^2 + r_c^2 - 2 s r_c \cos\ell\cos b$, where $\ell$ and $b$ are the Galactic longitude and latitude and $r_c=\SI{8.5}{\kilo\parsec}$ is the distance between Earth and the center of the Galaxy.

The flux ratio as computed by eq.~\ref{eqn:roi-conversion} between our region and a $40^\circ \times 40^\circ$ square region without the Galactic disk mask is 1.9. Similarly, the ratio is 1.8, 1.5, 1.3, 1.1, and 0.92 for regions centered on the Galactic center of side length $30^\circ \times 30^\circ $, $20^\circ \times 20^\circ$, $15^\circ \times 15^\circ$, $10^\circ \times 10^\circ$, and $7^\circ \times 7^\circ$ respectively, all without a disk mask. The $15^\circ \times 15^\circ$ square is the ROI used by Ref.~\cite{Ajello:2015kwa}, and the $7^\circ \times 7^\circ$ square is used by Ref.~\cite{Gordon13}. For Ref.~\cite{Ajello:2017opo}, we reproduce their ROI from the 3FGL PS catalog \cite{Fermi-LAT:2015bhf} and obtain a flux ratio of 0.56.\footnote{This ROI is a 10$^\circ$-radius disk, pixellated with pixel size $a\approx 0.47^\circ$. We place $\approx 1^\circ + a \sin(\pi/4)$-radius circular masks around the 200 brightest 3FGL sources in the sky.} Finally, Ref.~\cite{Abazajian:2014fta} uses the same $7^\circ \times 7^\circ$ ROI as Ref.~\cite{Gordon13}, but with $\gamma=1.1$. The ratio for this ROI to our ROI with $\gamma=1.1$ is also 0.56.

We perform the ROI conversion by taking the flux emitted from the entire ROI $\Omega'$ of the study in question and multiplying by $\frac{F_\Omega}{F_{\Omega'}}$ to get the flux in our ROI $\Omega$. To make figure \ref{fig:all-spectra}, we perform this multiplication for the flux in every energy bin.

\section{Broken power law fits to GCE spectra}
\label{app:spectra-fits}
Section \ref{sec:total-flux} relied on broken power law fits to nine energy spectra found in the literature. Those fits were done via $\chi^2$ minimization, with
\begin{equation*}
    \chi^2 = \sum_{i} \begin{cases}
        \frac{(y_i-y*_i)^2}{\sigma_{+,i}^2} & y_i^* \geq y_i\\
        \frac{(y_i-y*_i)^2}{\sigma_{-,i}^2} & y_i^* \leq y_i\\
    \end{cases}
\end{equation*}
where $y^*$ is the model data given the fit parameters, $y_i$ is the observed data, and $\sigma_{\pm, i}$ are the upper and lower error bar lengths.

\begin{figure}
    \centering
    \includegraphics[width=\textwidth]{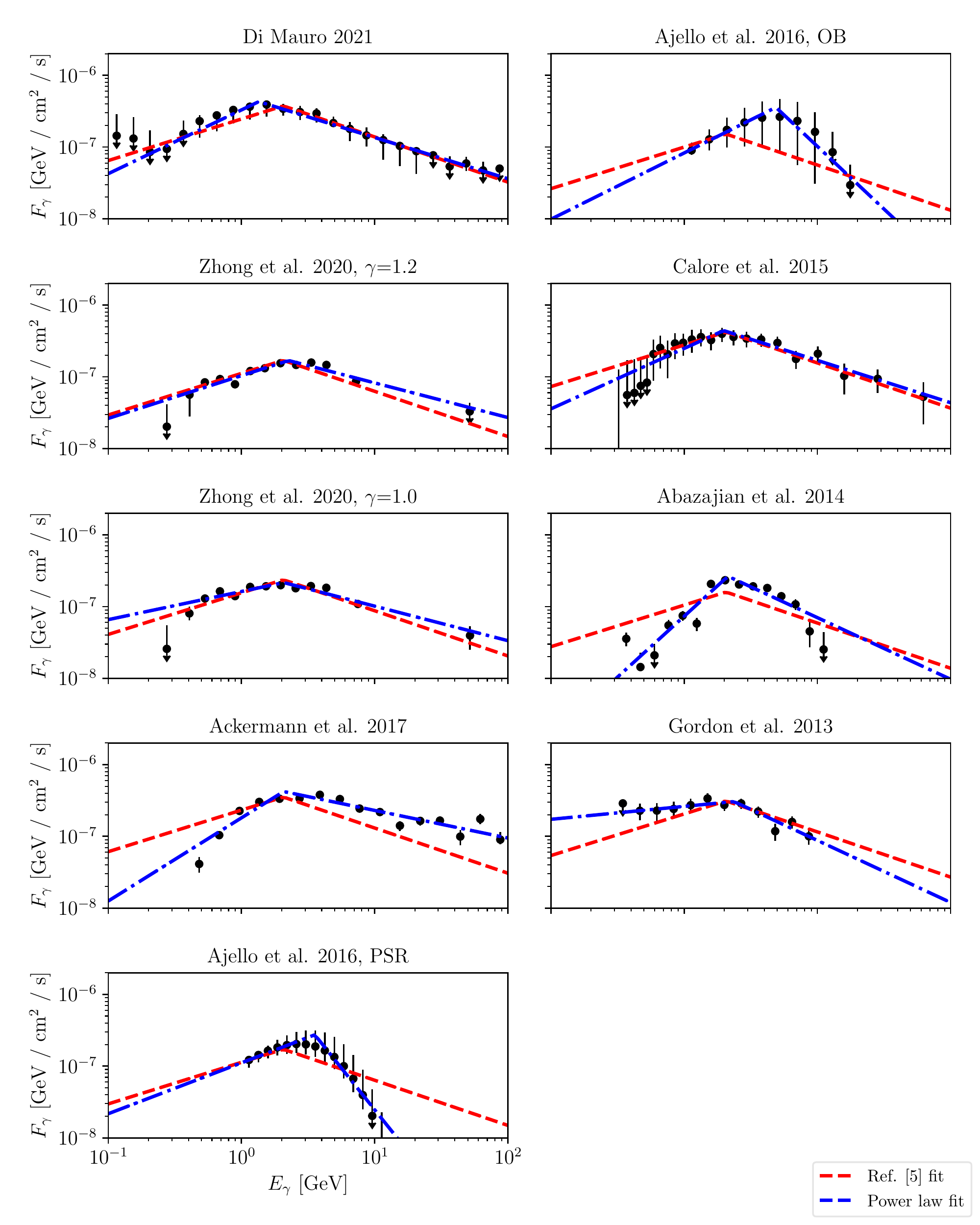}
    \caption{Broken power law fits to the nine spectra studied in section \ref{sec:total-flux} to extract the total GCE luminosity. Arrows on error bars denote bars that would descend past the lower edge of the plot. See text for which functions are being fitted. Fits are performed approximating the uncertainties at different energies as uncorrelated, which is why the results of the two fits differ slightly for the ``Calore 2015'' spectrum taken from Ref.~\cite{Calore:2014xka}; Ref.~\cite{Calore:2014xka} took the full covariance matrix into account.}
    \label{fig:spectrum-fits}
\end{figure}

As mentioned in section \ref{sec:total-flux}, the fit is done to a one-parameter broken power law model with $n_1, n_2, L_\text{b}$ fixed at parameters determined by Ref.~\cite{Calore:2014xka}, as well as a four-parameter broken power law with all parameters floated.  The resulting fits are shown in figure \ref{fig:spectrum-fits}. 

The best-fit results for the floated parameters do not agree perfectly between this analysis of data from Ref.~\cite{Calore:2014xka} and the best-fit parameters reported in that work, although they are similar.  This is likely due to the fact that this work approximates the error bars as uncorrelated and adds systematic and statistical uncertainties (where they are separated) in quadrature, whereas Ref.~\cite{Calore:2014xka} used a full covariance matrix.

Several of the studies report multiple spectra for the GCE. For the spectrum labeled ``Abazajian 2014'', we take the GCE spectrum extracted from the ``2FGL + 2PS + I + MG + ND + GCE'' fit, which provides the best fit \cite{Abazajian:2014fta}. For the spectrum labeled ``Ajello 2017'', we use the ``Sample'' spectrum and recreate the ROI from the 4FGL catalog \cite{Ajello:2017opo}. For the spectrum labeled ``Ajello 2016'', we use the plotted spectra in Figure 13 of Ref.~\cite{Ajello:2015kwa}.

\section{Fits to GCE and AIC luminosity functions}
\label{app:lum-func-fit}
We claimed in section \ref{sec:lum-funcs} that the AIC and GCE luminosity functions can be reasonably well-described by a log-normal function, which motivates treating the log-normal form as a plausible parameterization for a range of physically reasonable luminosity functions. In this appendix, we explain how the parameters were derived for this log-normal approximation to the relevant luminosity functions.

\begin{figure}
    \centering
    \begin{subfigure}[b]{0.49\textwidth}
        \centering
        \includegraphics[width=\textwidth]{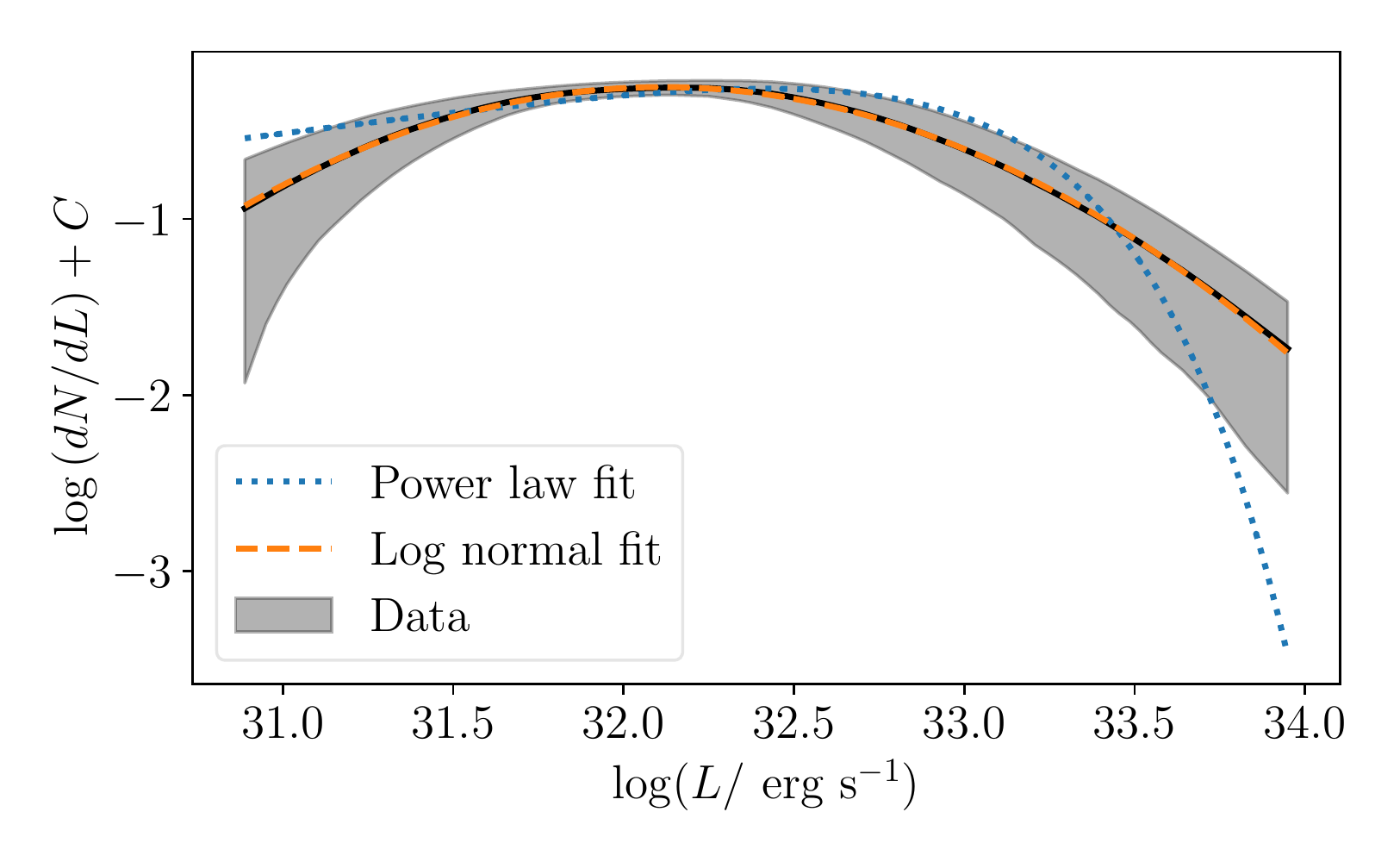}
        \caption{Fit to the GCE luminosity function}
        \label{fig:fit-to-gce-lf}
    \end{subfigure}
    \hfill
    \centering
    \begin{subfigure}[b]{0.49\textwidth}
        \centering
        \includegraphics[width=\textwidth]{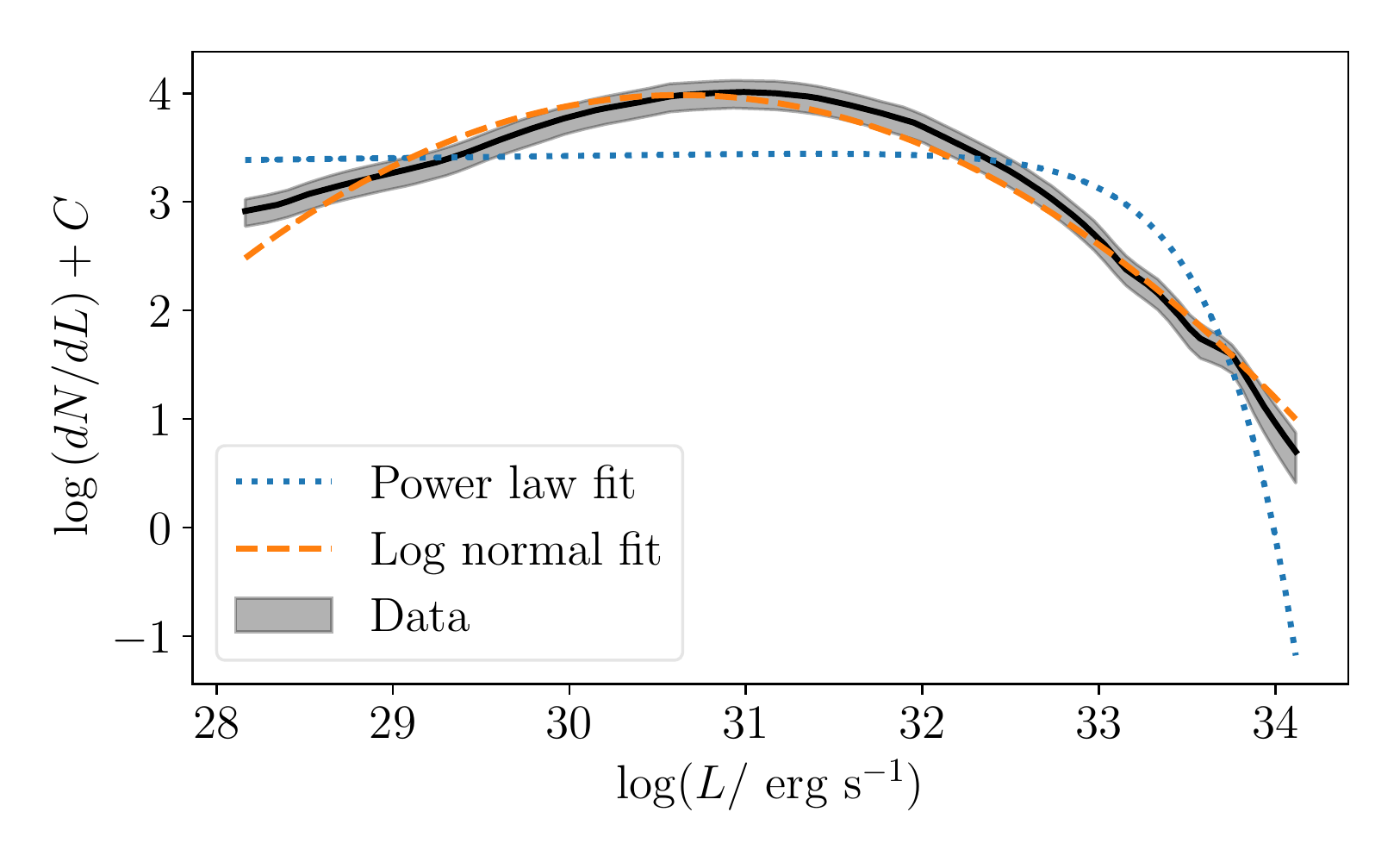}
        \caption{Fit to the AIC luminosity function}
        \label{fig:fit-to-aic-lf}
    \end{subfigure}
    \caption{Log normal fits to two numerical luminosity functions: the boxy bulge luminosity function from Ref.~\cite{Ploeg:2020jeh} and the luminosity function from \cite{Gautam:2021wqn}. Uncertainties are assumed to be uncorrelated and represent one standard deviation from the mean. The best fit for a broken power law with an exponential cutoff is also shown.}
\end{figure}

In both cases, the numerical luminosity functions were extracted from the papers where they were presented: Refs.~\cite{Ploeg:2020jeh} and \cite{Gautam:2021wqn} respectively. The functions are not normalized, which is why the vertical axes are marked with an arbitrary additional term $+C$. We sampled the functions at 100 evenly log-spaced values of luminosity spanning the domain of the original functions (this number was chosen arbitrarily), and used the width of the band at each luminosity value to estimate an uncertainty on the value. We then performed a least-squares fit to these extracted data for  log normal (eq.~\ref{eqn:log-normal}) and cutoff power law (eq.~\ref{eqn:power-law}) luminosity function models. The best-fitting models are shown in Figures \ref{fig:fit-to-gce-lf} and \ref{fig:fit-to-aic-lf} respectively; we observe that for the GCE luminosity function the log-normal fit very accurately tracks the numerical result, and for the AIC luminosity function the log-normal model lies consistently within the uncertainty band except at very low and high luminosities. This is not the case for the cutoff power law fit in either case. 

\section{Conversion between GCE luminosity and flux}
\label{app:lum-to-flux}
This analysis requires luminosity functions to be expressed as probability distributions as a function of luminosity. Yet several papers referenced in this study express luminosity functions as a function of flux, and sometimes also assume that the flux distribution is constant along all lines of sight (rather than assuming a consistent luminosity distribution, which translates into different flux distributions depending on the distribution of distances to the sources). This appendix describes how the conversion to luminosity is done.

As discussed in section \ref{sec:spatial-distro}, we represent the MSP population as distributed according to an gNFW squared distribution with $\gamma = 1.2$. One way to convert a function of flux to luminosity would be to use the luminosity function which, when integrated over the gNFW squared spatial distribution, would reproduce the observed function of flux. But this method would change the functional form of the luminosity function so that, for example, a luminosity function that is log normal when written in terms of flux would no longer be log normal when written in terms of luminosity; it would also require a spatially-dependent luminosity function. For source populations distributed through the Galaxy, this difference is crucial; an intrinsically faint source may be very visible simply because it is close to the Earth. However, for the GCE population, we expect most sources to be rather close to the GC and thus to have an approximately consistent distance from Earth, suggesting we may be able to approximate flux and luminosity as being proportional. This is the tacit assumption made in studies which treat the flux distribution of GCE sources as being the same along all lines of sight (as in general this would require a peculiar coincidence in the spatial evolution of the luminosity function).

Consequently, we simply convert the flux value of every bin to a luminosity according to the following method. Suppose the entire population of MSPs only contains pulsars with luminosity $L$; then we can compute the average flux $F$ per pulsar. Integrated over the region of interest $\Omega$, this yields a constant flux to luminosity ratio of
\begin{equation}
    \frac{F}{L} = \frac{1}{4\pi}\brackets{\int_{\Omega}d\Omega\int_0^\infty ds \rho_\text{GCE} (r)}\brackets{\int_{\Omega}d\Omega \int_0^\infty s^2 ds \rho_\text{GCE} (r)}^{-1} = \SI{1.11e-46}{\per\centi\meter\squared}.
    \label{eqn:f-to-l}
\end{equation}
Here, $s$ represents the distance from Earth to the point of integration, and $r$ represents the distance from the GC to the point of integration. They are related by the law of cosines: $r^2 = s^2 + r_c^2 - 2s r_c \cos b\cos \ell$, where $\ell$ is the Galactic longitude. The numerical value reported was computed for $r_c = \SI{8.5}{\kilo\parsec}$. It is slightly lower than the na\"ive value of $\frac{F}{L} = \frac{1}{4\pi r_c^2} = \SI{1.16e-46}{\per\centi\meter\squared}$, which assumes that all the MSPs are at the Galactic center, and therefore does not rely on a choice of $\gamma$. The similarity between the flux-to-luminosity ratio at the GC and the average flux-to-luminosity ratio makes it seem probable that it is quite a good approximation to treat the sources as being close to the GC, and the error due to this approximation is likely to be small. However, when computing the number of resolvable sources, which has a strong dependence on each source's distance to the GC, we do not use this approximation.

\section{Calculation of photon energy}
\label{app:photon-energy}
The break flux of the NPTF luminosity function benchmark of Ref.~\cite{Lee:2015fea} is given with flux units of photons per square centimeter per second ($F^\text{c}_\text{b}=\SI{1.76e-10}{\photon\per\centi\meter\squared\per\second}$), evaluated in the $1.893-\SI{11.943}{\giga\electronvolt}$ energy range. Our analysis requires a break luminosity $L_\text{b}$ in units of ergs per second, in the $0.1-\SI{100}{\giga\electronvolt}$ energy range.
To do this conversion, we use the GCE spectrum inferred from Ref.~\cite{DiMauro:2021raz} by floating all four broken power law parameters as described in section \ref{sec:total-flux} and plotted in figure \ref{fig:di-mauro-example}.

The conversion between $F^\text{c}_\text{b}$ and $L_\text{b}$ is then achieved by
\begin{equation}
    L_\text{b} = F^\text{c}_\text{b}\brackets{\int_{\SI{0.1}{\giga\electronvolt}}^{\SI{100}{\giga\electronvolt}}E_\gamma N(E_\gamma) dE_\gamma}
    \brackets{\int_{\SI{1.893}{\giga\electronvolt}}^{\SI{11.943}{\giga\electronvolt}} N(E_\gamma) dE_\gamma}^{-1} \frac{L}{F}
    \label{eqn:nptf-break-flux}
\end{equation}
where $N(E_\gamma) = dN/dE$ is the spectrum of the GCE, and the fraction $\frac{F}{L}$ is given in appendix \ref{app:lum-to-flux}. This calculation gives $L_\text{b} = \SI{2.5e+34}{\erg\per\second}$ (in the 0.1--100 GeV energy band relevant for our analysis).

\section{Analysis of $L_\text{th}$ for the simplified sensitivity model}
\label{app:step-thresh}
The simplified sensitivity model outlined in section \ref{sec:sensitivity} represented all PSs with luminosity $L>L_\text{th}$ as resolved, and all with $L<L_\text{th}$ as unresolved, where $L_\text{th}$ is pixel-independent. The key quantities $N_\text{GCE}$, $R_r$, and $N_{r}$ of a population of MSPs necessary to reproduce the GCE were then given by eq.~\ref{eqn:observables-sens-1}.

The value $L_\text{th}=\SI{e34}{\erg\per\second}$ was used in Ref.~\cite{Zhong:2019ycb}, and therefore was used in this paper for the simplified sensitivity model. However, the fact that the $N_\text{GCE}$, $R_r$, and $N_{r}$ values produced by the simplified sensitivity model are almost always larger than those produced by the more detailed standard sensitivity model (table \ref{tab:specific-results}) indicates that  $L_\text{th}=\SI{e34}{\erg\per\second}$ is an underestimate.
The fact that the none of the 47 PSs in the subpopulation identified by Ref.~\cite{Zhong:2019ycb} have $L$ below or near $\SI{e34}{\erg\per\second}$ (figure \ref{fig:flux-distro}) also demonstrates that the true average threshold is larger; we would expect some PSs to be observed at or even below the true threshold due to uncertainty in the threshold and the fact that some regions of the sky have greater sensitivity than others.

A more accurate estimate for a constant $L_\text{th}$ could be gained from an average over the per-pixel threshold sensitivities provided by Refs.~\cite{Fermi-LAT:2019yla, Ballet:2020hze} (figure \ref{fig:sensitivity}). We weight our average by the amount of flux predicted to emanate from each pixel with a gNFW-squared-distributed population of PSs. The new weighted-average flux threshold is then
\begin{equation}
F_\text{th} = \brackets{\sum_{\mathcal{P}(\ell, b)}L_\text{th}(\ell, b) \int_0^\infty ds  \rho_\text{GCE}(r)\int_{\mathcal{P}(\ell, b)} d\Omega}\brackets{\sum_{\mathcal{P}(\ell, b)}\int_0^\infty ds \rho_\text{GCE}(r)\int_{\mathcal{P}(\ell, b)} d\Omega}^{-1}
\label{eqn:avg-threshold}
\end{equation}
where $\mathcal{P}(\ell, b)$ represents the pixel at galactic coordinates $(\ell, b)$.
As before, $r$ represents the distance from the line of sight distance $s$ to the center of the Galaxy, defined by $r^2 = {s^2 + r_c^2 - 2 s r_c \cos(\ell)\cos(b)}$. We assume here that the average luminosity emitted from a PS is not correlated with its position, so that the luminosity emanating from a region of space is proportional to the number density $\rho_\text{GCE}$ of PSs in that region. Then the $s^2$ of the volume element and $s^{-2}$ required to convert luminosity to flux cancel out.

The result of eq.~\ref{eqn:avg-threshold} is $L_\text{th} = \SI{3.8e-12}{\erg\per\cm\squared \per \second}$, which corresponds to $L_\text{th} = \SI{3.4e34}{\erg\per \second}$ via the flux-luminosity conversion defined in appendix \ref{app:lum-to-flux}. Note that a few of the 47 wavelets have flux below this value of $F_\text{th}$ (figure \ref{fig:flux-distro-fix-flux}), demonstrating that it is a better candidate for \textit{Fermi}'s sensitivity.

\begin{figure}
    \centering
    \includegraphics[width=0.49\textwidth]{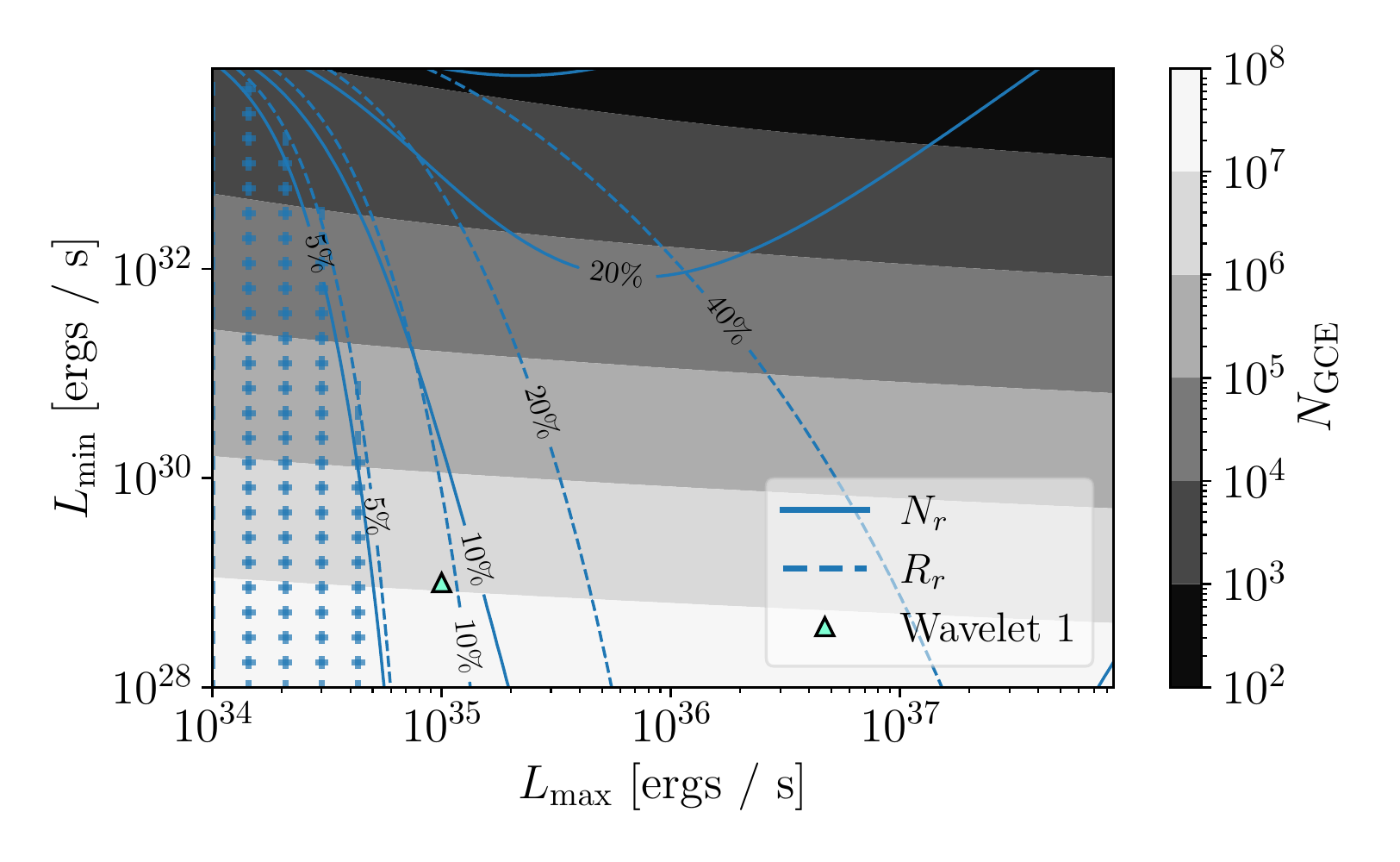}
    \includegraphics[width=0.49\textwidth]{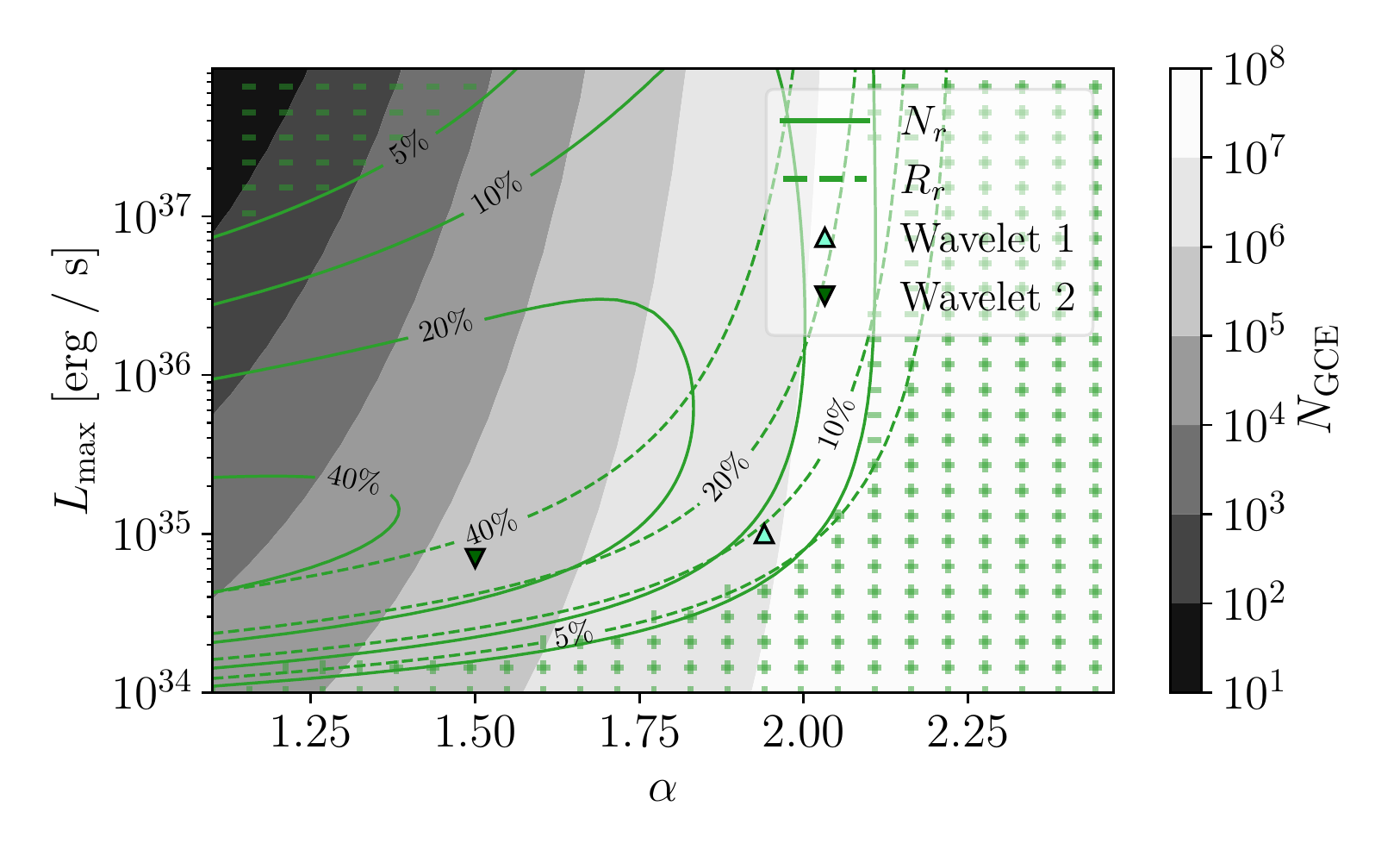}
    \includegraphics[width=0.49\textwidth]{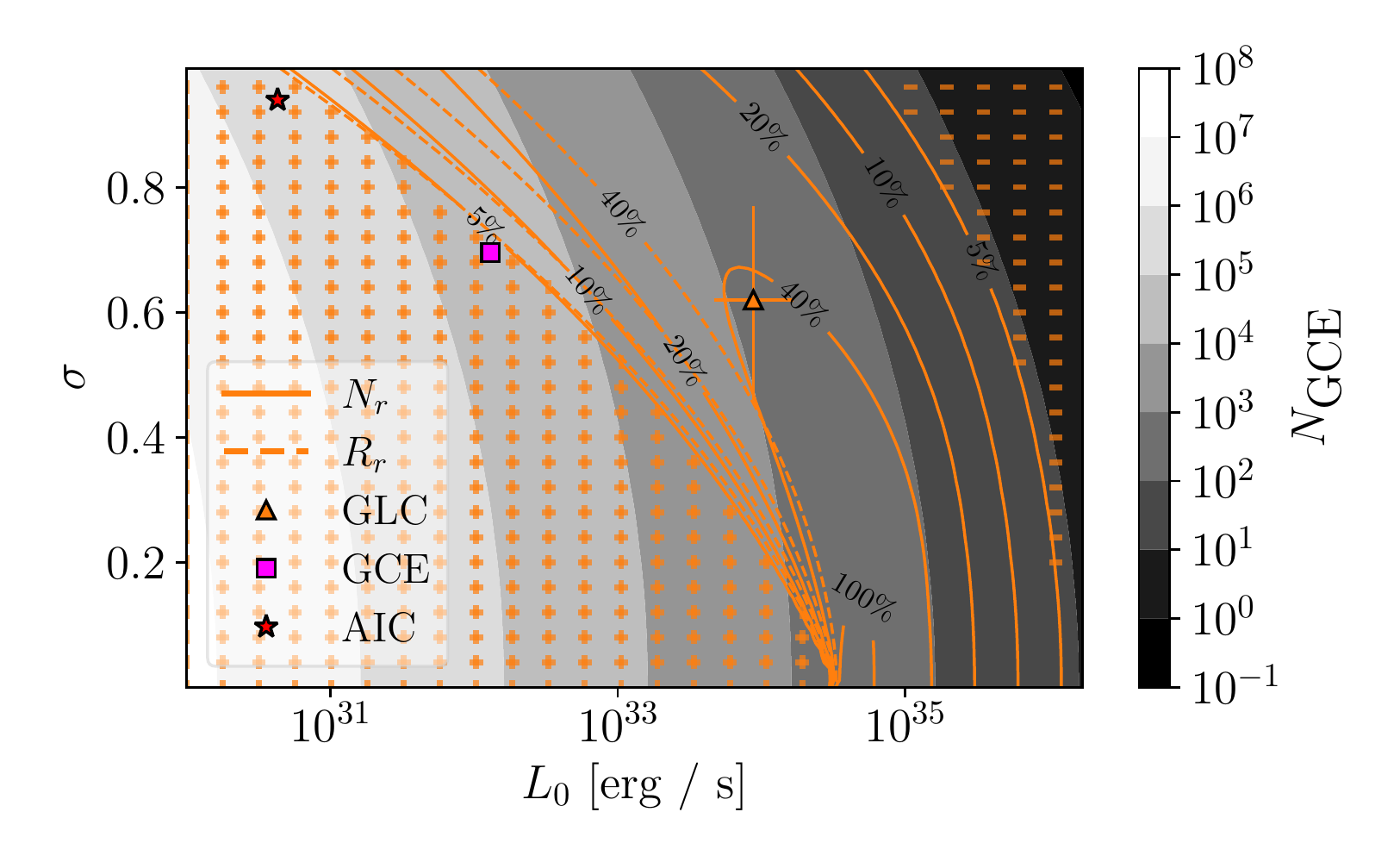}
    \caption{As figure \ref{fig:step-and-pos}, but using the simplified sensitivity model with the average flux threshold of $L_\text{th} = \SI{3.4e34}{\erg\per\second}$.}
    \label{fig:step-function-new}
\end{figure}

Regenerating previous results with this new $L_\text{th} = \SI{3.4e34}{\erg\per\second}$ value, figure \ref{fig:step-function-new} displays the configurations of luminosity functions with the power law or log normal functional form that obey the observational constraints, according to the simplified sensitivity model with the new threshold. It is the analog of figure \ref{fig:step-function}. Table \ref{tab:step-function-new-results} shows $N_r$, $R_r$, and $N_\text{GCE}$ for specific benchmarks drawn from other works, again with the simplified model and the new threshold. It is the analog of table \ref{tab:step-function-results}.

\begin{table}
    \centering
    \begin{tabular}{|p{4cm} | >{\centering\arraybackslash}p{2cm} >{\centering\arraybackslash}p{2cm} >{\centering\arraybackslash}p{2cm}|}
        \hline
        Luminosity function & $N_r$ & $R_r$ & $N_\text{GCE}$ \\ \hline\hline
        Wavelet 1 & 22 & 0.089 & $\num{8.5e6}$\\
        Wavelet 2 & 79 & 0.32 & $\num{2.2e5}$\\
        GLC & 113 & 0.68 & 660 \\
        GCE & 8.6 & 0.030 & $\num{3.4e4}$ \\
        AIC & 5.9 & 0.023 & $\num{3.6e5}$ \\
        NPTF & 0.47 & $\num{1.1e-3}$ & 960 \\
        Disk & 19 & 0.10 & $\num{2.5e4}$ \\ \hline
    \end{tabular}
    \caption{Number of resolved PSs, ratio of resolved flux to total flux, and total number of PSs predicted to make up the GCE based on the new $L_\text{th}=\SI{3.56e34}{\erg\per\second}$ simplified sensitivity threshold produced by a flux-weighted average of per-pixel sensitivity provided by Refs.~\cite{Fermi-LAT:2019yla, Ballet:2020hze}. Compare to table \ref{tab:step-function-results}, which produces higher values.}
    \label{tab:step-function-new-results}
\end{table}

We see from table \ref{tab:step-function-new-results} that the new step-function threshold values $N_r$ and $R_r$ for $L_\text{th} = \SI{3.4e34}{\erg\per\second}$ are generally closer to the standard values than the $L_\text{th}=\SI{e34}{\erg\per\second}$ were, indicating that the averaged threshold  $L_\text{th} = \SI{3.4e34}{\erg\per\second}$
is a better estimate than $L_\text{th}=\SI{e34}{\erg\per\second}$. The values for $N_\text{GCE}$ do not change because they are not affected by the threshold. The NPTF luminosity function is still poorly represented by the new value of $L_\text{th}$ because its strong peak just below the threshold sensitivity makes it behave very differently in the standard sensitivity model, where the peak is widened by the spatial distribution of the MSPs.

The abrupt change caused by the new value of $L_\text{th}$ (compare Figs.~\ref{fig:step-function} and \ref{fig:step-function-new}) demonstrates the large impact of $L_\text{th}$ on the range of allowed luminosity function configurations. In particular, with the more accurate, averaged value of $L_\text{th}$, the observational limits are satisfied at larger $L_\text{min}$ than with the old averaged value, which allows $N_\text{GCE}$ to be lower.  For the log normal case, slightly higher $L_0$ values are allowed by the averaged $L_\text{th}$, which corresponds to lower $N_\text{GCE}$.

\section{Flux histograms for the smoothed sensitivity model}
\label{app:smooth-hists}

The smoothed sensitivity model was not used in section \ref{sec:future-sensitivity} because its parameters were extracted to represent the probability that a PS is resolved as an MSP \cite{Ploeg:2020jeh, Gautam:2021wqn}, whereas in this study we focus on the probability for a PS to be resolved at all. Therefore, the resolved PS flux distributions using the smoothed sensitivity model cannot be compared directly to the current observational data. However, since the smoothed sensitivity model is more detailed and accounts for uncertainty in the thresholds, and might better represent the source populations we could positively identify with the GCE (via identifying the sources as pulsars), we present the flux distributions of resolved MSPs using the smoothed sensitivity model in this appendix for reference.

\begin{figure}
    \centering
    \begin{subfigure}[b]{0.49\textwidth}
        \includegraphics[width=\textwidth]{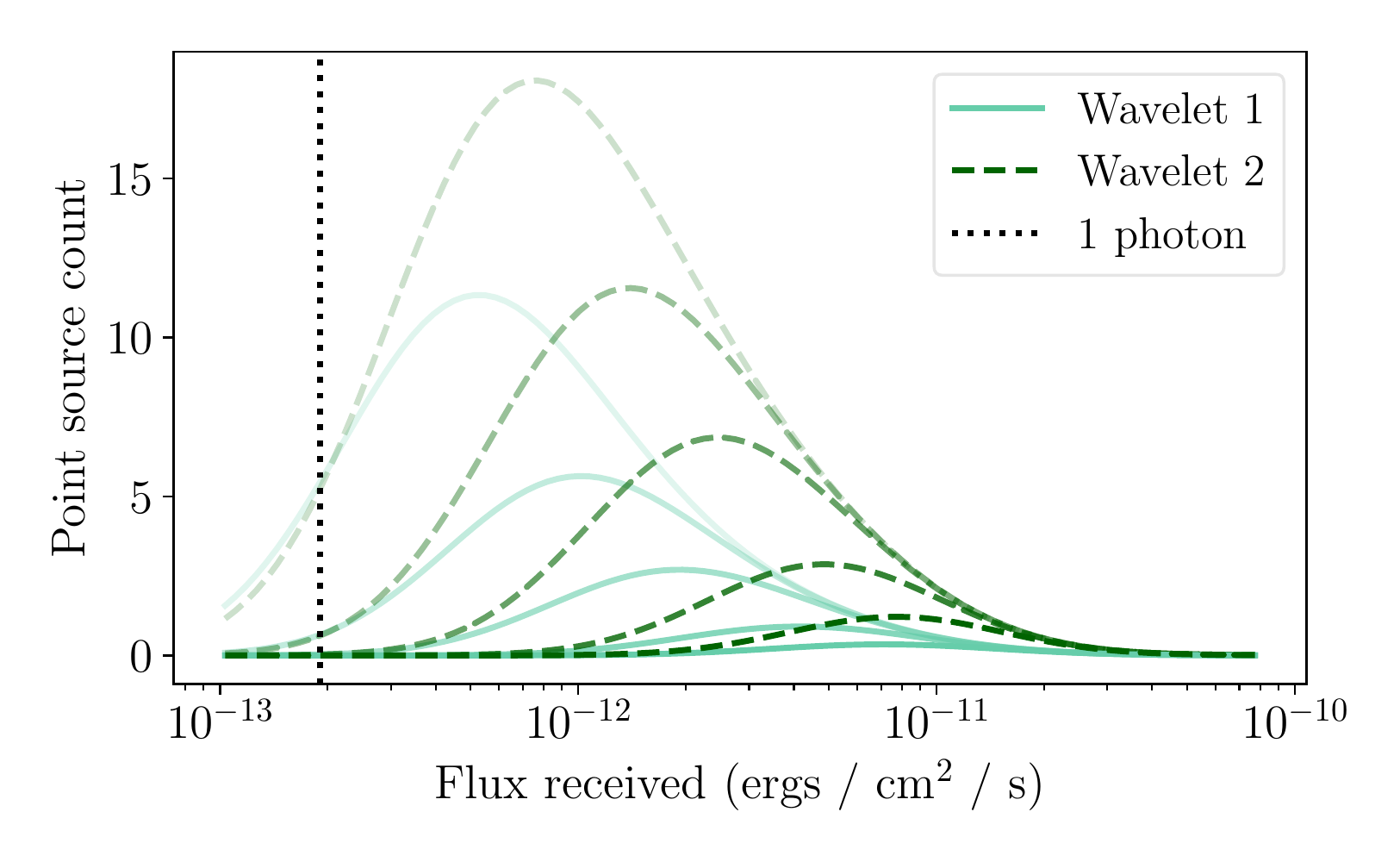}
        \caption{Power law luminosity function}
    \end{subfigure}
    \hfill
    \begin{subfigure}[b]{0.49\textwidth}
        \includegraphics[width=\textwidth]{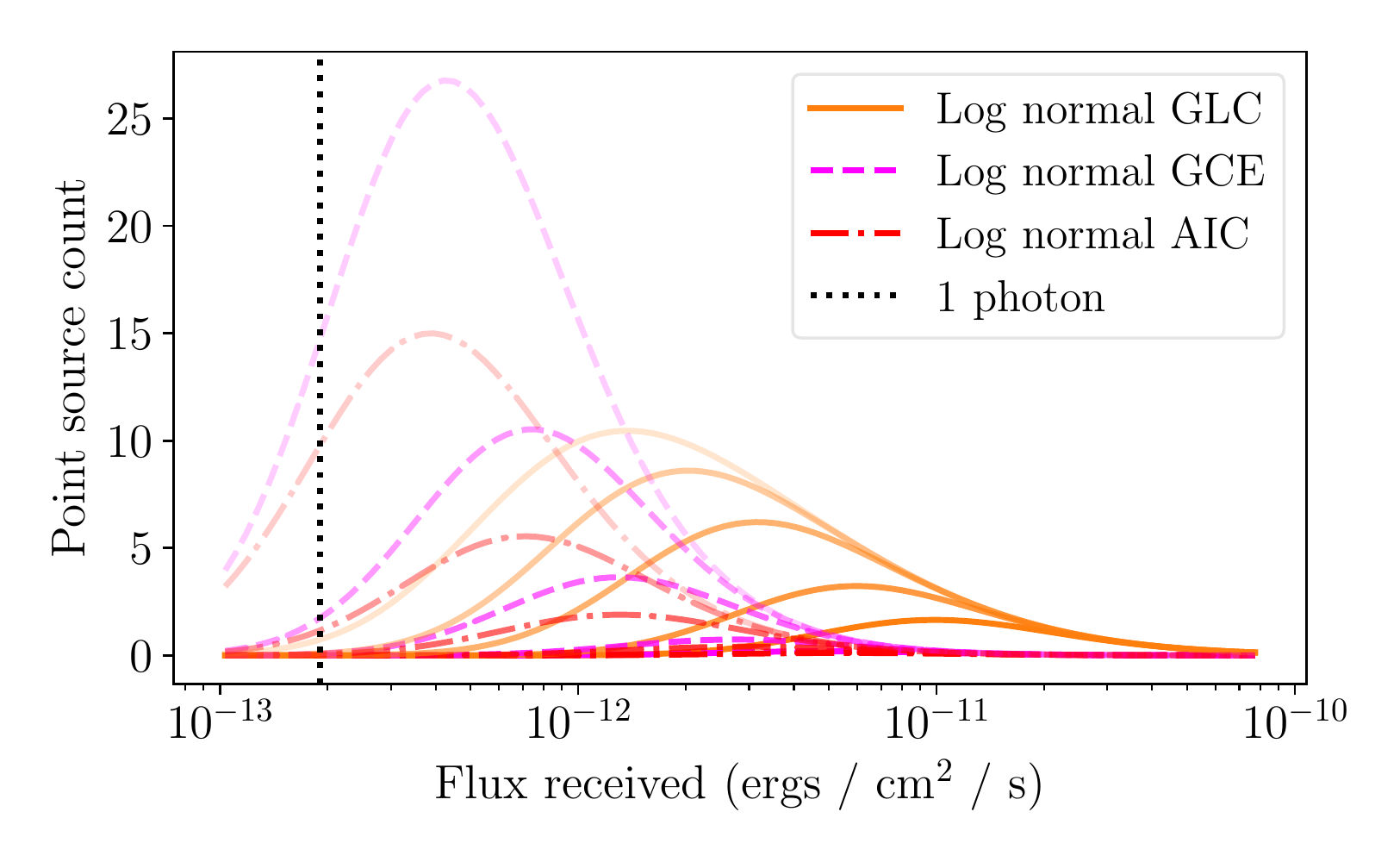}
        \caption{Log normal luminosity functions }
    \end{subfigure}
    \hfill
    \begin{subfigure}[b]{0.49\textwidth}
        \includegraphics[width=\textwidth]{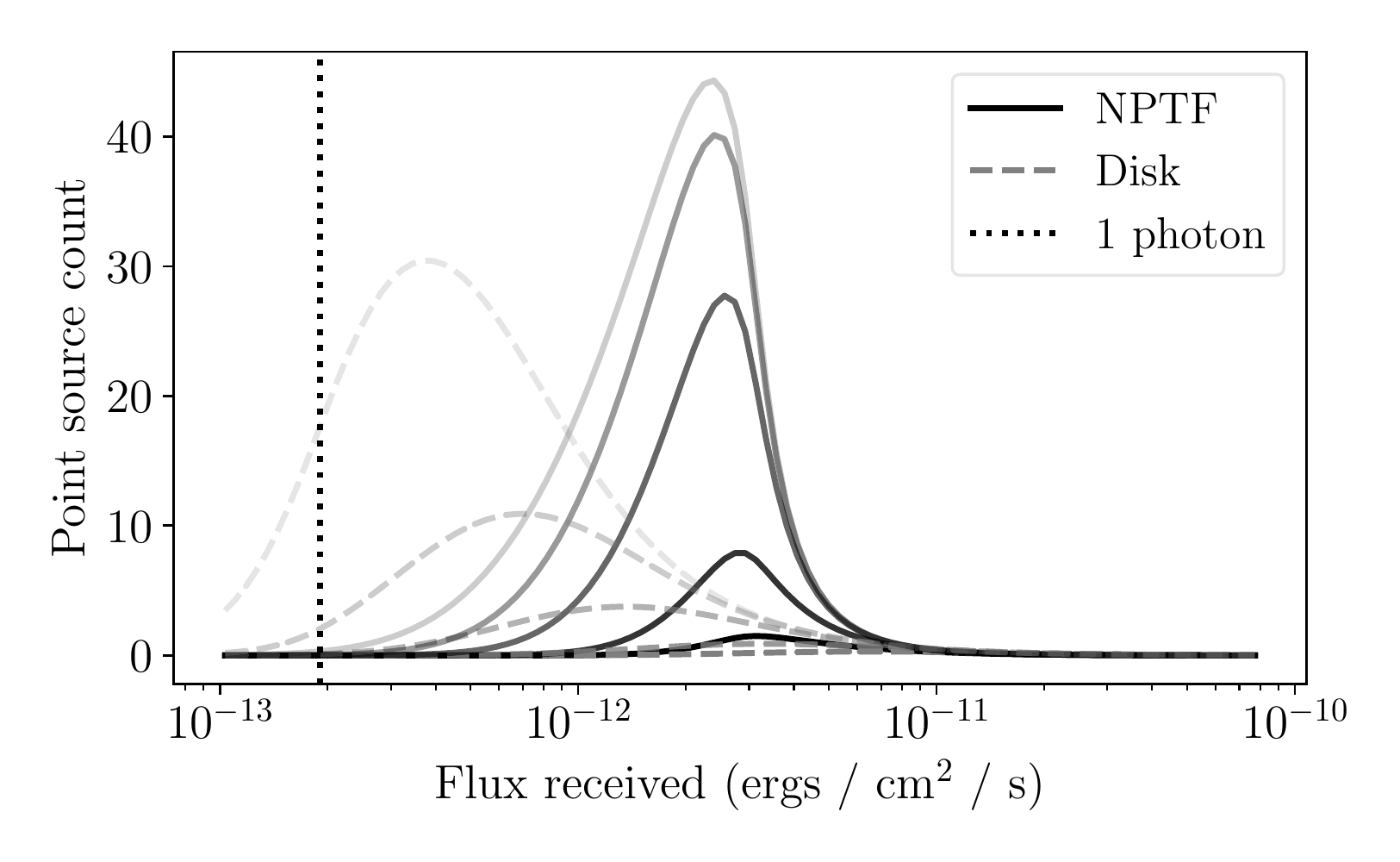}
        \caption{NPTF luminosity function}
    \end{subfigure}
    \caption{Histogram of the predicted number of resolvable pulsars required by each luminosity function benchmark to reproduce the GCE flux, for different models of \textit{Fermi} sensitivity. Darker lines indicate the current sensitivity; lighter lines represent increases in sensitivity by factors of two, five, ten, and twenty. See text for details. The bins are spaced evenly in log flux space, with width $\Delta F / F = 0.069$.}
    \label{fig:sensitivity-results-smoothing}
\end{figure}

Figure \ref{fig:sensitivity-results-smoothing} is the analog of figure \ref{fig:sensitivity-results}, showing the flux distributions at different sensitivities of resolved MSPs, assuming different luminosity functions, using the smoothed sensitivity model. Again, the histogram uses 100 flux bins distributed uniformly over the range $\num{1e-13}-\SI{8e-11}{\erg\per\second\per\centi\meter\squared}$

\begin{figure}
    \centering
    \includegraphics[height=0.8\textheight]{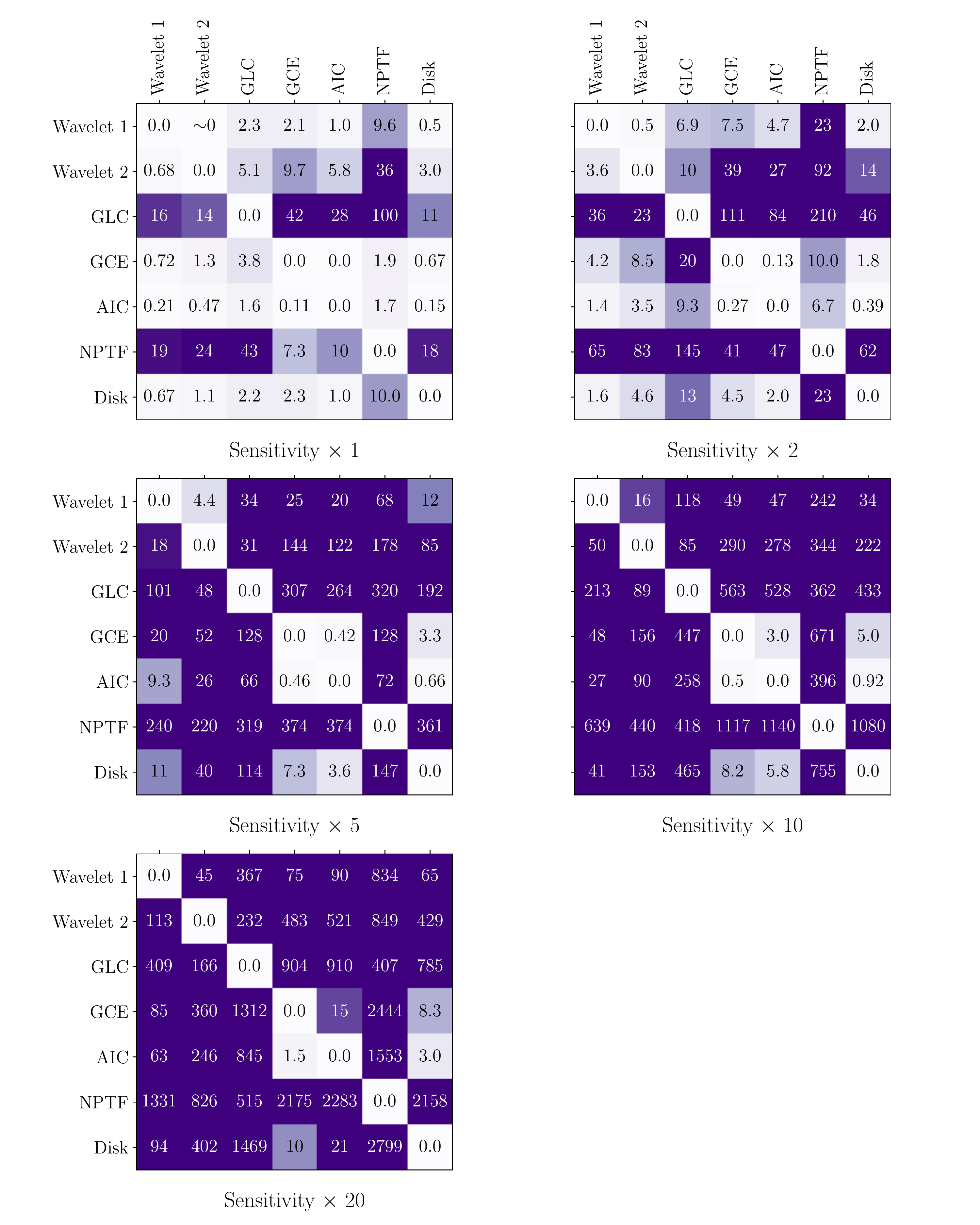}
    \caption{Map of the expected value of $\lambda$ resulting from assuming one luminosity function (horizontal axis) as the hypothesis while the other (vertical axis) is the true luminosity function. Each luminosity function is normalized to produce the same number of resolved PSs as the assumed-true luminosity function. Each plot represents a different multiplicative increase in the \textit{Fermi} sensitivity level. Diagonals indicate the fit of a luminosity function to itself. The smoothed sensitivity model is used.}
    \label{fig:hist-unbinned-sensitivity-smoothing}
\end{figure}

Figure \ref{fig:hist-unbinned-sensitivity-smoothing} is the analog of figure \ref{fig:hist-fitting-sensitivity}, showing $\lambda$ values (defined in eq.~\ref{eqn:lambda}) attained from fitting the distributions to each other. $N_r$ has been re-scaled to be equal to the $N_r$ predicted by the assumed-true luminosity function. The pairs of luminosity functions that were similar for figure \ref{fig:hist-fitting-sensitivity} are also similar for figure \ref{fig:hist-unbinned-sensitivity-smoothing}.

As expected, figure \ref{fig:sensitivity-results-smoothing} shows that the smoothed sensitivity model visibly smooths the low-flux falloff of the histograms of MSPs resolved. This smoothing nearly always makes flux distributions more similar to each other, as can be seen by comparing Figs.~\ref{fig:hist-fitting-sensitivity} and \ref{fig:hist-unbinned-sensitivity-smoothing} and noting which figure contains the higher value of $\lambda$ for each pair. This is likely due to the fact that the smoothed sensitivity model produces fewer MSPs than the standard model, thereby increasing the relative size of error bars on the smoothed model data.

\end{document}